\documentstyle[12pt]{article}
\documentstyle{epsf}

\newcommand{\ie}{{\em i.e.}}	
\newcommand{\eg}{{\em e.g.}}

\newcommand{\Eg}{{\em E.g.}}
\def\wt{\widetilde}
\def\dfrac#1#2{{\displaystyle{#1\over#2}}}
\def\erf{\mathop{\rm erf}\nolimits}
\def\sgn{\mathop{\rm sgn}\nolimits}
\def\tr{\mathop{\rm tr}\nolimits}
\def\eps{\epsilon}
\newcommand{\ourvec}[1]{{\mbox{\boldmath $#1$}}}
\def\ov{\ourvec}

\newcommand{\mtext}{\hbox}
\newcommand{\matr}[1]{{\mbox{\boldmath $#1$}}}
\def\defeq{\equiv}  
\def\const{\mtext{const}}
\def\simlt{\mathrel{\stackrel{{\displaystyle <}}{\sim}}}

\date{}

\begin{document}

\title{{\LARGE\sf\bf A Scaling Theory of Bifurcations in the Symmetric
Weak-Noise Escape Problem}}  
\author{
{\bf Robert S. Maier}\thanks{Partially supported by the 
National Science Foundation under grants NCR-90-16211 and DMS-95-00792.}\\
{\small \tt rsm\,@\,math.arizona.edu}\\
{\small \sl Dept.\ of Mathematics}\\
{\small \sl University of Arizona}\\
{\small \sl Tucson, AZ 85721, USA}
\and
{\bf Daniel L. Stein}\thanks{Partially supported by the 
U.S.~Department of Energy under contract DE-FG03-93ER25155.}\\
{\small \tt dls\,@\,physics.arizona.edu}\\
{\small \sl Dept.\ of Physics}\\
{\small \sl University of Arizona}\\
{\small \sl Tucson, AZ 85721, USA}
}
\maketitle
\begin{abstract}
We consider the overdamped limit of two-dimensional double well systems
perturbed by weak noise.  In~the weak noise limit the most probable
fluctuational path leading from either point attractor to the separatrix
(the~most probable escape path, or~MPEP) must terminate on the saddle
between the two wells.  However, as the parameters of a symmetric
double~well system are varied, a unique MPEP may bifurcate into two equally
likely MPEP's.  At~the bifurcation point in parameter space, the activation
kinetics of the system become non-Arrhenius.  We~quantify the
non-Arrhenius behavior of a system at the bifurcation point, by using the
Maslov-WKB method to construct an approximation to the quasistationary
probability distribution of the system that is valid in~a boundary layer
near the separatrix.  The approximation is a formal asymptotic solution of
the Smoluchowski equation.  Our analysis relies on the construction of a new
scaling theory, which yields `critical exponents' describing weak-noise
behavior at the bifurcation point, near the saddle.
\end{abstract}
\vfil

\bigskip
\noindent
{\em Archive Paper Number\/}: cond-mat/9506097
\eject

\section{ Introduction}
In this paper we build on previous work~\cite{MaierC,MaierB,MaierD} by
analysing an unusual bifurcation phenomenon in the theory of
noise-activated transitions.  We~study its appearance in the overdamped
limit of two-dimensional double well systems, with nongradient dynamics.
In~this context, the new phenomenon is a bifurcation of the most probable
transition path (in~the limit of weak noise) between the two wells, as~a
system parameter is varied.

In many ways, the behavior of a system whose most probable transition path
is just beginning to bifurcate resembles that of a system undergoing a {\em
phase transition\/}.  In~particular, double well systems which are
`at~criticality' in~the bifurcation sense will exhibit non-Arrhenius
behavior.  This means that the growth of the mean time between inter-well
fluctuations, \ie, the growth of the mean time needed for the system to
hop from one well to the other, will {\em not be pure exponential\/} in the
weak-noise limit.  In~double well systems at~criticality,
relaxation due to activation will proceed (in~the limit of weak noise) at
an anomalous, in~fact anomalously large,~rate.

To~treat the previously unnoticed phenomenon of bifurcation, we~need to
develop a new approach for treating transitions induced by weak noise, when
a `soft mode' appears in the dynamics of transverse fluctuations around the
most probable transition path.  Since this is analogous to a phase
transition, we~introduce a~{\em scaling theory\/}.  In~the context of
double well systems, our scaling theory is a theory of behavior near the
saddle point between the two wells, since the saddle is where the most
probable inter-well transition path begins to bifurcate.  We~shall
demonstrate that the theory explains the weak-noise behavior,
at~criticality, of a large universality class of double well systems.

The scaling theory will reveal a striking feature of the bifurcation
phenomenon, which is that in any `critical' double well system there
appears (in~the weak-noise limit) a~{\em nongeneric singularity\/} in the
stationary probability distribution, located at the saddle point.
As~Berry~\cite{Berry76} discusses, a~singularity is nongeneric if it
arises, in an appropriate WKB sense, from a catastrophe of unusual type;
\ie, one of {\em infinite codimension\/}.  The stationary distribution near
the saddle point is  described, in the limit of weak noise, by
an unusual (non-canonical) diffraction function.  The familiar special
functions of WKB theory (Airy functions, Pearcey functions,~etc.)\ do not
suffice.  The singularity at the saddle, and the diffraction function with
which it~is `clothed,' can be viewed as the mathematical source of the
non-Arrhenius weak-noise asymptotics.

We begin with three largely qualitative sections.  In
Section~\ref{sec:prelims} we review the physical relevance of overdamped
models with non-gradient dynamics, and in Section~\ref{sec:doublewell}
explain how the weak-noise behavior of any double well model of this type
is determined by its flow field of {\em instanton trajectories\/}
(most~probable fluctuational paths).
In~Section~\ref{sec:bifurcation} we sketch the gross features of the
bifurcation phenomenon, including features such as further bifurcations and
universality.  In~Section~\ref{sec:MAE}, our treatment becomes more
quantitative.  We~first review the matched asymptotic approximations technique
we~have employed elsewhere~\cite{MaierC,MaierB,MaierD}, and begin
extending~it to handle models with singularities.
In~Section~\ref{subsec:MAE25} we explain why the bifurcation transition
deserves to be called a phase transition.  In~particular, we explain how
behavior near criticality is described by {\em critical exponents\/}, which
characterize the rate of divergence of measurable quantities (\eg,~the
pre-exponential factor in the weak-noise asymptotics of the mean inter-well
fluctuation time).  In~Section~\ref{subsec:MAE3} we explain how to
determine whether any given double well model is `critical.'  The {\em
transverse Jacobi operator\/} is the differential operator appearing in the
second variation of the Onsager-Machlup action functional, when one varies
about the most probable inter-well transition path.  The onset of
bifurcation occurs when this operator acquires a zero eigenvalue.

In~Section~\ref{sec:Maslov} we explain the use we shall make of Maslov's
geometric theory of wave asymptotics~\cite{Berry76,Littlejohn92,Maslov81}.
In~Section~\ref{sec:real} we introduce the concept of a scaling theory, by
developing a scaling theory of weak-noise behavior near generic (cusp)
singularities.  We~show how the scaling theory justifies the
Ginzburg-Landau approximation used in this context by Dykman {\em
et~al.}~\cite{Dykman94}.  In~Section~\ref{sec:nascent} we develop an
analogous scaling theory for the nongeneric singularity associated with the
onset of bifurcation.  We~compare our theory with numerical data, and
examine its predictions for non-Arrhenius behavior and the stationary
distribution near the saddle point.  

In~Section~\ref{sec:discussion} we discuss our results.  The reader may
wish to glance ahead at Fig.~\ref{fig:arrhenius}, which is an Arrhenius
plot of the inter-well hopping rate of any double well system
at~criticality.  The non-Arrhenius behavior shown there, in~particular the
`logarithmic bend,' is the key result of this paper.

\section{{ Preliminaries}}
\label{sec:prelims}
Statistical physics and chemical physics include many examples of
stochastically perturbed dynamical systems.  It~is often the case that the
state of such a system is modelled as a particle moving in an
$n$-dimensional force field~$\ov F(\ov x)$, and subject to
additional random perturbations (`noise').  Since our interest is in the
modelling of nonequilibrium systems, we~shall not assume (as~is usually
done) that this force field is conservative.

If~the motion of the particle is isotropically damped, with damping
constant~$\gamma$, in~the absence of noise the particle position~$\ov
x$ would obey the deterministic equation
\begin{equation}
m\ddot\ov x + \gamma m\dot\ov x = \ov F(\ov x).
\end{equation}
Adding a random force $\ov F_{\mtext{\scriptsize random}}(t)$ yields the Langevin
equation
\begin{equation}
\label{eq:preLangevin}
m\ddot\ov x + \gamma m\dot\ov x = \ov F(\ov x) + \ov
F_{\mtext{\scriptsize random}}(t). 
\end{equation}
In physical problems $\ov F_{\mtext{\scriptsize random}}(t)$ is often modelled as
Gaussian white noise with amplitude $\sqrt{2\gamma mk_B T}$,
where $T$~is the ambient temperature.  In~this case the associated partial
differential equation, which describes the time evolution of the
probability density of~$\ov x$ and its velocity, is known as the
(forward) Fokker-Planck equation.

A case particularly important in applications is the {\em overdamped\/}, or
{\em inertialess\/} case, when $\gamma \gg {t_0^{-1}}$, for $t_0$~the
physical time scale.  In~this case the $m\ddot\ov x$ term
in~(\ref{eq:preLangevin}) can be dropped, and the Langevin equation becomes
first order in time.  If~time is rescaled by a factor~$\gamma m$
(\ie,~$t\leftarrow \gamma mt$), it~may be written in the normalized form
\begin{equation}
\label{eq:Langevin}
\dot\ov x = \ov u(\ov x) + \epsilon^{1/2}\dot\ov w(t).
\end{equation}
Here $\dot\ov w(t)$ is a standard $n$-dimensional Gaussian white noise
(the derivative of~$\ov w(t)$, a~standard $n$-dimensional Wiener
process), the `drift field'~$\ov u$ equals~$\ov F$, and
$\epsilon$~equals~${2k_B T}$.  The corresponding scalar advection-diffusion
equation for the probability density $\rho=\rho(\ov x,t)$ of~$\ov
x$,
\begin{equation}
\dot \rho = (\epsilon/2)\nabla^2\rho - \nabla \cdot (\rho\ov
u),
\end{equation}
is known as the (forward) Smoluchowski equation.  It~may be written as
$\dot \rho = - {\cal L}^*\rho$, where
\begin{equation}
{\cal L}^* \defeq -(\epsilon/2)\nabla^2 + \ov u \cdot \nabla +
\nabla\cdot\ov u.
\end{equation}
It~is often necessary to generalize the equation to include the effects of
anisotropic damping~\cite{Klosek89}, or state-dependent
noise~\cite{Landauer88}.  However, in~this paper we consider only
overdamped systems whose Langevin equation is of the
form~(\ref{eq:Langevin}).  Since we do~not require the deterministic forces
to be conservative, {\em we~do not require~$\ov u$ to be a gradient
field\/}.  This means that even in stationarity, the system may not display
detailed balance.  Equivalently, the stationary probability distribution for
the system may not be (in~the traditional sense) in thermal equilibrium.

Attractors of the drift field~$\ov u$, in~particular point attractors,
correspond to `metastable states': they are stable states of the underlying
deterministic dynamics, but the thermal noise may induce transitions
between them.  Of~great physical interest is the time needed for this to
occur.  For example, how long does it~take for the noise
in~(\ref{eq:Langevin}) to overcome the drift toward a specified stable
point~$S$, and drive the system state~$\ov x$ beyond the domain of
attraction of~$S$, toward another attractor?  The study of such
noise-activated transitions is known as the {\em stochastic exit
problem\/}, or the {\em escape problem\/}.  For general stochastic models
only numerical results can be obtained (see,~\eg, Ref.~\cite{Carmeli91}).
The Smoluchowski equation is particularly difficult to handle in the
$\epsilon\to0$ limit.  This is the weak-noise, or low-temperature limit,
in~which the mean first passage time (MFPT)~$\langle\tau\rangle$ from
$S$~to the boundary of its domain of attraction grows exponentially.
In~this limit a single escape path (the~most probable escape path, or~MPEP)
usually dominates.  Our approach to the weak-noise limit, which does not
rely on a numerical simulation of the Smoluchowski equation, will exploit
this asymptotic determinism quite heavily.

\section{{ Symmetric Double Well Models}}
\label{sec:doublewell}
\begin{figure}
\begin{center}
\epsfxsize=3.5in		
\leavevmode\epsfbox{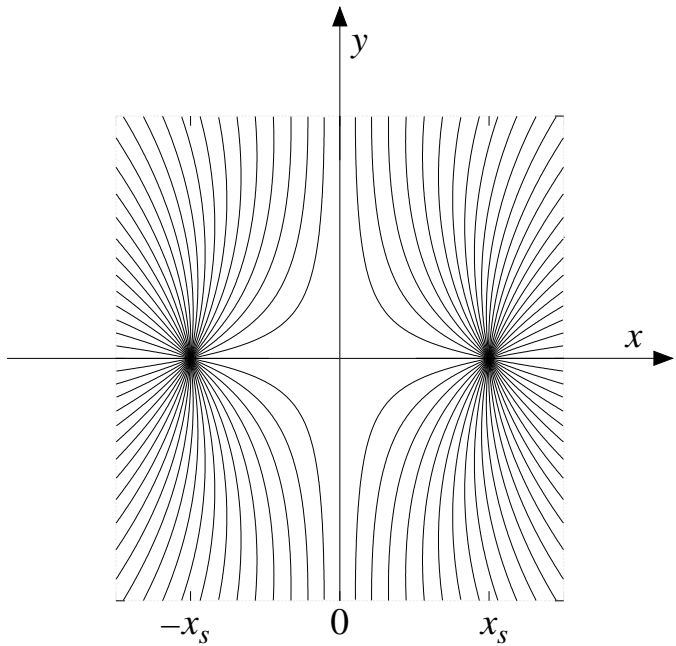}
\end{center}
\caption[The streamlines of a typical symmetric double well drift field.]
{The streamlines of a typical symmetric double well drift field,
indicating the path taken by the particle in the absence of noise.  There
are point attractors at $S=(x_s,0)$ and $S'=(-x_s,0)$, and a saddle point
at~$(0,0)$.}
\label{fig:drift}
\end{figure}

As in two of our earlier papers on the stochastic exit
problem~\cite{MaierC,MaierB}, we~shall focus on two-dimensional `double
well' systems, with smooth drift field $\ov u=(u_x,u_y)$ of the symmetric
form shown in Fig.~\ref{fig:drift}.  If~$\ov x=(x,y)$ is the
two-dimensional state variable, $u_x(x,y)$ is taken to be odd in~$x$ and
even in~$y$, while for $u_y(x,y)$ the reverse is true.  There is assumed to
be a linearly stable point attractor $S=(x_s,0)$ whose domain of attraction
is the entire open right-half plane.  By~symmetry, its reflection
$S'=(-x_s,0)$ attracts the open left-half plane.  There is also assumed to
be a single {\em saddle\/}, or hyperbolic point, on the $y$-axis separatrix
between the two domains of attraction.  It~must be at the origin,
by~symmetry.  Nongradient drift fields with this topology arise in
statistical and chemical physics, and also in theoretical biology, \eg,~in
stochastic competition models of population dynamics~\cite{Mangel93}.

One expects that as~$\epsilon\to0$, exit from either of the two domains of
attraction will occur preferentially over the saddle.  The drift field~$\ov
u$ is assumed to have a nondegenerate linearization at the saddle.
So~$\lambda_x=\partial u_x/\partial x(0,0) >0$, and~$\lambda_y = \partial
u_y/\partial y(0,0) <0$.  We~shall see that the character of the
abovementioned bifurcation phenomenon depends strongly on the
quotient~$\mu\defeq|\lambda_y|/\lambda_x$.

A typical (and not necessarily gradient) symmetric double well drift field,
which we have used elsewhere for purposes of illustration and shall examine
further below, is 
\begin{eqnarray}
u_x(x,y)&=&x-x^3-\alpha xy^2,\nonumber\\
u_y(x,y)&=&-\mu (1+x^2)y,
\label{eq:standard}
\end{eqnarray}
in which $\mu$~appears as a parameter.  We~shall call this drift field the
`standard' double well model.  For any choice of~$\mu>0$, its~structure is
that of Fig.~\ref{fig:drift}, with $S=(x_s,0)=(1,0)$.  It~is not a gradient
field unless the parameter $\alpha$~equals~$\mu$.  If~$\alpha>0$ it~has a
very significant additional property, which we shall require of all our
double well models.  This is the property that $\partial^2 u_x/\partial
y^2(x,0)$, which by symmetry is an odd function of~$x$, is strictly
negative for all~$x$ between $0$~and~$x_s$.  If~this is the case, the drift
from the saddle toward~$S$ `softens' as one moves away from the $x$-axis.
The off-axis softening, for the standard model, increases as $\alpha$~is
increased.

We remind the reader of our approach to the weak-noise limit of
stochastically perturbed dynamical systems.  (We~review the mathematical
aspects in Section~\ref{subsec:MAE1} and~\ref{subsec:MAE2}.)  Suppose that
such a system has a unique stationary probability density~$\rho_0$, which
satisfies the time-independent Smoluchowsk equation ${\cal L}^*\rho_0=0$.
Typically, as the noise strength~$\epsilon\to0$, $\rho_0$~takes~on an
asymptotic WKB~form.  In~fact
\begin{equation}
\label{eq:WKB}
\rho_0(\ov x) \sim K(\ov x)\exp [-W(\ov x)/\epsilon], \quad
\epsilon\to0,
\end{equation}
for certain functions $W$ and~$K$ whose smoothness properties we shall
leave unspecified; $K$,~in particular, may have singularities.  In~any
double well model, by~convention $W=0$ and~$K=1$ at~$\ov x=S$ and~$S'$.
Moreover, $W>0$ at~all points~$\ov x$ other than $S$ and~$S'$.  $W$~is
called the {\em nonequilibrium potential\/} of the model~\cite{Graham89}.
If~the drift~$\ov u$ equalled the negative gradient of a potential~$\Phi$,
then $W$~would equal~$2\Phi$, $K$~would reduce to a constant, and the WKB
form~(\ref{eq:WKB}) would reduce to a Maxwell-Boltzmann distribution.  For
systems with nongradient dynamics, the computation of $W$~and~$K$ is more
complicated.

In~general $W$~has an alternative interpretation as a {\em classical action
function\/}.  As~we review in Section~\ref{subsec:MAE1}, this is because
the WKB approximation~(\ref{eq:WKB}) is determined by a flow field of
`classical' trajectories, or WKB characteristics, emanating from the
attractors of the deterministic dynamics (\eg, $S$~and~$S'$).  These
classical trajectories (sometimes called {\em instanton\/}
trajectories~\cite{McKane1,MaierC}, or {\em optimal\/}
trajectories~\cite{Dykman94}) have a physical interpretation as {\em most
probable fluctuational paths\/}.  In~the double well case, the
exponentially rare fluctuations from~$S$ (resp.~$S'$) to any 
point~$\ov x$ in its domain of attraction become increasingly
concentrated around the classical trajectory extending from~$S$
(resp.~$S'$) to~$\ov x$.  Equivalently, the most probable `prehistory'
of any fluctuation passing through~$\ov x$ extends back toward
$S$~or~$S'$ along this trajectory~\cite{Dykman92}.  The trajectories are
determined by a classical Lagrangian (the Onsager-Machlup Lagrangian), and
$W(\ov x)$~is obtained by integrating this Lagrangian along the
classical trajectory terminating at~$\ov x$.  $W(\ov x)$~is
interpreted as the rate at~which fluctuations to the neighborhood
of~$\ov x$ are suppressed exponentially, as~$\epsilon\to0$.

In symmetric double well models the stationary density~$\rho_0$ (and
hence~$W$) must be even in~$x$.  In~the $\epsilon\to0$ limit the phenomenon
of noise-activated hopping between the two wells is governed by the closely
related {\em quasistationary\/} density~$\rho_1$, which is odd rather than
even.  The quasistationary density is the next lowest lying (\ie,~slowest
decaying) eigenmode of the Smoluchowski operator~${\cal L}^*$.  For any
choice of initial conditions the probability density $\rho=\rho(\ov x,t)$
necessarily satisfies
\begin{equation}
\rho(\ov x,t) \sim \rho_0(\ov x) + 
C\rho_1(\ov x) \exp(-\lambda_1 t), \quad t\to\infty
\end{equation}
for some constant~$C$, where $\lambda_1$~is the eigenvalue of~$\rho_1$.
The exponential decay of the quasistationary eigenmode is interpreted as
describing the equilibration of probability between the two wells due to
noise-activated hopping, or the absorption of probability on the
separatrix~\cite{Carmeli91}.  $\rho_1$~of~course satisfies Dirichlet
(absorbing) boundary conditions on the separatrix.  Its~eigenvalue
${\lambda_1=\lambda_1(\epsilon)}$ normally falls to zero exponentially
as~$\eps\to0$.  The exponentially small splitting between the ground state
eigenvalue $\lambda_0\equiv0$ and the eigenvalue~$\lambda_1$ is analogous
to the exponentially small splitting (as~$\hbar\to0$) between the ground
state and first excited state of a quantum-mechanical Hamiltonian with
double well potential.  Both are WKB phenomena.  In~the $\eps\to0$ limit,
$\lambda_1$~is interpreted as the rate at~which noise-activated hopping
takes place.  Equivalently, it~is a {\em reciprocal MFPT\/}.

The techniques reviewed in Section~\ref{subsec:MAE1} and~\ref{subsec:MAE2}
permit a computation of the $\epsilon\to0$ asymptotics of the
eigenvalue~$\lambda_1$, and hence of the MFPT~$\langle\tau\rangle$, in most
symmetric double well models.  Our basic approach is similar to that of
Kramers~\cite{Kramers40}.  In~the limit of weak noise we~approximate
$\rho_1(x,y)$ by $\rho_0(x,y)\sgn(x)$ except in a `boundary layer' of width
$O(\epsilon^{1/2})$ near the $x=0$ separatrix, and compute~$\lambda_1$ as
the rate at~which probability is absorbed on the separatrix.  Performing
this computation requires the construction of a boundary layer
approximation to~$\rho_1$, valid near the saddle, and matching to the
`outer' approximations on either side~\cite{Caroli80}.  Normally, we~find
$\langle \tau\rangle\sim A\exp[+W(0,0)/\eps]$, where $A\propto
K(0,0)^{-1}$.  So~the asymptotic MFPT growth rate in the limit of weak
noise is simply $\Delta W = W(0,0) - W(S) = W(0,0)$, the height of the
`action barrier,' or activation barrier, between the two wells.  And~the
MFPT generally displays a pure exponential (Arrhenius) growth, with an
explicitly computable ($\eps$-independent) prefactor.  We~shall see,
however, that the bifurcation phenomenon may induce more complicated
(non-Arrhenius) weak-noise asymptotics for the MFPT.

\section{{ The Bifurcation Phenomenon: Qualitative Features}}
\label{sec:bifurcation}
\begin{figure}
\vskip-0.75in
\vbox{
\hfil
\epsfxsize=2.6in		
\epsfbox{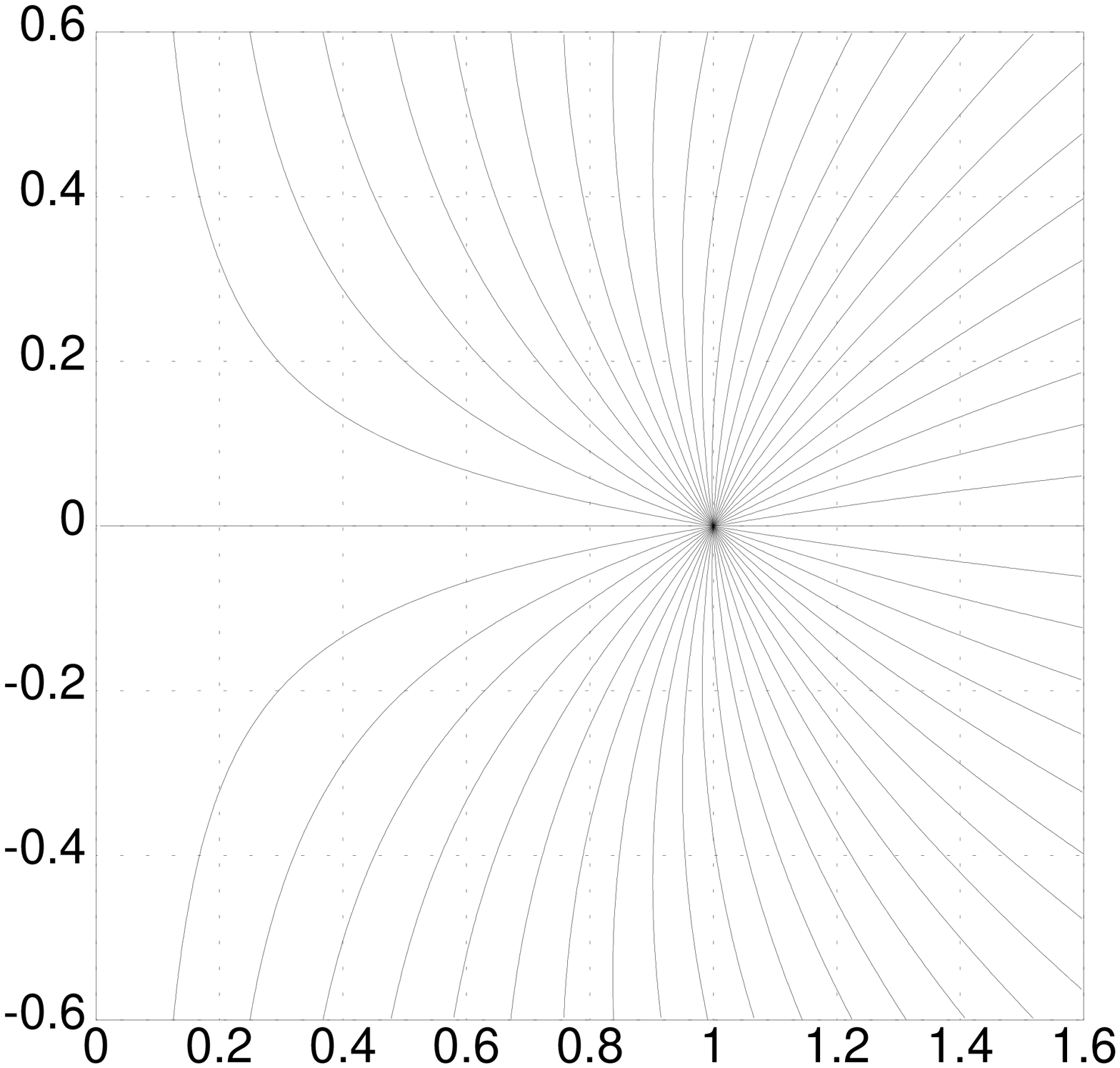}
\hfil
\epsfxsize=2.6in		
\epsfbox{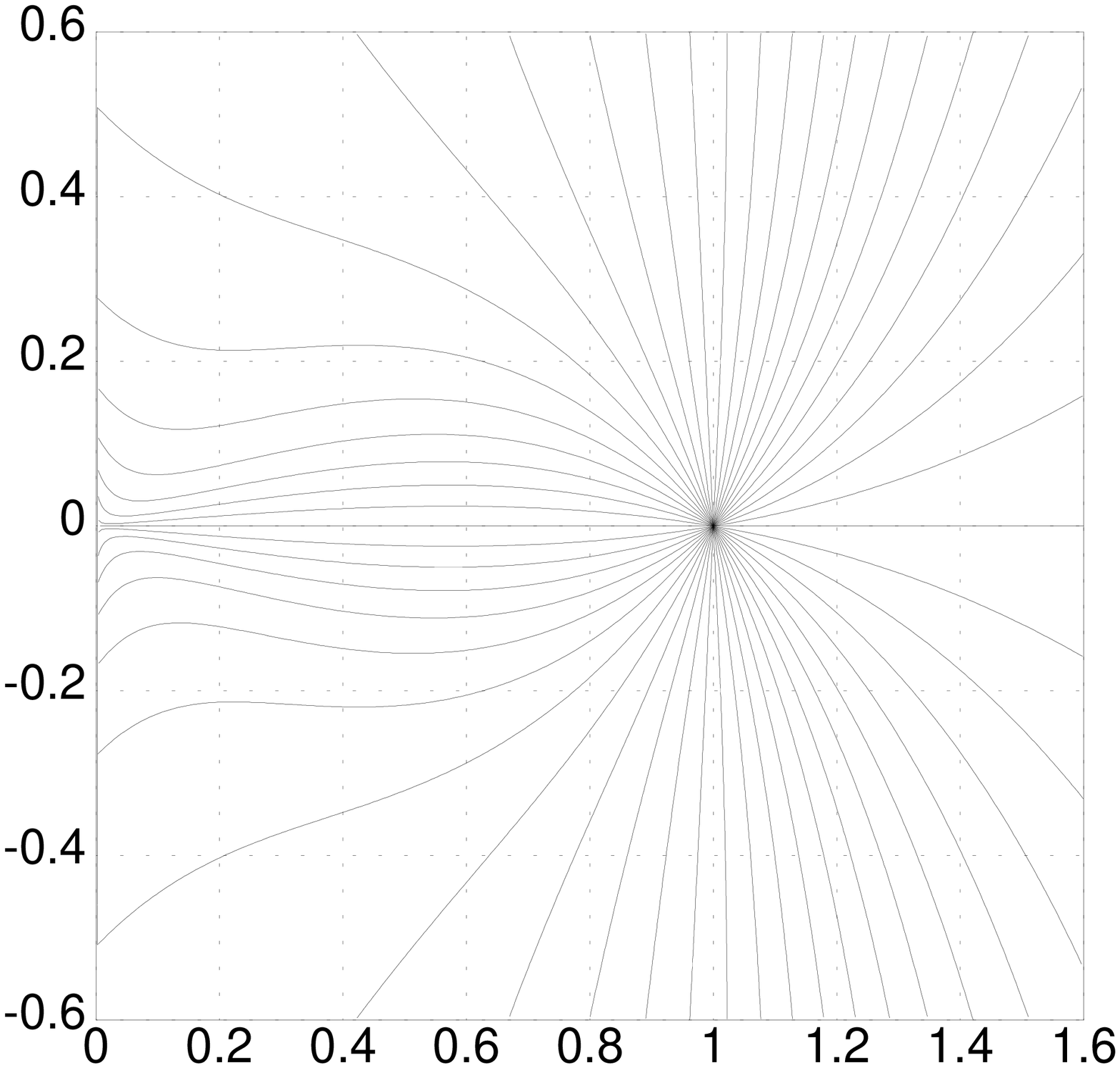}
}
\vskip0.5in
\vbox{
\hfil
\epsfxsize=2.6in		
\epsfbox{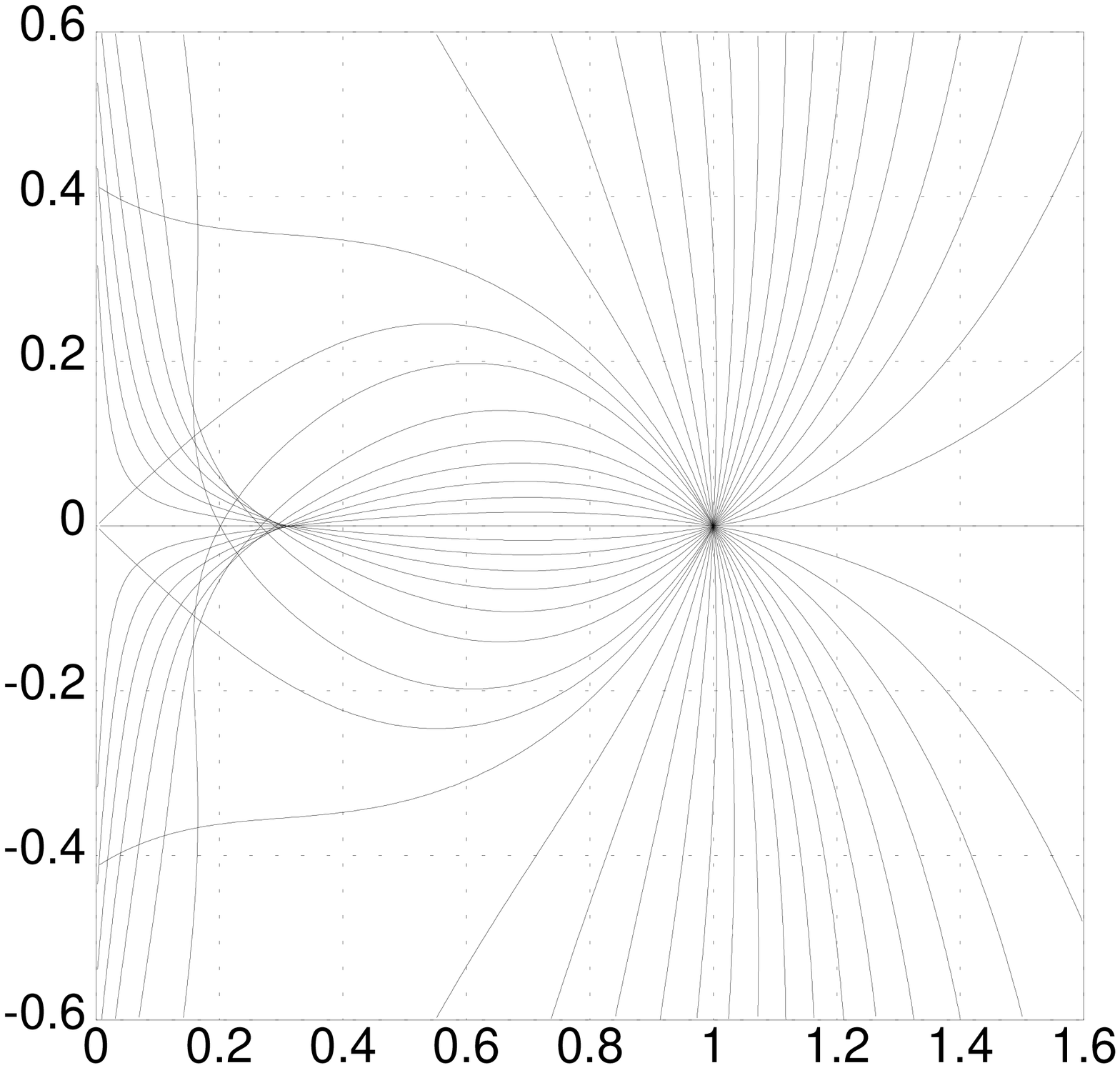}
\hfil
\epsfxsize=2.6in		
\epsfbox{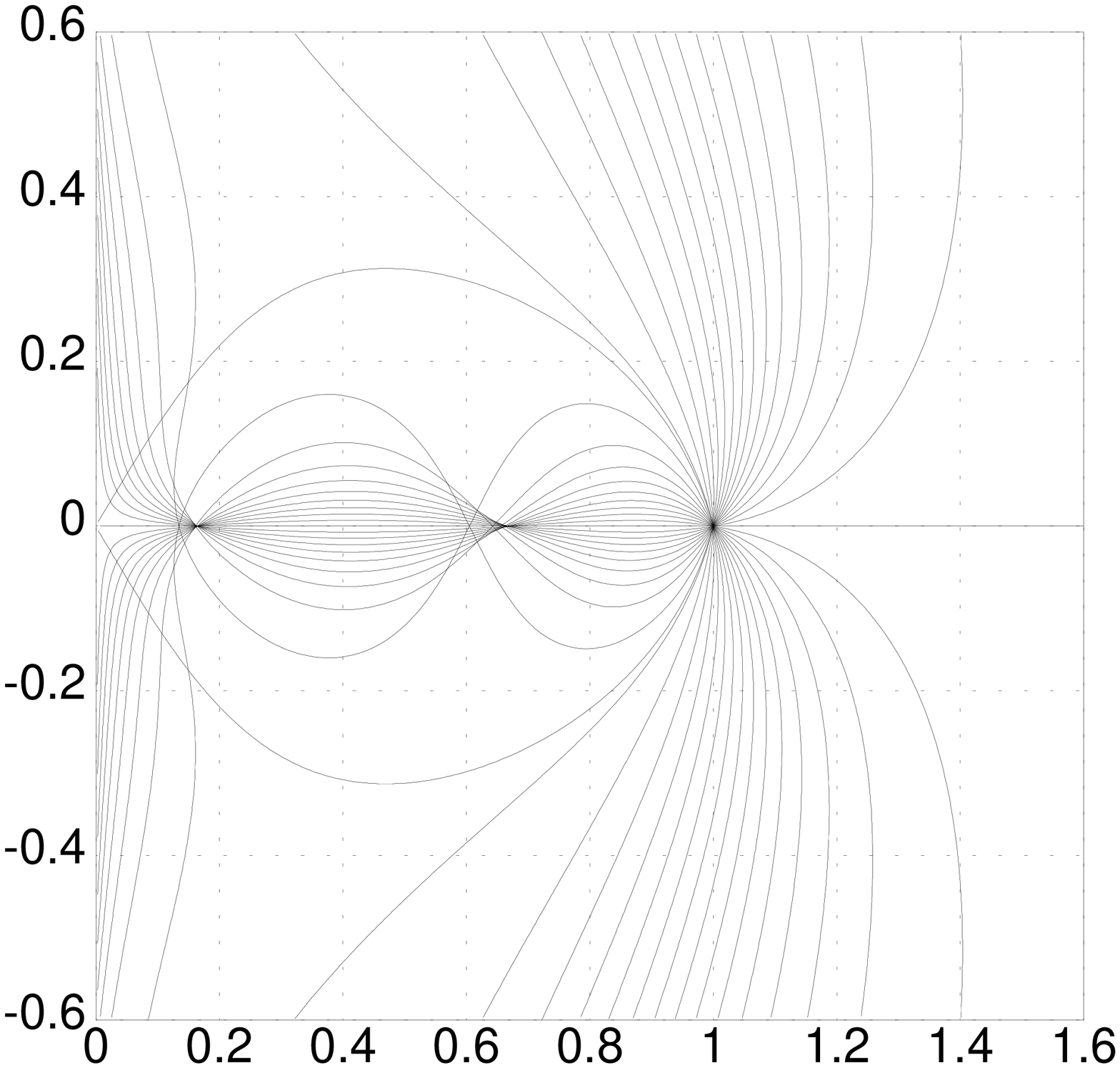}
}
\vskip0.25in
\caption[The flow field of outgoing instanton trajectories emanating from
the stable point~$S$ of the standard double well model.]{The flow field of
outgoing instanton trajectories (\ie,~most probable fluctuational paths, in
the weak-noise limit) emanating from the stable point~$S=(1,0)$ of the
standard double well model~(\ref{eq:standard}).  Here~$\mu=1$, and parts
(a), (b), (c),~(d) of~the figure illustrate the cases $\alpha=1,4,5,10$.
The $\alpha=4$ model is `critical' in the bifurcation sense.  Increasing
$\alpha$ above~$4$ causes the instanton trajectories to focus, and the MPEP
to bifurcate.}
\label{fig:flow}
\end{figure}

We pointed~out in Ref.~\cite{MaierC} that a bifurcation phenomenon
may occur in double well models as their parameters are varied.  
Figure~\ref{fig:flow}
displays the flow of instanton trajectories (\ie,~most probable
weak-noise fluctuational paths) emanating from the stable point $S=(1,0)$ in the
standard model~(\ref{eq:standard}) with~$\mu=1$, at several values of the
parameter~$\alpha$.  When ${0<\alpha<4}$ the general picture resembles
Fig.~\ref{fig:flow}(a): the line segment from~$(1,0)$ to the saddle~$(0,0)$ is 
the only
instanton trajectory from $S$ to the saddle.  This line segment is
interpreted as the most probable escape path~(MPEP).  In~the weak-noise
limit, the (exponentially rare) fluctuations from the right-half plane to
the left-half plane proceed preferentially along~it.
To~leading order, activation kinetics reduce to instanton dynamics.

As $\alpha$ is increased, there is a qualitative change, akin to a phase
transition, in the behavior of the instanton trajectories.  This takes
place at the critical value~$\alpha=4$, as~shown in Fig.~\ref{fig:flow}(b).
When~$\alpha>4$ as in Fig.~\ref{fig:flow}(c), they {\em focus\/} at a
point~$(x_f,0)$ on the x-axis, with~$x_f>0$.  $x_f$~converges to zero
as~$\alpha\to4^+$, so one may speak~of the focal point `being born'
at~criticality, and `emerging from the saddle' as $\alpha$~is increased
above its critical value.  In~geometrical optics the focal point would be
called a~{\em cusp\/}.  From~it there extends a {\em fold\/}, or~{\em
caustic\/} (an~envelope of crossing trajectories, with $\Delta y \propto
(\Delta x)^{3/2}$).

Each point in the sharp-tipped region within the fold is reachable from~$S$
via {\em three\/} instanton trajectories.  Of~these three trajectories,
only the one(s) with minimum action are `physical,' and can be interpreted
as most probable fluctuational paths.  For~example, on-axis points $(x,0)$
with $0\le x<x_f$ are reachable via an on-axis (straight) trajectory, and
via two additional symmetrically placed off-axis (curved) trajectories.
Computation shows that the off-axis trajectories have lesser action, and
are dominant.  The true (`least action') MPEP's in Fig.~\ref{fig:flow}(c) are
accordingly the symmetrically placed pair of off-axis trajectories, one
above and one below the $x$-axis, that terminate on the saddle.  Note that
beyond the cusp (\ie,~at $x<x_f$), the physical action~$W$ is no~longer
differentiable through the $x$-axis.  This nondifferentiability arises from
different dominant off-axis trajectories being selected as~$y\to0^+$
and~$y\to0^-$.

The transition at~$\alpha=4$ can be interpreted as a {\em bifurcation of
the MPEP\/}, corresponding to a sort of symmetry breaking.  At~larger
values of~$\alpha$, the drift field~$\ov u$ and the Langevin
equation~(\ref{eq:Langevin}) remain symmetric about the $x$-axis, but each
of the two MPEP's is~not.  The line segment from $S$~to the saddle,
formerly the (unique) MPEP\hbox{}, in no way contributes to the leading
weak-noise asymptotics for escape.  (It~remains an extremum of the
Onsager-Machlup action functional, but is no~longer the minimum.)

The occurrence of a bifurcation in the standard model at sufficiently
high~$\alpha$ (when~$\mu=1$, at~$\alpha=4$) is due to the fact that by
increasing~$\alpha$, one softens the resistance to motion toward the
separatrix in the vicinity of the $x$-axis (though not on the $x$-axis
itself).  This enhances the probability of escape trajectories that deviate
from the axis.  Of~course it~is only in the limit, as~$\eps\to0$, that
well-defined MPEP's appear.  And the existence of a sharp, well-defined
transition when $\alpha$~equals some critical value~$\alpha_c$ is not
at~all obvious!

When $\alpha$ is increased beyond~$\alpha_c$, further bifurcations of the
on-axis instanton trajectory will occur.  In Section~\ref{subsec:MAE3} we
explain how the critical values of~$\alpha$ are determined by a {\em Jacobi
equation\/}, with a classical mechanical interpretation.  It~turns out that
in the standard model with~$\mu=1$ the $j$'th bifurcation occurs at
$\alpha=\alpha_c^{(j)}=(j+1)^2$.  Figure~\ref{fig:flow}(d) shows the
situation at~$\alpha=10$, when a second focus $(x_f^{(2)},0)$ has emerged,
with its~own caustic.  Beyond the first focus $(x_f,0)$ each~point on the
$x$-axis is reached from~$S$ by {\em three\/} instanton trajectories;
beyond the second focus, each such point is reached by~{\em five}.  The
MPEP's in Fig.~\ref{fig:flow}(d), however, remain the symmetrically placed
pair of off-axis trajectories that terminate on the saddle.  Computation
shows that the oscillatory trajectories from~$S$ to the saddle arising from
the second, third,...\ bifurcations have higher actions, and are
accordingly not
physical.  

\begin{figure}
\begin{center}
\epsfxsize=3.0in		
\leavevmode\epsfbox{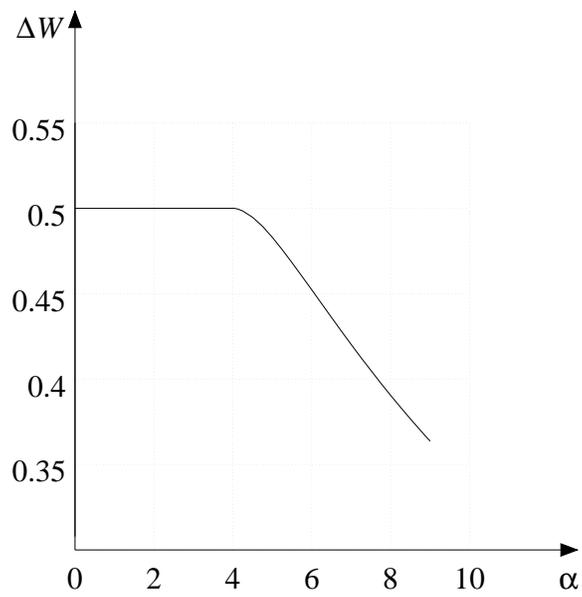}
\end{center}
\caption[The activation barrier between the two wells of the standard
double well model, as~a function of the off-axis softening parameter.]{The
activation barrier $\Delta W=W(0,0)-W(S)=W(0,0)$ between the two wells of
the standard double well model~(\ref{eq:standard}), as~a function of the
off-axis softening parameter~$\alpha$.  Here~$\mu=1$.  The lowering of the
activation barrier beyond $\alpha=\alpha_c=4$ is due~to the bifurcation of
the MPEP\hbox{}, along which the action difference~$\Delta W$ is computed.}
\label{fig:barrierheight}
\end{figure}

That caustics can occur in the flow pattern of the most probable
fluctuational paths has been known for some time~\cite{Day87,Jauslin87},
but our Ref.~\cite{MaierC} was the first to consider the effects on exit
phenomena.  We~shall see that what occurs at the first critical value
of~$\alpha$ has much in common with a critical point characterizing a {\em
phase transition\/} in a condensed matter system.  This is suggested by
Fig.~\ref{fig:barrierheight}, which plots the activation barrier ${\Delta
W=W(0,0)}$, as~determined by the true MPEP or MPEP's, as a function
of~$\alpha$ for the standard model with~$\mu=1$.  (Recall that ${\Delta
W=W(0,0)}$ is the exponential growth rate of the MFPT as the noise strength
tends to zero.)  $W(0,0)$ decreases above ${\alpha=\alpha_c=4}$ as the
bifurcating MPEP's move away from the $x$-axis.  The WKB prefactor $K(0,0)$
turns~out to be singular at the bifurcation transition; in
Section~\ref{subsec:MAE25} we note that it diverges
as~$\alpha\to\alpha_c^-$.  As~a consequence, in the standard model
at~least, {\em the weak-noise MFPT asymptotics at criticality cannot be of
a pure Arrhenius form\/}.

There is in fact a set of critical exponents describing the behavior of the
$\epsilon\to0$ asymptotics of the standard model as $\alpha$~tends to its
($\mu$-dependent) first critical value~$\alpha_c=\alpha_c^{(1)}$, and as
${x\to0}$,~${y\to0}$.  It~is a reasonable conjecture that behavior near
criticality is {\em universal\/} in the sense that it does not depend on
the details of the stochastic model exhibiting the bifurcation phenomenon.
To~analyse the critical behavior and demonstrate universality, in Sections
\ref{sec:real} and~\ref{sec:nascent} we~begin the construction of a {\em
scaling theory\/} of the bifurcation phenomenon.  Our treatment extends
from the standard model~(\ref{eq:standard}) to any symmetric double well
model with a similar `off-axis softening parameter'~$\alpha$, and a first
critical value~$\alpha_c$.  We~first identify the singular behavior, for
any double well model at criticality, of the action~$W$ and the WKB
prefactor~$K$ at the saddle point.  We~then show that at~criticality, the
stationary density~$\rho_0$ and the quasistationary density~$\rho_1$ may be
approximated on~an appropriate ($\epsilon$-dependent) length scale near the
saddle point by certain `diffraction functions,' which have explicit
integral representations.  The technique for constructing these
representations is due to Maslov~\cite{Maslov81}, and ultimately to
Keller~\cite{Keller60}.  It~was Maslov who first worked~out, in~the context
of wave fields, the diffraction functions that `clothe' generic
singularities other than cusps and folds.

A very important discovery, from a mathematical point of view, will be that
when $\alpha$~equals the critical value~$\alpha_c$ where the MPEP begins to
bifurcate, the saddle point~$(0,0)$ acquires a certain nonzero {\em
singularity index\/}.  What this means is best understood by comparing the
singularity at the saddle (when~$\alpha=\alpha_c$) with the cusp and fold
singularities present when~$\alpha>\alpha_c$.  The terminology of
geometrical optics~\cite{Berry76} is appropriate.  The cusp at~$(x_f,0)$ is
a structurally stable singularity (or~{\em catastrophe\/}, in the language
of Thom), with codimension~$2$.  The fold extending from~it, though not
`physical' in the above least-action sense, is a catastrophe of
codimension~$1$.  For points~$\ov x$ in the vicinity of the cusp, the
WKB approximation~(\ref{eq:WKB}) for the value of the stationary
density~$\rho_0(\ov x)$ breaks~down.  The proper treatment of points
near the cusp and the fold is similar to the short-wavelength treatment of
wave fields near caustics~\cite{Berry76,Dykman94}.  The cusp is said to have
singularity index~$1/4$, and points on the fold would (if~it were physical)
have singularity index~$1/6$.  This means that at~these singular points the
prefactor in the WKB approximation to~$\rho_0$, which formally diverges,
if~properly constructed would acquire a factor~$\eps^{-1/4}$
(resp.~$\eps^{-1/6})$.  There is a non-WKB (but~uniformly valid)
approximation to~$\rho_0(\ov x)$ in the vicinity of each such singular
point, in~terms of canonical diffraction functions.  We~shall re-derive
these facts in Section~\ref{sec:real}, in~terms of scaling functions.

We shall show in Section~\ref{sec:nascent} that the singularity index of the
point singularity appearing at the saddle, in critical models, depends in a
universal way on~$\mu$, \ie, on the ratio of the eigenvalues of the
linearization of the drift~$\ov u$ at the saddle.  It~turns out to
equal~${(\mu+1)/6}$.  Moreover, the approximations to $\rho_0$ and~$\rho_1$
near the saddle are given by {\em non-canonical\/} diffraction functions.
By~using the non-canonical approximation to~$\rho_1$ to compute the rate
at~which probability is absorbed on the separatrix we~shall quantify the
{\em universal non-Arrhenius behavior\/} of the weak-noise MFPT
asymptotics.  We~shall show that in symmetric double well models
at~criticality,
\begin{equation}
\langle\tau\rangle \sim \const\times \eps^{s} \exp[+\Delta W/\eps],
\quad \eps\to0,
\end{equation}
where $s=s(\mu)=(\mu+1)/6$ is the index of the singularity at the saddle.
We~shall also derive scaling corrections, in the weak-noise limit, to the
normal distribution of exit location points near the saddle.  The preceding
results will hold for all~$\mu$ satisfying $3/4<\mu<3$; the weak-noise
asymptotics of models with $\mu\le3/4$ and~$\mu\ge3$ are still under 
investigation.

\begin{figure}
\begin{center}
\epsfxsize=4in		
\leavevmode\epsfbox{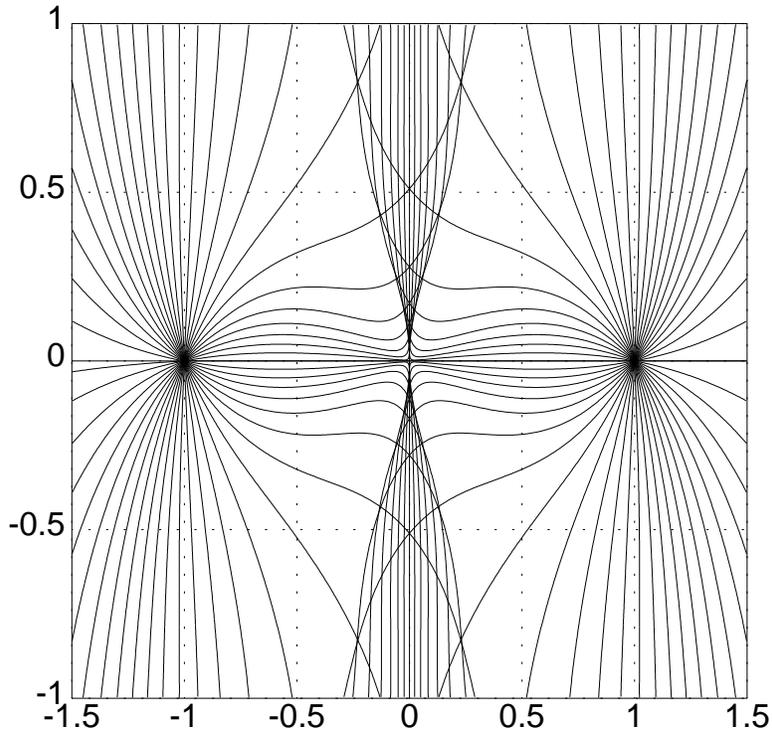}
\end{center}
\caption[The flow field of instanton trajectories emanating from both
stable fixed points, in~a critical version of the standard double well
model.]{The flow field of the instanton trajectories emanating from both
stable fixed points $S$~and~$S'$, in~a critical version of the standard
double well model~(\ref{eq:standard}).  This figure reveals that
at~criticality, a two-sided caustic extends sideways from the saddle point.
Although a universal phenomenon, this  caustic is nongeneric in
the sense of singularity theory.  Here $\mu=1$, and the parameter~$\alpha$
is set equal to the corresponding first critical value~$\alpha_c=4$, as~in
Fig.~\ref{fig:flow}(b).}
\label{fig:nascent}
\end{figure}

The point singularity appearing in critical models at the saddle point,
which may be termed a {\em nascent cusp\/}, is {\em nongeneric\/}.  It~is
not a member of the well known family of singularities that includes folds,
cusps, swallowtails,~etc.  This becomes clear if one plots the flow field
of the instanton trajectories emanating from both $S$~and~$S'$ in the
standard model at criticality ($\mu$~varying, and $\alpha$~set equal to its
$\mu$-dependent first critical value).  At~least when~$3/4<\mu<3/2$,
one~finds that at~criticality {\em a two-sided caustic extends
transversally from the saddle point itself\/}.  (Cf.~Dykman {\em
et~al.}~\cite{Dykman94}.)  Figure~\ref{fig:nascent}, which is an extended
version of Fig.~\ref{fig:flow}(b), shows the flow field when $\mu=1$ and
${\alpha=\alpha_c=4}$.  The caustic is clearly visible.  It~is not
`physical,' since it~is formed by high-action instanton trajectories that
have crossed the separatrix.  But the `nascent' cusp is clearly a cusp in
its own right, of~an unusual~sort.  Numerically one finds that the
two-sided caustic extending from~it is located at
\begin{equation}
|x|\simlt \const\times |y|^{(3/2-\mu)^{-1}}, \quad y\to0.
\end{equation}
A~conventional (generic) caustic would have an exponent of~$3/2$.  The
continuously varying exponent $(3/2-\mu)^{-1}$, which turns~out to be
universal and which we shall derive in Section~\ref{sec:nascent} from our
scaling theory, signals that the two-sided caustic is nongeneric.  The
nascent cusp from which it~extends is itself nongeneric in the sense of
singularity theory.

As~Berry~\cite{Berry76} has emphasized, nongeneric singularities arise from
{\em catastrophes of infinite codimension\/}.  It~is remarkable that a
singularity of such complexity is a universal feature of singly
parametrized symmetric double well models with non-gradient dynamics.

\section{{ Quantitative Semiclassical Asymptotics}}
We now begin a quantitative treatment of the weak-noise asymptotics for
escape.  We~first recast our earlier results in a form that facilitates the
analysis of singularities.  In~Section~\ref{subsec:MAE1} we discuss
geometric aspects of the WKB approximation, and
in~Section~\ref{subsec:MAE2} we discuss our matched asymptotic
approximations technique for computing MFPT asymptotics.
In~Section~\ref{subsec:MAE25} we use the standard model~(\ref{eq:standard})
to illustrate the nature of the nascent cusp appearing at the saddle at
criticality, and the ways in~which bifurcation can be viewed as a phase
transition.  In~Section~\ref{subsec:MAE3} we explore the bifurcation
phenomenon from a classical mechanical point of view, and relate~it to the
appearance of a {\em transverse soft mode\/}.  We~explain how its
appearance is governed by a Jacobi equation, and how this equation
determines whether or not a given double well model is at criticality.
\label{sec:MAE}
\subsection{{ The WKB Approximation and Classical Mechanics}}
\label{subsec:MAE1}
The time-independent forward
Smoluchowski equation ${\cal L}^*\rho_0=0$ may be written as
\begin{equation}
\label{eq:Heqn}
H({\ov x}, -\epsilon{\nabla})\rho_0=0
\end{equation}
where
\begin{equation}
\label{eq:Hamiltonian}
H(\ov x,\ov p) = {\ov p}^2/2 +\ov u(\ov x)\cdot\ov p
\end{equation}
is the so-called Wentzell-Freidlin Hamiltonian~\cite{VF}, whose dual is the
Onsager-Machlup Lagrangian~\cite{Onsager53}
\begin{equation}
\label{eq:Onsager}
L(\ov x,\dot\ov x)=|\dot\ov x-\ov u(\ov x)|^2/2.
\end{equation}
In equation~(\ref{eq:Heqn}) we have adopted an operator ordering convention
according to which the action of~$\nabla$ precedes that
of~$\ov x$.  

In the weak-noise ($\epsilon\to0$) limit the stationary density~$\rho_0$
and the quasistationary density~$\rho_1$ are given in the interior of each
well by a WKB, or semiclassical form.  A~full WKB expansion for~$\rho_0$
would be of the form
\begin{equation}
\rho_0(\vec x) \sim[K^{(0)}(\ov x) +\epsilon K^{(1)}(\ov x)
+\cdots] \exp[-W(\ov x)/\epsilon], \qquad \epsilon\to0,
\end{equation}
as in geometrical optics.  By~substituting this formal series
into~(\ref{eq:Heqn}) and examining the coefficients of each power
of~$\epsilon$, one obtains equations for $W$~and the~$K^{(m)}$.  That is
what we shall~do, though we shall work only to leading order: our~WKB
Ansatz will be $\rho_0(\ov x)\sim K(\ov x)\exp[-W(\ov
x)/\epsilon]$.  Notice that since the eigenvalue
$\lambda_1=\lambda_1(\epsilon)$ of~$\rho_1$ is exponentially small
as~$\epsilon\to0$, the asymptotic expansions (in~powers of~$\epsilon$) for
$\rho_1$~and~$\rho_0$ will be the same.  To~see the difference between
them, which is significant only near the separatrix between the two wells,
one would have to go `beyond all orders' in the WKB expansion.

The eikonal equation for~$W$ is the time-independent Hamilton-Jacobi equation
\begin{equation}
H(\ov x,\nabla W)=0
\end{equation}
so that $W$ is a {\em classical action at zero energy\/}.  For any
point~$\ov x$ in either well, it~may be computed by integrating the
Lagrangian along the zero-energy classical trajectory extending from~$S$
(resp.~$S'$) to~$\ov x$.  Each such trajectory, which satisfies the
Euler-Lagrange equations, is interpreted as a most probable fluctuational
path in the $\epsilon\to0$ limit.  These trajectories are the `instanton
trajectories' of the last section; the~term is justified by analogy with
the semiclassical limit in quantum mechanics and quantum field
theory~\cite{Auerbach85}.  In~the language of
Gutzwiller~\cite{Gutzwiller90}, the points at~which the instanton
trajectories focus would be called {\em zero-energy conjugate points.}

It is convenient to work in the Hamiltonian picture, according to which the
classical trajectories of~interest lie~on a zero-energy surface in a
nonphysical phase space, coordinatized by position~$\ov x$ and
momentum~$\ov p$.  The flow in this phase space ($2n$-dimensional, if
configuration space is $n$-dimensional) is determined by Hamilton's
equations and the Wentzell-Freidlin Hamiltonian.  From this point of view
the instanton trajectories of Figs.~\ref{fig:flow} and~\ref{fig:nascent}
are mere images of phase-space trajectories, projected `down' to
configuration space by the map $(\ov x,\ov p)\mapsto\ov x$.
The phase-space trajectories emanate from~$(S,\ov 0)$
(resp.~$(S',\ov 0)$).  In~WKB theory the projected trajectories are
traditionally called {\em characteristics\/}, and the phase space
trajectories {\em bicharacteristics\/}.  Characteristics may intersect, as
in Figs.~\ref{fig:flow}(c) and~\ref{fig:flow}(d), but bicharacteristics may
not.

It is easy to verify, using Hamilton's equations, that $(S,\ov 0)$
and~$(S',\ov0)$ are hyperbolic fixed points of the Hamiltonian flow.
And~the unstable manifold of~$(S,\ov 0)$, for~example, comprises all
points $(\ov x,\ov p)$ that lie on one of the bicharacteristics
emanating from~$(S,\ov 0)$.  The unstable manifolds of~$(S,\ov 0)$
and~$(S',\ov 0)$ are {\em Lagrangian\/}~\cite{Littlejohn92}: they are
invariant under the Hamiltonian flow.  By~the term `Lagrangian manifold' we
shall refer to either of these two unstable manifolds, or their union.
We~denote by~${\cal M}$ this union, \ie, the set of all points~$(\ov
x,\ov p)$ that lie on a bicharacteristic emanating from either
$(S,\ov 0)$ or~$(S',\ov 0)$.
If~configuration space is $n$-dimensional, ${\cal M}$ will be an
$n$-dimensional manifold.

Each point ${{P}}=(\ov
x,\ov p)$ on~${\cal M}$ has a value for the zero-energy action~$W$
associated with~it, computed by
\begin{equation}
W({{P}}) = \int \ov p\cdot d\ov x,
\end{equation}
the line integral being taken along the bicharacteristic terminating
at~${P}$.  If~due to intersecting characteristics, or the crossing of
characteristics from one well to the other, there are several manifold
points ${{P}}_i=(\ov x,\ov p^{(i)})$ `above' some
point~$\ov x$, then $W(\ov x)$~and its gradient~$\ov p=\ov
p(\ov x)$ will in a mathematical sense be multivalued.  As~a function
of~$\ov x$, $W$~may in~fact have branch points, branch lines
(cuts),~etc.  But the {\em physical\/} action~$W(\ov x)$ appearing in
the WKB approximation will be single-valued: it~will equal the minimum of
the values~$W({{P}}_i)$ at the manifold points above~$\ov x$.
This `least action' computation determines which instanton trajectories are
physical. 

The WKB prefactor~$K$ satisfies an easily derived transport equation.
(Cf.~Talkner~\cite{Talkner87}.)  If~one uses the fact that $\dot\ov x=\ov
p+\ov u(\ov x)$ (which is one of Hamilton's equations), the transport
equation takes~on the comparatively simple form
\begin{equation}
\label{eq:origKeqn}
\dot K = - [ \nabla\cdot \ov u + \nabla^2 W/2] K,
\end{equation}
the time derivative referring to instanton transit time, \ie, to motion
along a characteristic or bicharacteristic.
  Similarly to~$W$, $K$~may be regarded as a
function on~${\cal M}$ rather than on configuration space.  Integration of
the equation~(\ref{eq:origKeqn}) requires knowledge of the second spatial
derivatives of~$W$ along the characteristic.  But $(\partial^2 W/\partial
x_i\partial x_j)(\ov x)$ equals $(\partial p_i/\partial x_j)(\ov
x)$, which is a measure of the `slope' of the manifold above the
point~$\ov x$.  By~differentiating Hamilton's equations it~is easy to
show that the Hessian matrix $\matr{Z}=(Z_{ij})$ whose elements are the
partial slopes $\partial p_i/\partial x_j$ satisfies the matrix Riccati
equation
\begin{equation}
\label{eq:Riccati}
\dot\matr Z = -{\matr Z}^2 - \matr Z\matr B - {\matr B}^t\matr Z
-\sum_l p_l {\matr Y}^{(l)}
\end{equation}
along any characteristic.  (Cf.~Ludwig~\cite{Ludwig75}.)
Here $\matr B=(\partial u_i/\partial x_j)$ and
$\matr Y^{(l)}=(\partial u_l/\partial x_i\partial x_j)$ are auxiliary
matrices.  Since $\nabla^2W=\tr\matr Z$, the computation of~$K$ by
numerical integration is straightforward.

It is interesting to compare these results with those of
Littlejohn~\cite{Littlejohn92} on the WKB prefactor for the solutions of
the Schr\"odinger equation in the semiclassical ($\hbar\to0$) limit.
He~introduces a Lagrangian manifold, and a similar integration along
characteristics.  But because he analyses the {\em time-dependent\/}
Schr\"odinger equation, he~finds that the transport equation for his
analogue of~$K$ can be integrated explicitly, yielding a Van~Vleck
determinant.  Matters are not so simple in the time-independent case, for
the Schr\"odinger equation as~well as for the Smoluchowski equation.  Our
WKB analysis of the weak-noise limit of the stationary density actually has
more in common with the work of Gutzwiller on the semiclassical
approximation of fixed-energy quantum-mechanical Green's
functions~\cite{Gutzwiller71,Gutzwiller90,Schulman} than it does with the
semiclassical approximation of time-dependent quantum-mechanical
propagators.  The prefactor~$K$ is analogous to the prefactor of a
semiclassical Green's function (at~fixed energy).  It~can in~fact be related
to the {\em density of bicharacteristics\/} on the Lagrangian manifold.
This resembles Gutzwiller's interpretation of the prefactor of a
semiclassical Green's function in~terms of the density of classical
trajectories on an energy surface~\cite{Gutzwiller90}.

\subsection{{ Matched Asymptotic Approximations and the MFPT}}
\label{subsec:MAE2}
We now specialize to two-dimensional double well models with the structure
of Fig.~\ref{fig:drift}.  On~account of symmetry and smoothness we may
expand the drift $\ov u=(u_x,u_y)$ thus:
\begin{eqnarray}
u_x(x,y) &=& v_0(x) + v_2(x)y^2+\cdots\nonumber\\
u_y(x,y) &=& u_1(x)y + u_3(x)y^3+\cdots.
\label{eq:velexp}
\end{eqnarray}
By assumption $v_0(x)>0$ for all~$x$ between $0$~and~$x_s$, and $u_1(x)<0$
for all~$x$ between $0$~and~$x_s$ inclusive.  If~the symmetry through the
axis is unbroken, $W$~and~$K$ (both~of them computed by integration along
instanton trajectories emanating from~$S$) will have similar expansions
\begin{eqnarray}
\label{eq:fooa}
W(x,y) &=& w_0(x) + w_2(x)y^2/2! + \cdots\\
K(x,y) &=& k_0(x) + k_2(x)y^2/2! + \cdots.
\label{eq:foob}
\end{eqnarray}
Here $w_{2m}(x) \defeq \partial^{2m}W/\partial y^{2m}(x,0)$ and $k_{2m}(x)
\defeq \partial^{2m}K/\partial y^{2m}(x,0)$.  Since $W$ can be viewed as a
classical action, the functions~$w_{2m}$ can be expressed in~terms of the
momentum $\ov p=\ov p(\ov x)$ of the instanton trajectories passing through
near-axis points~$\ov x$.  For~example, $w_0'(x)=p_x(x,0)$ and
$w_2(x)=\partial p_y/\partial y(x,0)$.  Substituting the WKB Ansatz into
the Smoluchowski equation ${\cal L}^*\rho_0=0$, and examining the
coefficients of each power of~$\epsilon$ and~$y$, will yield equations for
the various coefficient functions in~(\ref{eq:fooa}) and~(\ref{eq:foob}).
One finds in~particular that $w_0'=p_x=-2v_0$, or
\begin{equation}
\label{eq:holder}
w_0(x) = 2\int_{x}^{x_s} v_0(x')\,dx'.
\end{equation}
Therefore the Hamilton equation $\dot x=p_x + v_0(x)$, which follows from
the Wentzell-Freidlin Hamiltonian~(\ref{eq:Hamiltonian}), implies that
$\dot x$~must equal~$-v_0(x)$ at all points between $S$~and the saddle.
The instanton trajectory on the $x$-axis moves with a speed equal to the
local value of the drift speed, but in the direction opposite to the drift.

Examining coefficients also yields the two equations
\begin{eqnarray}
\label{eq:Keqn}
\dot k_0 &=& - [u_1 + w_2/2] k_0\\
\dot w_2 &=& -w_2^2 - 2u_1w_2 + 4v_0 v_2
\label{eq:w2eqn}
\end{eqnarray}
where we have changed the independent variable from $x$ to~$t$ by writing
$\dot k$ for $-v_0k_0'$, and $\dot w_2$ for~$-v_0w_2'$.  Equations
(\ref{eq:Keqn}) and~(\ref{eq:w2eqn}) could equally well be deduced
from~(\ref{eq:origKeqn}) and~(\ref{eq:Riccati}).  For later reference we
note that
\begin{eqnarray}
\dot w_4 &=& -v_0 w_4'\nonumber \\
&=& -4(w_2 + u_1)w_4 - 3[ (w_2')^2 + 4v_2 w_2' + 8u_3 w_2] + 48v_0v_4
\label{eq:w4eqn}
\end{eqnarray}
is the equation satisfied by the fourth derivative $w_4=\partial^4
W/\partial y^4=\partial^3 p_y/\partial y^3$ on the $x$-axis.

The physical interpretation of the functions $k_0$ and~$w_2$ is
straightforward.  The WKB Ansatz implies that
\begin{equation}
\label{eq:tube}
\rho_0(x,y) \sim k_0(x) \exp\left\{ -\left[w_0(x) +
w_2(x)\dfrac{y^2}{2!}\right]/\epsilon\right\}, \quad \epsilon\to0.
\end{equation}
So when $\epsilon$ is small, $w_2$~governs the {\em small transverse
fluctuations\/} about the $x$-axis.  At~any time when the system
state~$\ov x$ has fluctuated leftward from $x_s$ to~$x$, the
distribution of the transverse component~$y$ will (provided
that~$w_2(x)>0$) be approximately Gaussian, with
variance~$\sim\epsilon/w_2(x)$.  Of~course such fluctuations are exponentially
rare, on~account of the $\exp[-w_0(x)/\epsilon]$ factor.

The Riccati equation~(\ref{eq:w2eqn}) therefore gives the position
dependence of the width of the `WKB tube' of probability density
surrounding the MPEP\hbox{}, when this MPEP is in fact the line segment
between $S$~and the saddle.  Moreover this equation captures the essence of
the bifurcation phenomenon, as we shall see in Section~\ref{subsec:MAE3}.
For the moment we note only that it may readily be integrated from
$t=-\infty$ (when~the instanton trajectory formally emerges from~$S$)
to~$t=+\infty$ (when~the trajectory, obeying $\dot x=-v_0(x)$, reaches the
saddle).  Since $v_0\left(x(t)\right)\to0$ as~$t\to\pm\infty$, we~see
from~(\ref{eq:w2eqn}) that $w_2$~must converge as~$t\to+\infty$
(resp.~$t\to-\infty$) to one of the two zeroes of the quadratic polynomial
\begin{equation}
-w_2^2 - 2u_1w_2 = -w_2(w_2 - 2|u_1|),
\end{equation}
where
$u_1$~signifies $u_1(0)$ (resp.~$u_1(x_s)$).  On~physical grounds one
expects that usually (`generically') the WKB tube will have a finite
variance at both endpoints, \ie, as~$t\to-\infty$ and~$x\to x_s$, and
as~$t\to+\infty$ and~$x\to0$.  So~$w_2(x_s)$ should equal~$2|u_1(x_s)|$,
and $w_2(0)$ should equal~$2|u_1(0)|$.

If these endpoint (`turning point') conditions hold, it is easy to match
the tube approximation~(\ref{eq:tube}) to auxiliary, non-WKB approximations
valid near the endpoints: the stable points and the saddle.  On~physical
grounds, $\rho_0$~and~$\rho_1$ may be approximated on the
$O(\epsilon^{1/2})$ lengthscale near~$S$ by a Gaussian function of the
system state~$\ov x$.  Let~us write $\nu_x$~and~$\nu_y$ for $\partial
u_x/\partial x(S)$ and~$\partial u_y/\partial y(S)$, the two (negative)
eigenvalues of the linearization of the drift~$\ov u$ at~$S$.  Then
\begin{equation}
\label{eq:nearS}
\rho_0(x,y) \sim \const\times e^{-|\nu_x|(x-x_s)^2/\epsilon}
e^{-|\nu_y|y^2/\epsilon}, \quad \epsilon\to0
\end{equation}
near~$S$, the same being true of~$\rho_1$.  Since $\nu_y=u_1(x_s)$, this
will match to the tube approximation if~$w_2(x_s)=2|u_1(x_s)|$.
Similarly, on the $O(\epsilon^{1/2})$ lengthscale near the saddle,
$\rho_0$~may be approximated by the
inverted Gaussian 
\begin{equation}
\label{eq:nearH}
\rho_0(x,y) \sim \const\times e^{+\lambda_xx^2/\epsilon}
e^{-|\lambda_y|y^2/\epsilon}, \quad \epsilon\to0.
\end{equation}
Since $\lambda_y=u_1(0)$, the tube approximation will match
to~(\ref{eq:nearH}) if~$w_2(0)=2|u_1(0)|$ and $k_0(0)$~is finite and
nonzero.  It~is easy to verify that the approximations
(\ref{eq:nearS})~and~(\ref{eq:nearH}) satisfy the time-independent
Smoluchowski equation on the $O(\epsilon^{1/2})$ lengthscale near their
respective turning points.

The appropriate (generic) approximation to the quasistationary
density~$\rho_1$ near
the saddle is slightly more complicated; it~is an error function
approximation of the sort first used by Kramers~\cite{Kramers40}.
We~have~\cite{MaierB}
\begin{eqnarray}
\label{eq:erfintegral}
\rho_1(x,y) &\sim& \const \times
\left\{\epsilon^{-1/2}\int_0^\infty \exp\left[-(xp_x +
p_x^2/4\lambda_x)/\epsilon\right]\,dp_x \right\}  e^{-|\lambda_y|y^2/\epsilon}
\\
&=&\const \times \erf(\lambda_x^{1/2} x/\epsilon^{1/2})
e^{+\lambda_xx^2/\epsilon}e^{-|\lambda_y|y^2/\epsilon}
\label{eq:erf}
\end{eqnarray}
on the $O(\epsilon^{1/2})$ lengthscale.  This `boundary layer'
approximation agrees with the inverted Gaussian
approximation~(\ref{eq:nearH}) in the far field, \ie,
as~$x/\epsilon^{1/2}\to+\infty$.  So~under the same conditions, the tube
approximation~(\ref{eq:tube}) will match to~it.

We have now approximated $\rho_1(\ov x)$ at all points~$\ov x$ in
the vicinity of the line segment joining $S$~and the saddle.  We~must
emphasize that the validity of this procedure depends on two assumptions:
\begin{itemize}
\item That the physical values of $W(\ov x)$ and~$K(\ov x)$ at all
points~$\ov x$ along the axis arise from integration along the {\em
on-axis\/} instanton trajectory extending from $S$ to~$\ov x$.
\item That the WKB tube surrounding the axis is well behaved as the saddle
is approached, so that the error function approximation to the
quasistationary density is valid near the saddle.  This requires that
${w_2\to2|u_1(0)|}$, and that $k_0$ tend to a finite, nonzero limit.
\end{itemize}
The first assumption breaks~down when the MPEP has bifurcated, and we shall
see that the second assumption breaks down at the onset of bifurcation.
But if both assumptions hold, it~is easy to compute the weak-noise
asymptotics of the quasistationary eigenvalue~$\lambda_1$ and its asymptotic
reciprocal, the MFPT~$\langle\tau\rangle$.  The time-dependent equation
$\dot\rho=-{\cal L}^*\rho$ may be written as
\begin{equation}
\label{eq:continuity}
\dot\rho + \nabla \cdot [-(\epsilon/2)\nabla\rho +
\rho\ov u] = 0.
\end{equation}
Equation~(\ref{eq:continuity}) is a continuity equation, and $\ov j =
-(\epsilon/2)\nabla\rho + \rho\ov u$ can be viewed as a
probability current density.  Since $\lambda_1$~is the decay rate of the
eigenmode~$\rho_1$, it~may be computed as the rate at~which probability is
absorbed on the separatrix~\cite{Kramers40,Naeh90}.
Necessarily
\begin{equation}
\lambda_1= \int_{-\infty}^\infty [-j_x(0,y)]\,dy
\left/
\int_0^\infty\int_{-\infty}^\infty
\rho_1(x,y)\,dy\,dx
\right.
\end{equation}
where $\ov j = (j_x,j_y)$ is computed from~$\rho_1$.  The numerator
(an~absorption rate) is computed from~(\ref{eq:erf}), and the normalization
factor in the denominator from the Gaussian approximation~(\ref{eq:nearS}).
If~the constant prefactors of these two approximations are chosen to ensure
consistency with the intermediate WKB tube approximation~(\ref{eq:tube}), the
quotient will acquire a factor $k_0(0)\exp[-w_0(0)/\epsilon]$, \ie,
$K(0,0)\exp[-W(0,0)/\epsilon]$.

This computation, if carried through, yields a so-called Eyring formula for
the weak-noise asymptotics of the quasi-stationary eigenvalue, \ie, the
weak-noise asymptotics of the rate of noise-activated
hopping~\cite{Caroli80,Glasstone41}:
\begin{equation}
\label{eq:Eyring}
\lambda_1(\epsilon)\sim
K(0,0)
\left[{{\sqrt{\lambda_x|\nu_x|}\sqrt{|\nu_y|/|\lambda_y|}}\over\pi}\right]
\exp[-\Delta W/\epsilon], \qquad \epsilon \to0.
\end{equation}
Here the presence of the `frequency factor' $K(0,0)$ is attributable to the
non-gradient dynamics; it~will equal unity if the drift~$\ov u$ is a
gradient.  The formula is otherwise familiar.  Since
$\langle\tau\rangle\sim \lambda_1^{-1}$ as~$\epsilon\to0$, this formula
predicts a {\em pure Arrhenius\/} growth of the MFPT in the weak-noise
limit.  But as noted, this conclusion depends crucially on the validity of
the Kramers-type error function approximation to the quasistationary
density near the saddle.  This approximation will prove not to be valid
in double well models undergoing a bifurcation.

\subsection{{ Indications of a Phase Transition}}
\label{subsec:MAE25}
We~shall now explain how the bifurcation transition displays characteristic
features of a phase transition, such as power-law divergences governed by
critical exponents.  We~begin by using the standard double well
model~(\ref{eq:standard}), and the transport equations of the last section,
to reveal the nature of the `nascent cusp' singularity appearing at the
saddle point, at~criticality.

For the standard model,
the stable point~$S$ is located at~$x=x_s=1$, and the
coefficient functions (drift velocity derivatives) in the transport equations
are of the form
\begin{eqnarray}
\label{eq:standardfirst}
v_0(x) &=& u_x(x,0) = x-x^3  \\
2v_2(x) &=& \partial^2 u_x/\partial y^2(x,0) = -2\alpha x\\
u_1(x) &=& \partial u_y/\partial y(x,0) = -\mu(1+x^2).
\label{eq:standardlast}
\end{eqnarray}
These may be substituted into the Riccati equation~(\ref{eq:Riccati}) for
the transverse second derivative $w_2(x)=\partial^2 W/\partial y^2(x,0)$,
and the equation numerically integrated.  As~noted, the appropriate initial
condition is $w_2(x=x_s)=2|u_1(x_s)|$, \ie, $w_2(x=1)=4\mu$.  Consider the
case~$\mu=1$ (the~subject of~Fig.~\ref{fig:flow}), in~particular.  One finds for
all~$\alpha$ in the range $0< \alpha<4$ that $w_2$~is positive on the
line segment between $x=x_s$ and the saddle at~$x=0$.  Since the WKB tube
centered on the axis, which is formed by small transverse fluctuations
about the MPEP\hbox{}, has variance~$\sim\epsilon/w_2(x)$, this positivity implies
that the tube is everywhere well-defined.  One also finds that $w_2\to2$,
\ie, $w_2\to 2|u_1(0)|$, as the saddle is approached.  Moreover, by
integrating the transport equation~(\ref{eq:Keqn}) one finds that
$k_0(x)=K(x,0)$ tends to a finite, nonzero limit as~$x\to0$.  As~we
explained above, these two conditions are precisely what is needed to
ensure Arrhenius weak-noise asymptotics, with an MFPT prefactor
proportional to~$K(0,0)^{-1}$.

\begin{figure}
\begin{center}
\epsfxsize=2.5in		
\leavevmode\epsfbox{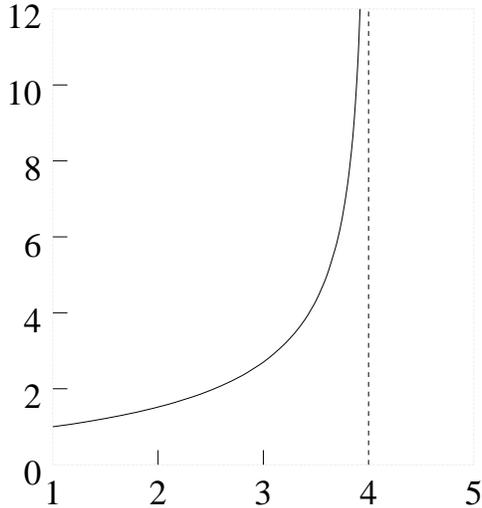}
\end{center}
\caption[A~plot of~$K(0,0)$, to which the weak-noise activation rate
prefactor is proportional, as a function of the off-axis softening
parameter~$\alpha$.]{A~plot of~$K(0,0)$, to which the weak-noise activation
rate prefactor is proportional, as a function of the off-axis softening
parameter~$\alpha$.  As~shown, $K(0,0)$ diverges as~$\alpha\to\alpha_c^-$.
This is for the standard double well model~(\ref{eq:standard}),
with~$\mu=1$, for which~$\alpha_c=4$.}
\label{fig:prefactorblowup}
\end{figure}

The bifurcation transition present in the $\mu=1$ standard model
at~$\alpha=4$ is reflected in the behavior of $w_2$ and~$k_0$ as~$x\to0$.
When~$\alpha=\alpha_c=4$, equations (\ref{eq:Riccati}) and~(\ref{eq:Keqn}) can be
solved exactly; one finds
\begin{eqnarray}
\label{eq:baz}
w_2(x) &=& \partial^2 W/\partial y^2(x,0) = 4x^2,\\
k_0(x) &=& K(x,0) = 1/x.
\label{eq:bazdognascent}
\end{eqnarray}
We know from the Eyring formula that the activation rate prefactor, in the
limit of weak noise, is proportional to~$K(0,0)$.  The fact that
$k_0(x=0)=K(0,0)$ is {\em infinite\/} here strongly suggests that
at~criticality the activation rate, \ie, the rate at~which the
quasistationary density is absorbed on the separatrix, is {\em anomalously
large\/}.  Equivalently, it~suggests that at~criticality the weak-noise
behavior of the MFPT (which~is asymptotically equal to~$\lambda_1^{-1}$) is
{\em non-Arrhenius\/}, with a pre-exponential factor that tends to zero
as~$\epsilon\to0$.  There is an even stronger piece of evidence that this
is the case.  It~is not difficult to show, by analysing the transport
equation~(\ref{eq:Keqn}), that
$K(0,0)=k_0(x=0)\sim(\alpha_c-\alpha)^{-1/2}$ as~$\alpha\to\alpha_c^-$.
Figure~\ref{fig:prefactorblowup} shows the result of a numerical
computation when~$\mu=1$.  {\em The activation rate prefactor diverges
as~$\alpha\to\alpha_c^-$.}  Equivalently, the MFPT prefactor tends to zero.
The natural deduction is that at~criticality the activation rate prefactor,
and its reciprocal the MFPT prefactor, become $\epsilon$-{\em dependent\/}.
This blends nicely with the behavior {\em above\/} the transition, since
(as~shown in~Fig.~\ref{fig:barrierheight}) the exponential growth rate of
the MFPT (the~action barrier~$\Delta W$, \ie,~$W(0,0)$) begins to decrease
as $\alpha$~increases beyond~$\alpha_c$.  An~$\epsilon$-dependent
activation rate
prefactor at criticality, containing a negative power of~$\epsilon$, would
unify the exit behavior both below and above criticality.

We~shall show in Section~\ref{sec:nascent} that in critical double well
models (\eg,~in the standard model with~$\alpha=\alpha_c$), the weak-noise
activation rate~$\lambda_1=\lambda_1(\epsilon)$ indeed has 
asymptotics
\begin{equation}
\lambda_1(\epsilon) \sim \const\times \epsilon^{-s}
\exp\left[-\Delta W/\epsilon\right], \qquad \epsilon\to0,
\end{equation}
where~$s>0$ is the singularity index mentioned in
Section~\ref{sec:bifurcation}.  The computation of the singularity index is
nontrivial.  Since the MFPT $\langle\tau\rangle$ satisfies
$\langle\tau\rangle\sim\lambda_1^{-1}$, the $\epsilon^{-s}$~prefactor
in~$\lambda_1$ gives rise to an~$\epsilon^s$ MFPT prefactor.  In~critical
double well models, in the weak-noise limit the growth of the MFPT is {\em
slower than pure exponential\/}.

At criticality, the action~$W$ as well as the prefactor~$K$ displays
unusual behavior at the saddle.  We~shall see that the behavior
of~(\ref{eq:baz}), \ie, that $w_2$~tends to zero quadratically as~$x\to0$,
is universal.  Since the WKB tube has variance $\sim \eps/w_2(x)$,
this implies that the tube {\em splays~out\/} as the
saddle is approached.  To~leading order, it~splays out to infinite width.
This is an indication that the transverse fluctuations around the MPEP\hbox{},
on~the $O(\epsilon^{1/2})$ lengthscale, at~criticality become very strong
near the saddle.  In~fact, that $w_2(x=0)=0$ causes some
difficulty in the interpretation, near the saddle, of the WKB approximation
to $\rho_0$ and~$\rho_1$.  One might expect that even though $w_2(x=0)=0$,
the quartic tube approximation
\begin{equation}
\label{eq:improvedtube}
\rho_0(x,y) \sim k_0(x) \exp\left\{ -\left[w_0(x) +
w_2(x)\dfrac{y^2}{2!} + w_4(x)\dfrac{y^4}{4!}
\right]/\epsilon\right\}, \qquad \epsilon\to0
\end{equation}
would suffice for an understanding of the behavior of the WKB approximation
near the saddle.  If $w_2(x=0)$ were zero but $w_4(x=0)$ were finite and
nonzero, transverse fluctuations around the saddle would be,
by~(\ref{eq:improvedtube}), of~magnitude $O(\epsilon^{1/4})$ rather
than~$O(\epsilon^{1/2})$.  However, explicit solution of
eq.~(\ref{eq:w4eqn}), the transport equation for $w_4=\partial^4 W/\partial
y^4=\partial^3 p_y/\partial y^3$, shows that in the $\mu=1$ standard model
at~$\alpha=\alpha_c=4$,
\begin{equation}
\label{eq:f4}
w_4(x)\sim(4/5)x^{-4}+(16/5)x^{-2}+8+\cdots, \qquad x\to0^+.
\end{equation}
The fact that $w_4(x=0)$~is {\em infinite\/}, coupled with the fact that
$w_2(x=0)$ is zero, suggests that at~criticality, the transverse
fluctuations near the saddle {\em have no natural scale\/}.  In~any event,
at~criticality the standard matched asymptotic approximations technique of
the last section breaks down.  We~shall need to construct an approximation
to the quasistationary density~$\rho_1$ near the saddle which (i)~is valid
at criticality, and (ii)~matches to the WKB
approximation~(\ref{eq:improvedtube}), despite its singular character.

{\em Critical exponents\/}, as~we define them, describe the weak-noise
behavior of a parametrized double well model with a singularity at some
point~${\ov x}_0$, as~${\ov x}\to{\ov x}_0$ and as the
parameters of the model tend to the values for which the singularity
appears at~${\ov x_0}$.  In~particular, they characterize the behavior
at~and near the bifurcation transition, and at~and near the saddle point,
of the functions $W$~and~$K$ appearing in the WKB approximation to $\rho_0$
and~$\rho_1$.  At~criticality, the divergence rates of the WKB
prefactor~$K$ as~$x\to0$ and~$y\to0$ supply two such exponents; the scaling
form which we shall use to approximate $W$ near a nascent cusp (which
involves fractional powers) will supply others.  There are also critical
exponents describing what happens as one moves off criticality.  As
$\alpha$ is increased above~$\alpha_c$ (\ie,~above~4, in the $\mu=1$
standard model), the MPEP bifurcates.  There is a critical exponent
describing the separation rate of the two resulting MPEP's, as
Fig.~\ref{fig:barrierheight} makes clear.  There is also a critical
exponent describing the divergence rate of $K(0,0)$
as~$\alpha\to\alpha_c^-$, which as~we have already noted equals~$1/2$.
The singularity index~$s$ can be regarded as a critical exponent~too,
though of a different kind; to~compute~it, one must go beyond the WKB
approximation.

The `nascent cusp,' as~a singularity, is located in a space parametrized by
$x$,~$y$, and the parameter(s) of the drift field~$\ov u$.  But if one
restricts oneself to a single double well model, the only parameters are
$x$~and~$y$.  In~this case there is a natural analogy between the
Lagrangian manifold~${\cal M}$ in phase space, formed by bicharacteristics
emanating from $(S,\ov 0)$ and~$(S',\ov 0)$, and a {\em thermodynamic
surface\/}.  The action~$W$, as a function on the manifold, corresponds to
a thermodynamic potential, in~fact a Gibbs free energy.  The equation $\ov
p=\ov p(\ov x)=\partial W/\partial{\ov x}$ corresponds to a relation between conjugate
state variables, such as pressure and volume.  The singularities (points of
non-differentiability) of the physical action~$W(\ov x)$ therefore
correspond to {\em phase transitions\/}.  The order of such a phase
transition, in the traditional sense, is the lowest order of spatial
derivative (of~$W$) which fails to be continuous.
In~Section~\ref{sec:nascent} we shall compute the order of the nascent
cusp.

\begin{figure}
\vskip-0.75in
\vbox{
\hfil
\epsfxsize=2.4in		
\epsfbox{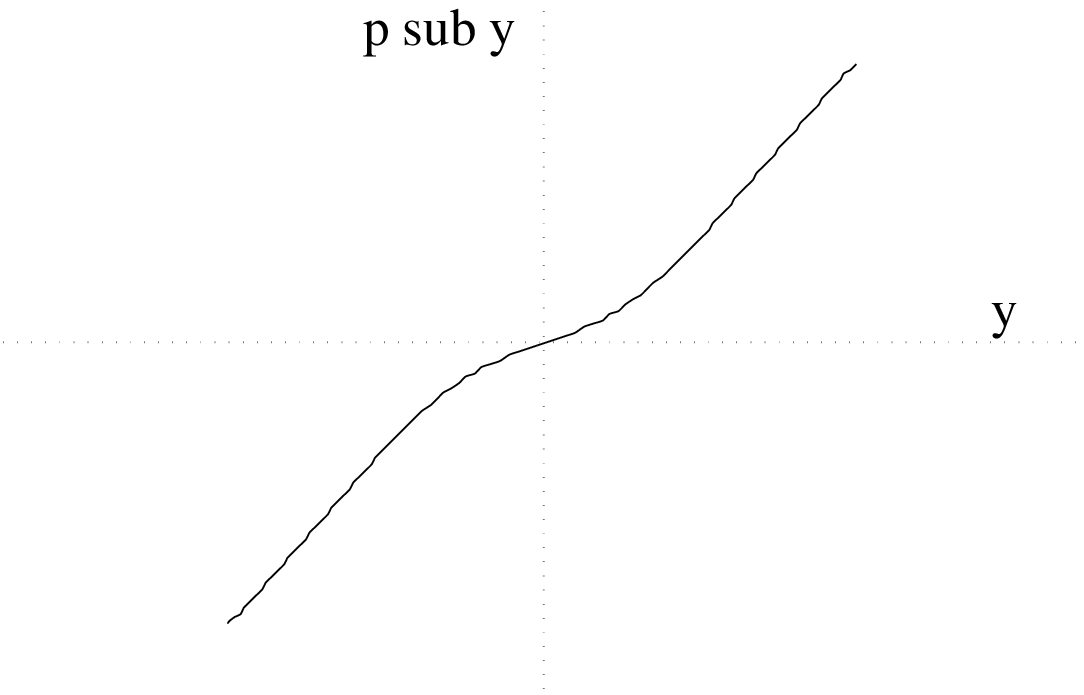}
\hfil
\epsfxsize=2.4in		
\epsfbox{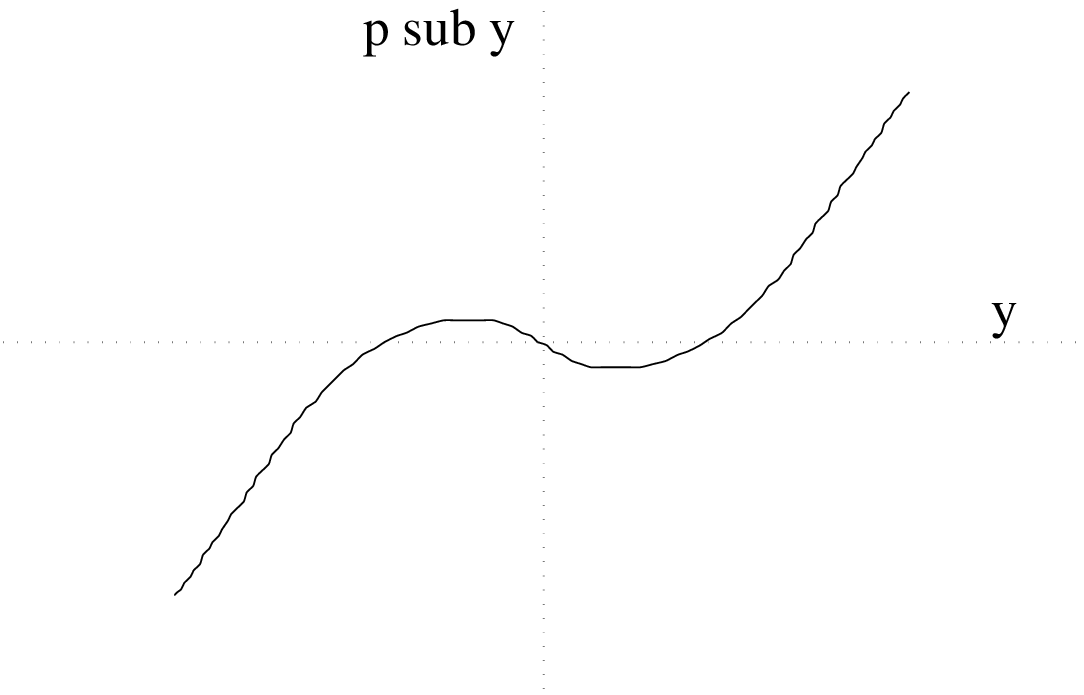}
}
\vskip0.5in
\vbox{
\hfil
\epsfxsize=2.4in		
\epsfbox{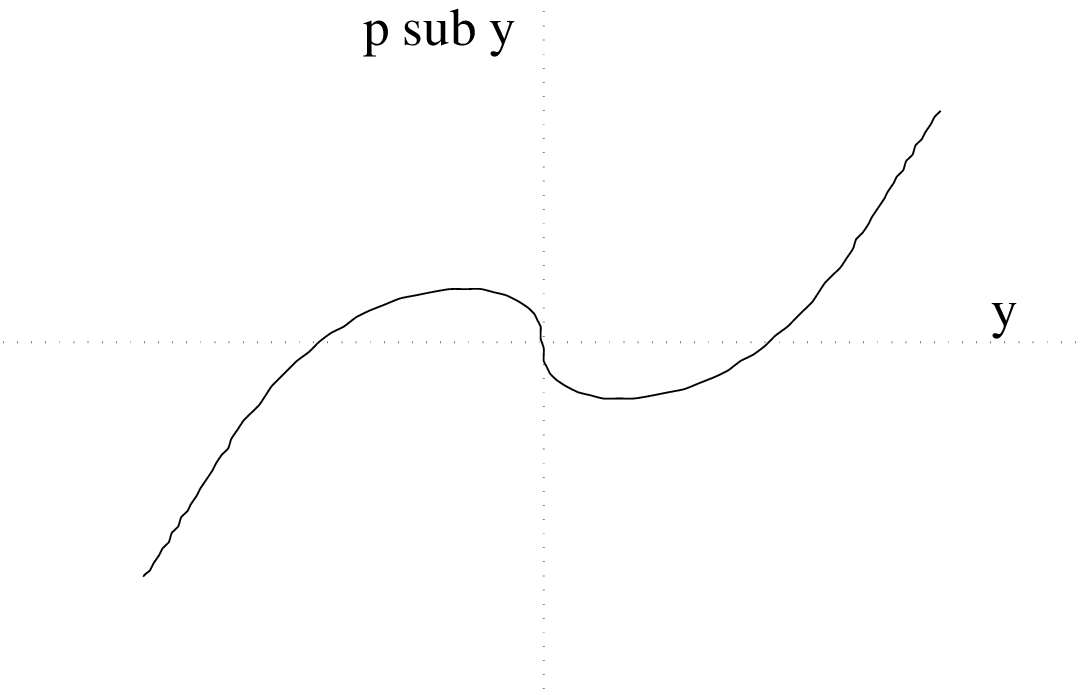}
\hfil
\epsfxsize=2.4in		
\epsfbox{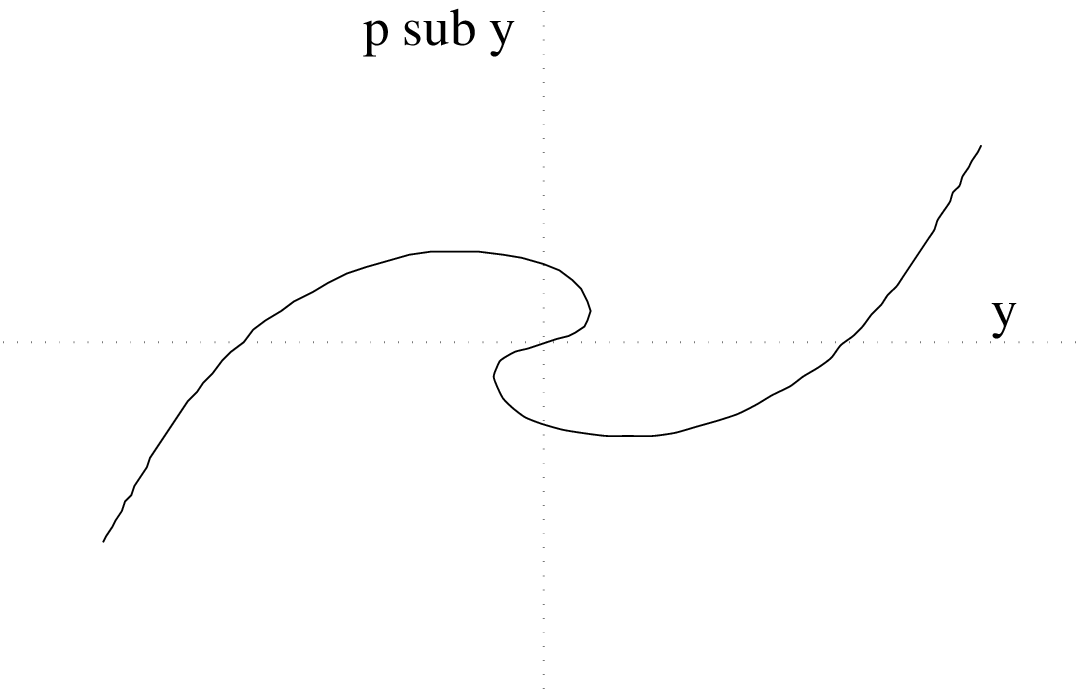}
}
\vskip0.5in
\vbox{
\hfil
\epsfxsize=2.4in		
\epsfbox{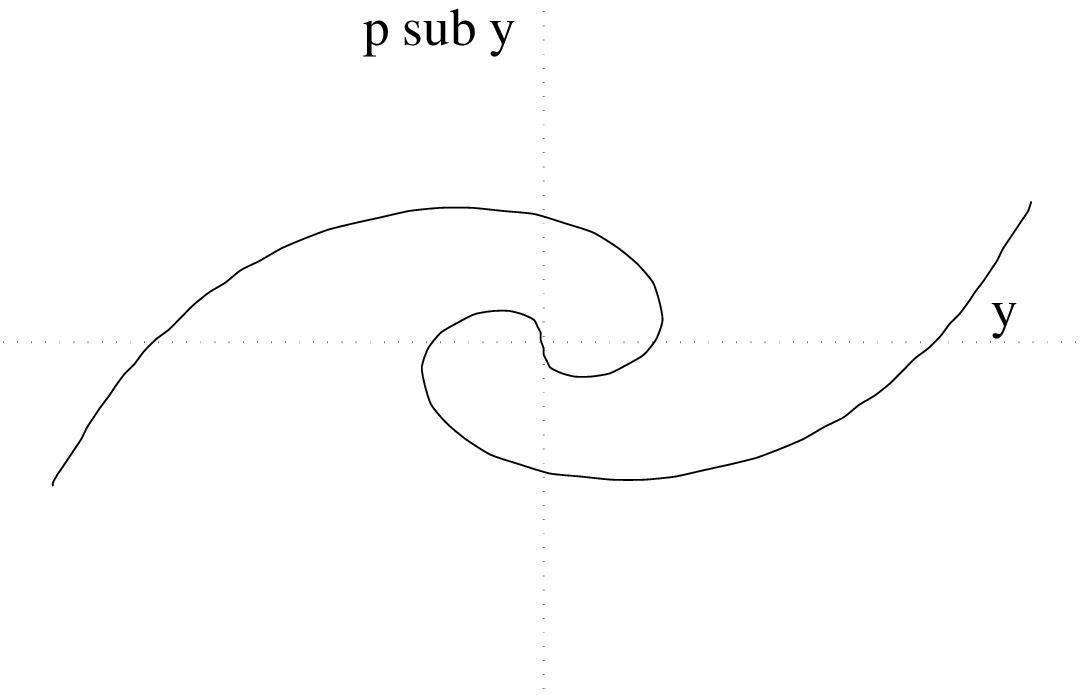}
\hfil
\epsfxsize=2.4in		
\epsfbox{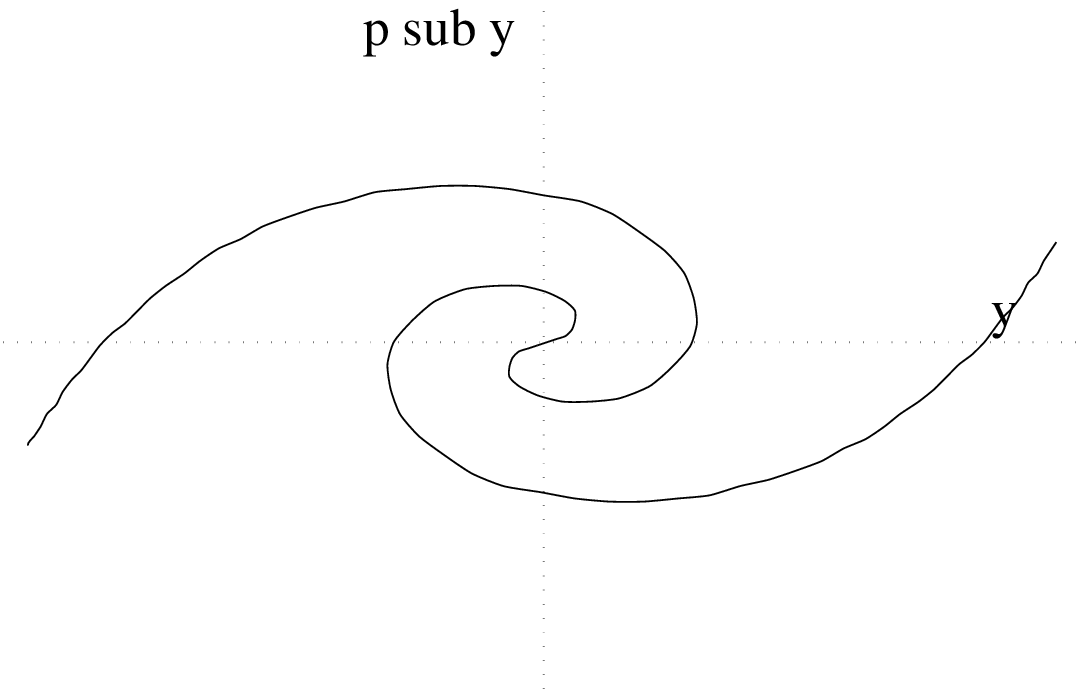}
\hfil
}
\vskip0.25in
\caption[Cross-sections through the Lagrangian manifold~$\protect\cal M$, revealing
the `whorling' that takes place as one passes through any on-axis
focus.]{Cross-sections through the Lagrangian manifold~$\cal M$, revealing
the `whorling' that takes place as one passes through any on-axis focus.
These sketches show the map $y\mapsto p_y$ at~successively decreasing
values of~$x$, as~one moves from~$S=(x_s,0)$, past two foci $(x_f^{(1)},0)$
and~$(x_f^{(2)},0)$, toward the saddle point~$(0,0)$.  Shown are the cases
(a)~$x_s>x>x_f^{(1)}$ and $x$~near~$x_s$, (b)~$x_s>x>x_f^{(1)}$ and
$x$~near~$x_f^{(1)}$, (c)~$x=x_f^{(1)}$, (d)~$x_f^{(1)}>x>x_f^{(2)}$,
(e)~$x=x_f^{(2)}$, and (f)~$x_f^{(2)}>x>0$.}
\label{fig:whorl}
\end{figure}

\subsection{{ The Bifurcation Transition and Classical Mechanics}}
\label{subsec:MAE3}
We now explain how the equations of Section~\ref{subsec:MAE2} allow the
bifurcation transition to be interpreted in~terms of classical mechanics,
and how one can predict whether or not any given double well model is at
criticality.  We~begin by considering models in which the MPEP has already
bifurcated, and the instanton trajectories emanating from the stable
point~$S$ focus along the axis, as in Figs.\ \ref{fig:flow}(c)~and~\ref{fig:flow}(d). 

Empirically, focusing occurs in models with a sufficiently large `off-axis
softening' parameter~$\alpha$.  In~such models, the action~$W$ in a
mathematical sense becomes multivalued near a portion of the axis.
Points~$(x,0)$ beyond the first focus are reached by multiple off-axis
instanton trajectories emanating from the stable point~$S$, and in~general
these trajectories will have different actions.  They will also have
different momenta $\ov p=\nabla W$ at~the time they
reach~$(x,0)$.

This multivaluedness has a geometric interpretation, in~terms of the shape
of the two-dimensional Lagrangian manifold (in~the four-dimensional phase
space) formed by the bicharacteristics emanating from the point $(\ov x,\ov
p)=(S,\ov 0)$.  As~$x$~decreases from~$x_s$ toward zero, the map $y\mapsto
p_y$ in the vicinity of~$y=0$ is at~first single-valued; the value $p_y=0$,
and no~other, corresponds to~$y=0$.  Beyond the first focus
$(x_f^{(1)},0)$, \ie, when~$x<x_f^{(1)}$, the map $y\mapsto p_y$ becomes
three-valued.  At~the second focus $(x_f^{(2)},0)$ it~becomes
five-valued,~etc.  The generic evolution is shown in
Figs.~\ref{fig:whorl}(a) through~\ref{fig:whorl}(f).  Up~to the first focus
$y=y(p_y)$ near~$p_y=0$ may be modelled as a {\em linear\/} function; beyond
the focus, as a {\em cubic\/}.  Beyond the second focus the global
description becomes more complicated, as is clear from the whorl in
Fig.~\ref{fig:whorl}(f).  A~cubic approximation is still appropriate in the
immediate vicinity of $(y,p_y)=(0,0)$, however.

Since the locus of all points~$(y,p_y)$ at constant~$x$ is obtained by
intersecting the Lagrangian manifold  with the hyperplane $x=\const$,
the manifold itself becomes increasingly `whorled' with each passage through a
focus.  The formation of convolutions in Lagrangian manifolds was first
considered by Berry and Balazs~\cite{Berry79} (in~a time-dependent
context), and the progression in Fig.~\ref{fig:whorl} resembles the figures
in their paper. 
Geometrically, the linear-to-cubic transition at each successive
focus corresponds to the creation of a {\em fold\/}~\cite{Dykman94}.  One
can fit the shape of the manifold near the $l$'th~focus, \ie, near
$(x,y,p_y)=(x_f^{(l)},0,0)$, by the phenomenological formula
\begin{equation}
\label{eq:phenomenological}
y=y(x,p_y)=-a_{0}^{(l)}p_y^3 - a_{1}^{(l)} \left(x-x_f^{(l)}\right)p_y
\end{equation}
where $a_{0}^{(l)}$ and~$a_{1}^{(l)}$ are certain positive constants.
So~each successive focus resembles a Ginzburg-Landau
{\em second-order phase transition\/} ($x$~corresponding to
temperature, the focus location~$x_f^{(l)}$ to a critical temperature,
$-y$~to a magnetic field, and $p_y$~to a magnetization).  We~shall say more
about the `equation of state'~(\ref{eq:phenomenological}) (which we stress
is {\em not\/} applicable near the `nascent cusp' appearing at~the saddle
point of critical models) in Sections \ref{sec:real} and~\ref{sec:nascent}.

For on-axis (\ie, $y=0$, $p_y=0$) trajectories, the derivative
$w_2(x)=\partial p_y/\partial y(x,y=0)$ satisfies the Riccati
equation~(\ref{eq:w2eqn}).  So~the appearance of a focus, and of multiple
foci, can be investigated analytically.  It~is clear from 
Figs.\ \ref{fig:whorl}(c)~and~\ref{fig:whorl}(e) that passage through a focus is
signalled by the tangent plane to the manifold (at~$y=0$, $p_y=0$) `turning
vertical'; equivalently, by $\partial y/\partial p_y$ passing through zero,
or its~reciprocal~$w_2$ (a~negative magnetic susceptibility, in this
context) passing through~$-\infty$.  To~study this, recall that the Riccati
equation
\begin{equation}
\label{eq:w2eqn2}
\dot w_2 = -w_2^2 - 2u_1w_2 + 4v_0 v_2
\end{equation}
involves a derivative with respect to instanton transit time, and that the
on-axis instanton trajectory (directed anti-parallel to the drift
toward~$S$) satisfies $\dot x=-v_0(x)$.  Solutions~$w_2$ can be regarded
either as a function of~$t$, for $-\infty<t<\infty$, or~of~$x$, for
$x_s>x>0$.  We~see from the form of the Riccati equation that $w_2$~can
indeed be driven to~$-\infty$ in finite time, \ie, at~some point~$x={x_f>0}$
to the right of the saddle.  In~fact one sees, if $t_f$~is the time when
this occurs (the~focus time), that as~$t\to t_f^-$, \ie, $x\to x_f^+$,
\begin{eqnarray}
w_2(x) = \partial^2 W/\partial y^2(x,0) &\sim& -(t_f-t)^{-1},\nonumber\\
&\sim& \const\times [-(x-x_f)^{-1}].
\label{eq:bazdogrealminus}
\end{eqnarray}
Here the constant multiplier equals $1/v_0(x_f)$, the reciprocal speed of the on-axis
instanton trajectory when it passes through the focus~$(x_f,0)$.
We note in passing that by the transport equation~(\ref{eq:Keqn}), this
blowup will induce a blowup of the on-axis WKB prefactor~$k_0$.  One finds
\begin{eqnarray}
k_0(x) =K(x,0) &\sim& \const\times (t_f - t)^{-1/2},\nonumber\\
&\sim& \const\times (x-x_f)^{-1/2}.
\label{eq:bazdogreal}
\end{eqnarray}
Equations (\ref{eq:bazdogrealminus})--(\ref{eq:bazdogreal}) contrast
markedly with eqs.\ (\ref{eq:baz})--(\ref{eq:bazdognascent}), which apply
to the $\mu=1$ standard model at criticality (where, in a formal sense,
$x_f=0$, since there is a {\em nascent\/} focus at the saddle).  Equations
(\ref{eq:bazdogrealminus})--(\ref{eq:bazdogreal}) are not restricted to the
standard model; they hold in greater generality.  But they apply only when
a bona~fide focus is present at some~$x_f>0$ (\ie,~before the saddle is
reached), and the MPEP has already bifurcated.

By examining the Riccati equation~(\ref{eq:w2eqn2}), we see that $w_2$~will
be driven to~$-\infty$, and a focus will be present, only if the
inhomogeneous term~$4v_0v_2$ on the right-hand side of~(\ref{eq:w2eqn2}) is
sufficiently negative.  (This is because $u_1<0$, by~assumption.)  But
$2v_2=\partial^2 u_x/\partial y^2(x,y=0)$, which we are taking to be
negative when~${0<x<x_s}$, measures the extent to which the drift
toward~$S$ softens as one moves off-axis.  So~our empirical observation is
confirmed analytically: a~sufficiently strong off-axis softening will
create a focus, and a bifurcation of the MPEP!

It is best to think of $w_2=\partial p_y/\partial y(y=0)$ as a {\em
slope\/}, as in Fig.~\ref{fig:whorl}.  As~such, it~may rotate repeatedly
through the point at infinity as $t$~increases, \ie, as $x$~decreases.
Each such rotation results in increased whorling of the Lagrangian
manifold, and also corresponds to a passage through a focus.  So~by
counting the number of singularities of the solution curve $w_2=w_2(t)$,
one may determine the number of foci present in any given double well
model.

\begin{figure}
\begin{center}
\epsfxsize=3.5in		
\leavevmode\epsfbox{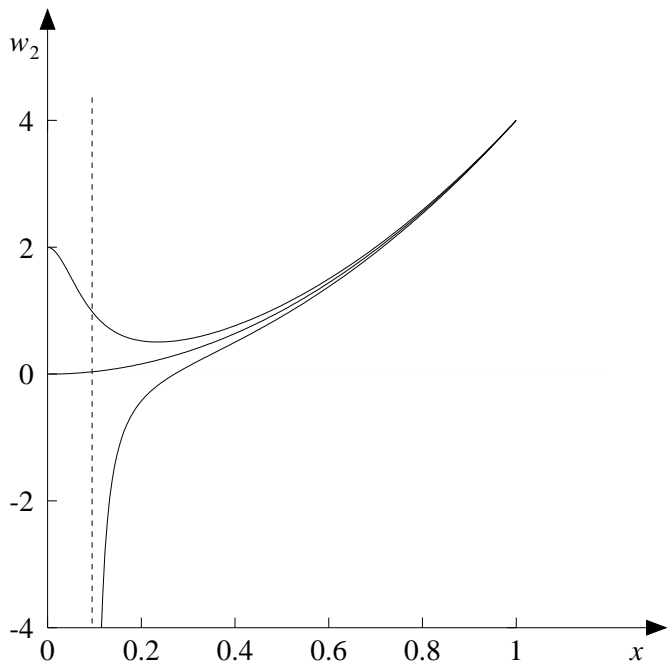}
\end{center}
\caption[The transverse action derivative~$w_2$ in the standard double well
model, in the vicinity of the bifurcation transition.]{The transverse
action derivative $w_2(x)\defeq\partial^2 W/\partial y^2(x,0)$ in the
$\mu=1$ variant of the standard double well model~(\ref{eq:standard}), in
the vicinity of the bifurcation transition.  The three curves, from top to
bottom, obtain when $\alpha=3.9$, $4.0$~(the critical value~$\alpha_c$),
and~$4.1$.  When~${\alpha<\alpha_c}$, $w_2\to 2|\lambda_y|$, \ie,~$2$,
as~$x\to0^+$.  When~${\alpha>\alpha_c}$, $w_2$~is driven negative
as~$x$~decreases, and passes through~$-\infty$.  The focal point $x=x_f$
where this occurs, in the model with~${\alpha=4.1}$, is indicated by a
dashed line.}
\label{fig:dls}
\end{figure}

The standard model~(\ref{eq:standard}) will serve as an example.  For the
reasons discussed in Section~\ref{subsec:MAE2} $w_2(t=-\infty)$, \ie,
$w_2(x=x_s)$, in the standard model always equals
$2|u_1(x_s)|=2|u_1(1)|=4\mu$.  Suppose that~$\mu=1$.  We~noted in the last
section that if~$0<\alpha<4$, $w_2$~is well-behaved and positive at all
times~$t$ between $-\infty$ and~$\infty$ inclusive, \ie, at all~$x$
satisfying $0\le x\le1$.  We~also explained what happens
at~$\alpha=\alpha_c=4$, when the nascent cusp appears at the saddle and the
MPEP begins to bifurcate.  At~criticality, $w_2\to0$ as~$t\to\infty$, \ie,
as~$x\to0^+$.  If~$4<\alpha<9$, $w_2$~is driven negative (as~$t$ increases
toward~$\infty$), and passes through~$-\infty$ before returning
(through~$+\infty$) to finite, positive values.  The change in behavior is
shown in Fig.~\ref{fig:dls}.  When $\alpha$~is raised above~$\alpha_c$, we
say that the graph of~$w_2$ acquires {\em unit winding number\/}, since it
winds once through the point at~infinity.  A~second transition occurs
at~$\alpha=\alpha_c^{(2)}=9$.  If~$9<\alpha<16$, $w_2$~passes
through~$-\infty$ {\em twice\/}, and its graph has winding number equal
to~$2$.  Except at the critical values $\alpha=\alpha_c^{(j)}=(j+1)^2$,
$w_2$~in this model converges to the generic value
$2|u_1(0)|=2\left|\lambda_y\right|=2$ as~$t\to\infty$, \ie, as~$x\to0^+$.

Since each passage of~$w_2$ through~$-\infty$ gives rise to a focus, the
sequence of near-axis instanton flow fields in the $\mu=1$ standard model,
as $\alpha$~is increased, displays an progressively larger number of~foci.
In~fact the progression is precisely as displayed in
Figs.~\ref{fig:flow}(a) through~\ref{fig:flow}(d).  It~is worth noting that
in models with one or more foci, the WKB tube centered on the axis becomes
ill-defined when $w_2$~goes negative, which takes place at a location on
the axis somewhat before the first focus is reached.
(See~Fig.~\ref{fig:dls}.)

\smallskip
In Section~\ref{sec:nascent} we shall determine exactly what happens at the
bifurcation transition of any singly parametrized symmetric double well model.
But we can now pose the question: What, physically, causes the above values
for~$\alpha$ to be critical?  If~in general the odd function
$2v_2(x)=\partial^2 u_x/\partial y^2(x,0)$ is negative between $x=0$
and~$x=x_s$ and is proportional to a parameter~$\alpha$, is there a
classical mechanical technique of predicting the values of~$\alpha$
at~which the on-axis instanton trajectory will bifurcate?  The answer to
this question is~`yes.'  Our technique relies on a {\em linear stability
analysis\/} of the on-axis instanton trajectory, and identifies the
critical values of~$\alpha$ as the values for which a transverse {\em soft
mode\/} is present in the zero-energy Hamiltonian dynamics.  This is
reminiscent of Langer's analysis of metastability in one-dimensional
models~\cite{Langer69,Schulman}.  But because we shall consider transverse,
rather than longitudinal, fluctuations around the instanton trajectory, our
stability analysis will be considerably simplified.

Let $\ov x={\ov x}^*(t) = \left(x^*(t),0\right)$ be the on-axis
instanton trajectory, where $x=x^*(t)$ is the solution of $\dot x=-v_0(x)$.
Near-axis instanton trajectories, \ie, near-axis zero-energy classical
trajectories emanating from~$S$, may to leading order be written as
\begin{equation}
\label{eq:xasymp}
\ov x = \ov x(t) = \left(x(t),y(t)\right) 
\sim \left(x^*(t),0\right) + \delta \left(0,Y(t)\right)
\end{equation}
where $\delta\ll1$, and where $Y=Y(t)$ is some model-dependent function
satisfying $Y(t=-\infty)=0$.  $Y(t)$, $-\infty<t<\infty$, is a {\em
normalized transverse deviation\/}.  Similarly, near-axis trajectories have
momenta
\begin{equation}
\label{eq:pxasymp}
\ov p = \ov p(t) = \left(p_x(t),p_y(t)\right) 
\sim \left(p_x^*(t),0\right) + \delta
\left(0,P_y(t)\right)
\end{equation}
for some unknown function $P_y=P_y(t)$ satisfying $P_y(t=-\infty)=0$.  Here
$\ov p = \ov p^*(t) = \left(p_x^*(t),0\right)$ is the momentum of
the on-axis instanton trajectory at instanton transit time~$t$.  We~noted
before eq.~(\ref{eq:holder}) that as a function of~$x$, $p_x^*$
equals~$-2v_0$.  So~$p_x^*(t)$ equals $-2v_0\left(x^*(t)\right)$.
We~necessarily have
\begin{equation}
w_2(t) = P_y(t)/Y(t),
\end{equation}
on account of $w_2$ equalling $\partial p_y/\partial y(y=0)$.

Substituting equations (\ref{eq:xasymp}) and~(\ref{eq:pxasymp}) into the
Hamilton equations derived from the Wentzell-Freidlin Hamiltonian~$H$, and
separating terms proportional to~$\delta$, yields the pair of equations
\begin{eqnarray}
\label{eq:foo1}
\dot Y &=&
{\dfrac {\partial^2 H}{\partial p_y\partial y} }
\left(\ov x^*(t),\ov p^*(t)\right)
Y+
{\dfrac {\partial^2 H}{\partial p_y^2} }
\left(\ov x^*(t),\ov p^*(t)\right)
P_y\\
\dot P_y &=&
-{\dfrac {\partial^2 H}{\partial y^2} }
\left(\ov x^*(t),\ov p^*(t)\right)
Y-
{\dfrac {\partial^2 H}{\partial p_y\partial y} }
\left(\ov x^*(t),\ov p^*(t)\right)
P_y.
\label{eq:foo2}
\end{eqnarray}
Due to the special form of the Hamiltonian~(\ref{eq:Hamiltonian}), and the
expansions~(\ref{eq:velexp}), this pair becomes
\begin{eqnarray}
\label{eq:bar1}
\dot Y &=& u_1\left(x^*(t)\right) Y + P_y\\
\dot P_y &=& 4v_0\left(x^*(t)\right)v_2\left(x^*(t)\right) Y 
-u_1\left(x^*(t)\right) P_y.
\label{eq:bar2}
\end{eqnarray}
So $w_2=w_2(t)$ can be represented as the quotient of two functions of
instanton transit time, which satisfy a pair of coupled {\em linear\/}
differential equations.  We~note in~passing that an analogous
representation is possible for solutions $\matr Z=\matr Z(t)$ of the matrix
Riccati equation~(\ref{eq:Riccati}).  The existence of such quotient
representations is well known in the theory of Riccati equations,
and has a geometric interpretation~\cite{Shayman91}.

Equations (\ref{eq:foo1})--(\ref{eq:foo2}), and
(\ref{eq:bar1})--(\ref{eq:bar2}), may be viewed as Hamilton's equations for
the {\em effective\/} (\ie,~time-dependent) {\em transverse
Wentzell-Freidlin Hamiltonian\/}
\begin{eqnarray*}
H_{\mtext{\scriptsize eff}}(Y,P_y,t) &= &
\dfrac12 \dfrac{\partial^2 H}{\partial p_y^2} P_y^2
+ \dfrac{\partial^2 H}{\partial p_y\partial y} Y P_y
+ \dfrac12 \dfrac{\partial^2 H}{\partial y^2} Y^2\\
&=& P^2_y/2 + u_1\left(x^*(t)\right) YP_y - 2 v_0\left(x^*(t)\right) v_2\left(x^*(t)\right)Y^2.
\end{eqnarray*}
To save space we have suppressed the arguments $\left({\ov
x}^*(t),{\ov p}^*(t)\right)$ of the partial derivatives.  This quadratic
Hamiltonian governs the small transverse fluctuations about the on-axis
instanton trajectory.  Its~Legendre transform $\dot
YP_y-H_{\mtext{\scriptsize eff}}$,
namely
\begin{eqnarray*}
L_{\mtext{\scriptsize eff}}(Y,\dot Y,t) &= &
\dfrac12 \dfrac{\partial^2 L}{\partial \dot y^2} \dot Y^2
+ \dfrac{\partial^2 L}{\partial \dot y\partial y} Y \dot Y
+ \dfrac12 \dfrac{\partial^2 L}{\partial y^2} Y^2\\
&=& |\dot Y - u_1\left(x^*(t)\right)Y|^2/2 + 2v_0\left(x^*(t)\right) v_2\left(x^*(t)\right)Y^2,
\end{eqnarray*}
is an {\em effective transverse Onsager-Machlup Lagrangian\/}.  Here $L$~is
the Onsager-Machlup Lagrangian~(\ref{eq:Onsager}), and we have suppressed
the arguments $\left({\ov x}^*(t),\dot{\ov x}^*(t)\right)$ of the
partial derivatives.  The corresponding Euler-Lagrange equation
for the normalized transverse deviation~$Y$, \ie,
\begin{equation}
\ddot Y + 
\left\{
-\dfrac{d}{dt} \left[ u_1\left(x^*(t)\right)\right]
- u_1^2\left(x^*(t)\right) - 4v_0\left(x^*(t)\right)
v_2\left(x^*(t)\right)
\right\}Y = 0,
\end{equation}
is called a (transverse) {\em Jacobi equation\/}~\cite{Schulman}.  It~may
be written as an equation for $Y=Y(x)$, $0\le x\le x_s$, by changing the
independent variable from $t$ to~$x$.  One gets
\begin{equation}
{\cal J}Y \defeq \dfrac{d}{dx}\left[v_0(x) \dfrac{dY}{dx}\right]
+\left[ u_1'(x) - \dfrac{u_1^2(x)}{v_0(x)} - 4v_2(x)
\right]Y=0
\label{eq:Jacobix}
\end{equation}
together with the boundary condition $Y(x=x_s)=0$.  This Jacobi equation,
which is in Sturm-Liouville form, governs the behavior of the instanton
trajectories near the on-axis trajectory (the~MPEP\hbox{}, if it has not
bifurcated).  So~it is responsible for the various behaviors shown in
Fig.~\ref{fig:flow}.  It~is clear from our derivation that the Jacobi operator~$\cal J$,
considered as a quadratic form, defines the transverse {\em second
variation\/} of the Onsager-Machlup action functional about the on-axis
trajectory.

Foci are by definition the points $(x,0)$ where the off-axis instanton
trajectories converge (to~leading order).  Equivalently, $(x,0)$~is a focus
only~if $Y(x)=0$.  But since $w_2=P_y/Y$, this implies that (unless
$P_y(x)=0$ also) $w_2(x)$~is infinite.  This is precisely the necessary
condition for a focus that we derived earlier.  If~$Y$ passes through zero
more than once, then $w_2$~will pass through the point at infinity more
than once.  This is the mechanism by~which, \eg, the two-focus flow field
of Fig.~\ref{fig:flow}(d) engenders the increasingly `whorled'
Lagrangian manifold of Figs.~\ref{fig:whorl}(a) through~\ref{fig:whorl}(f).

We can now give a simple criterion for determining whether or not a given
double well model is at criticality.  Suppose that the most probable escape
path (MPEP) extends along the axis from $S$ to the saddle, so that the
symmetry is as~yet unbroken.  We~know by the discussion in
Section~\ref{sec:bifurcation} that criticality is signalled by the
appearance of a nascent cusp at the saddle.  The nascent cusp itself is
not a focus, as Fig.~\ref{fig:flow}(b) makes clear.  But if the off-axis softening is
increased, the nascent cusp becomes a genuine cusp (\ie,~focus); it~moves
inward along the axis from the saddle toward~$S$.  This picture is
consistent with the interpretation of the near-axis instanton flow field
in~terms of the function~$Y(x)$, $0\le x\le x_s$, only~if the nascent cusp,
like a conventional on-axis focus, is a zero of~$Y$.

So the signal for criticality is $Y$ equalling zero at~$x=0$.  We~can
rephrase this as~follows.  {\em Critical double well models are those
models with unbroken symmetry for which the Jacobi equation ${\cal J}Y=0$
for the transverse deviation function~$Y$, equipped with boundary condition
$Y(x=x_s)=0$ and also with $Y(x=0)=0$, has a nontrivial (\/\ie,~nonzero)
solution.}  The nonzero solution $Y=Y_1(x)$, $0\le x\le x_s$, when it
exists, can be interpreted as a transverse soft mode of the zero-energy
Hamiltonian dynamics.  If~the off-axis softening is increased, the on-axis
MPEP will bifurcate.  Just beyond criticality, there will be two
symmetrically placed off-axis MPEP's from $S$ to the saddle.  They will be
of the form $\left(x^*(t),\pm\delta Y_1\left(x^*(t)\right)\right)$, for
some small~$\delta$.  This `motion in the direction of a soft mode' is a
standard bifurcation effect.  {\em At~criticality, the transverse soft mode~$Y_1$
describes the way in~which the two MPEP's separate.}

Suppose that the double well model is parametrized by an off-axis softening
parameter~$\alpha$, \ie, that $v_2=\alpha \hat v_2$ for some odd
function~$\hat v_2$, and that $v_0$~and~$u_1$ are independent of~$\alpha$.
Then by rewriting the Jacobi equation, one sees that the model will be at a
bifurcation point if and only if the Sturm-Liouville equation
\begin{equation}
\hat {\cal J} Y \defeq
\dfrac1{4\hat v_2(x)}
\dfrac{d}{dx}\left[ v_0(x) \dfrac{dY}{dx}\right] +
\dfrac1{4\hat v_2(x)} \left[ u_1'(x) - \dfrac{u_1^2(x)}{v_0(x)}\right] Y =
\alpha Y,
\end{equation}
equipped with Dirichlet boundary conditions $Y(x=0)=Y(x=x_s)=0$, has a
nonzero solution.  The Sturm-Liouville operator~$\hat{\cal J}$ may be
called a {\em normalized Jacobi operator\/}. 

We see that the set of critical values of~$\alpha$ is precisely the {\em
spectrum\/} of the normalized Jacobi operator!  Only the first critical
value (\ie,~lowest eigenvalue) $\alpha_c=\alpha_c^{(1)}$ will yield an
actual bifurcation of the MPEP\hbox{}. To~each higher critical
value~$\alpha_c^{(j)}$, $j=2,3,\ldots$, there corresponds a transverse
eigenmode~$Y_j$.  But after the first bifurcation, the on-axis instanton
trajectory is no~longer the physical MPEP\hbox{}. The higher
eigenmodes~$Y_j$, $j=2,3,\ldots$, which are oscillatory, govern the further
bifurcations of the on-axis instanton trajectory rather than the further
bifurcations (if~any) of the physical MPEP's, which have already moved
off-axis.

The case of the standard model~(\ref{eq:standard}) is instructive.  Substituting
from (\ref{eq:standardfirst})--(\ref{eq:standardlast}) one finds as
normalized Jacobi operator
\begin{equation}
\hat{\cal J} = -\dfrac1{4x}\dfrac{d}{dx}(x-x^3)\dfrac{d}{dx}
+\left[ \dfrac\mu2 + \dfrac{\mu^2(1+x^2)^2}{4x^2(1-x^2)}\right].
\end{equation}
It is easily verified that on the interval from $x=0$ to
$x=x_s=1$, this operator (when~equipped with Dirichlet boundary conditions) has 
spectrum
\begin{equation}
\label{eq:hardwon}
\alpha_c^{(j)} = j^2 + (3\mu-1)j + (2\mu^2-\mu), \qquad j=1,2,3,\ldots
\end{equation}
So in the standard model, the bifurcation of the physical MPEP occurs at
$\alpha_c=\alpha^{(1)}_c=2\mu(\mu+1)$.  Also, the standard model with
$\mu=1$ has $\alpha_c^{(j)}=(j+1)^2$, so the on-axis instanton
trajectory bifurcates at~$\alpha=4,9,16,\ldots$\ \ We~have several times
mentioned this curious progression of squares.  The eigenfunctions~$Y_j$
corresponding to the eigenvalues~$\alpha_c^{(j)}$, \ie, the transverse soft
modes appearing at~$\alpha=\alpha_c^{(j)}$, turn~out to be of the form
\begin{equation}
Y_j(x) = (x-x^3)^\mu q_j(x),
\end{equation}
where $q_j$ is an even polynomial of degree~$2j-2$.  Substituting into the
transverse Hamilton equation~(\ref{eq:bar1}) yields the analogous
transverse momentum deviations~$(P_y)_j$.  One gets
\begin{eqnarray*}
(P_y)_j(x) &=& -v_0(x) Y_j'(x) -u_1(x) Y_j(x)   \\
&=& (x-x^3)^{\mu} \left[
4\mu x^2 q_j(x) - (x-x^3)q_j'(x)
\right],
\end{eqnarray*}
so that in the standard model at~$\alpha=\alpha_c^{(j)}$,
\begin{equation}
w_2(x) = (P_y)_j(x)/Y_j(x) = 
4\mu x^2  - (x-x^3)q_j'(x)/q_j(x).
\end{equation}
If $j=1$ then $q_j$~reduces to a constant, and the second term is absent.
So~at the physical bifurcation point [\ie,~at
$\alpha=\alpha_c=2\mu(\mu+1)$], $w_2(x)$ equals~$4\mu x^2$.  Moreover,
$Y_1(x)=(x-x^3)^\mu$.  This transverse soft mode is seen clearly in 
Figs.\ \ref{fig:flow}(c)~and~\ref{fig:flow}(d), which show the behavior of the
$\mu=1$ standard model beyond the bifurcation point.  In~those figures the
off-axis MPEP's are roughly proportional to~$\pm(x-x^3)$, \ie, to~$\pm
Y_1$.  {\em As~$\alpha$~is increased above~$\alpha_c$, the MPEP's move in
the direction of the transverse soft mode.}

Recall that the profile of the WKB tube of probability density centered on
the $x$-axis is asymptotically Gaussian, and that at specified~$x$ this
Gaussian has variance~$\sim \epsilon/w_2(x)$.  But in the standard model,
at~the first (and~only physical) critical value
$\alpha=\alpha_c=2\mu(\mu+1)$, $w_2(x)$ equals~$4\mu x^2$.  That $w_2\to0$
as~$x\to0$ implies that at~criticality, the WKB tube {\em splays~out\/} as
the saddle is approached.  We~have already seen the $\mu=1$~case of this in
Section~\ref{subsec:MAE25}.  The splayout is what one would expect from our
picture of the bifurcation of the MPEP, which begins at the saddle, as a
phase transition.  It~simply says that on the $O(\epsilon^{1/2})$
transverse lengthscale, the Gaussian fluctuations about the MPEP grow
without bound as the nascent cusp is approached.

It is easy to see that this behavior is {\em universal\/}: it~occurs in any
critical double well model with a bifurcating MPEP\hbox{}.  If~the
(diagonal) linearization of the drift field~$\ov u$ at the saddle has
eigenvalues $(\lambda_x,\lambda_y)$, and $\mu$~is defined as~usual to
equal~$|\lambda_y|/\lambda_x$, then examination of the Jacobi equation
shows that the soft mode~$Y_1$ has asymptotics $Y_1(x)\sim Cx^\mu$,
$x\to0^+$, for some nonzero constant~$C$.  We~have mentioned this `approach
path' property elsewhere~\cite{MaierC}.  Also, examination of the Hamilton
equation for~$(P_y)_1$ shows that $(P_y)_1(x)\sim C' x^{\mu+2}$ for some
nonzero~$C'$.  So~at criticality, the quotient~$w_2(x)$ satisfies
(for~any~$\mu$)
\begin{equation}
\label{eq:z1}
w_2(x) =\partial^2 W/\partial y^2 (x,0)\sim \const \times x^2,
\end{equation}
as~$x\to0^+$, and the tube splayout always occurs.  Incidentally, it~follows
by integrating the transport equation~(\ref{eq:Keqn}) that
\begin{equation}
\label{eq:z2}
k_0(x) = K(x,0)\sim \const \times x^{-\mu},
\end{equation}
as~$x\to0^+$.  Equations (\ref{eq:z1})--(\ref{eq:z2}) summarize  the universal
behavior of the WKB tube near the saddle, in any critical double well
model.  They are the extension to arbitrary critical models of 
eqs.\ (\ref{eq:baz})--(\ref{eq:bazdognascent}), which applied only to the
critical variant ($\alpha=\alpha_c=4$) of the $\mu=1$ standard model.

We stressed in Section~\ref{subsec:MAE2} that a Kramers-type error function
approximation to the quasistationary density~$\rho_1$ near the saddle is
appropriate only if $w_2\to 2|u_1(0)|$ as the saddle is approached.
At~criticality, since $w_2\to0$ instead, in~order to apply the method of
matched asymptotic approximations we shall need to construct a different
boundary layer approximation.  This will give rise to the universal
non-Arrhenius MFPT asymptotics for models at criticality.

\section{{ Maslov-WKB Asymptotics}}
\label{sec:Maslov}
By building on the previous sections, we can analyse the weak-noise
behavior of double-well models with singularities.  We~have seen that
singularities may appear in the WKB approximation $K(\ov x)\exp[-W(\ov
x)/\epsilon]$ for the stationary density~$\rho_0$ and quasistationary
density~$\rho_1$.  The possible singular behaviors are summed~up in 
eqs.\ (\ref{eq:bazdogrealminus})--(\ref{eq:bazdogreal}), which apply to models
in~which the MPEP has already bifurcated, and 
eqs.\ (\ref{eq:z1})--(\ref{eq:z2}), which apply to models which are critical in
the sense of bifurcations.  Models in~which the MPEP has already bifurcated
have the property that the instanton trajectories emerging from~$S$ focus
at a point $(x_f,0)$ on the axis, with~${x_f>0}$.  The prefactor~$K$ of the
WKB approximation will diverge there.  In~critical models, there is no
actual on-axis focusing.  But the prefactor will nonetheless diverge at the saddle
point~$(0,0)$.

There is a standard procedure for extending the WKB approximation to such
singular points, by `glueing~in' auxiliary, non-WKB approximations.
It~originated with the work of Keller and Rubinow on short-wave
asymptotics~\cite{Keller60}, and has been most extensively developed by
Maslov~\cite{Maslov81}.  For a mathematically rigorous treatment, see
Duistermaat~\cite{Duistermaat74}.  See also Eckmann and
S\'en\'eor~\cite{Eckmann76}, for a partly pedagogical one-dimensional
treatment.  The procedure may be applied to the (formal) asymptotic
solutions of any partial differential equation of the form $H(\ov
x,-\epsilon\nabla)\rho=0$, where $H$~is a specified Hamiltonian.
Here we discuss its~application to the Smoluchowski equation, in arbitrary
dimensionality~$n$.

We know from Section~\ref{subsec:MAE1}  that mathematically, the
WKB approximation to $\rho_0$~and~$\rho_1$ is determined by (i)~a
Lagrangian manifold~${\cal M}$ in the $2n$-dimensional phase space, formed
by the bicharacteristics emanating from $(S,\ov 0)$ and~$(S',\ov
0)$, and (ii)~functions $W$~and~$K$ defined on this manifold, and
computable by integration along the bicharacteristics.  Of~the
points~${{P}}^{(i)}=(\ov x,\ov p^{(i)})$ `over' any
point~$\ov x$, only the one with least action is physical.  The values
there of~$W$ and~$K$ are the values $W(\ov x)$ and~$K(\ov x)$
appearing in the WKB approximation.

This geometric interpretation motivates the introduction of a new,
`diffraction integral' way of formulating the WKB approximation.  At~any
point ${{P}}=(\ov x,\ov p)$ on the Lagrangian manifold~$\cal
M$, we have
\begin{equation}
W({{P}})  = \int \ov p\cdot d\ov x,
\end{equation}
the line integral being taken along the bicharacteristic terminating
at~${P}$.  We~can define a Legendre transform~$\wt W$, satisfying
$\wt W = \ov x\cdot\ov p - W$, by
\begin{equation}
\wt W({{P}})  = \int \ov x\cdot d\ov p.
\end{equation}
It is natural to think of~$\wt W$ as a function of momentum~$\ov p$,
by projecting `sideways' onto momentum space.  Of~course $\wt W(\ov
p)$~is potentially multivalued, like~$W(\ov x)$.  For~$W$, it~is the
{\em least\/} of the possible values that is physical; for~$\wt W$,
it~is the~{\em most\/}.  But if one ignores the multivaluedness of~$\wt
W(\ov p)$, one can write
\begin{equation}
\label{eq:pointtointegraln}
K(\ov x)\exp[-W(\ov x)/\epsilon] \sim
\epsilon^{-n/2} \int \cdots \int \wt K(\ov p) \exp \left\{
\left[-\ov x\cdot \ov p + \wt W(\ov
p)\right]/\epsilon\right\}\,dp_1\cdots dp_n,
\end{equation}
where $K(\ov x)$ and~$\wt K(\ov p)$ are related by
\begin{equation}
\label{eq:alternativeall}
K(\ov x) \propto
\wt K(\ov p) \left/
\sqrt{\det\left[ - \dfrac{\partial^2 \wt W}{\partial p_i\partial
p_j}(\ov p)\right]}
\right.
=
\wt K(\ov p) \,
\sqrt{\det\left[ - \dfrac{\partial^2  W}{\partial x_i\partial
x_j}(\ov x)\right]},
\end{equation}
the correspondence between $\ov p$ and~$\ov x$ being given by
$\ov p(\ov x) = \partial W/\partial\ov x$, or $\ov
x(\ov p) = \partial \wt W/\partial\ov p$.  The asymptotic
equality in~(\ref{eq:pointtointegraln}), as~$\epsilon\to0$, is justified by
the method of steepest descent.  (It~may be necessary to cut~off the
integral at large momentum to ensure convergence.)  The method of steepest
descent automatically picks~out the point ${{P}}=(\ov x,\ov
p)$ `over'~$\ov x$ with the least action~$W({{P}})$.  We~shall
call~(\ref{eq:pointtointegraln}) a {\em diffraction integral
representation\/}, since (if~$\epsilon$~is pure imaginary) it~resembles the
diffraction integrals used in physical optics~\cite{Berry76}.

We have assumed that the Hessian matrix $\partial^2 W/\partial x_i\partial
x_j = \partial p_i/\partial x_j$, whose inverse is the matrix $\partial^2
\wt W/\partial p_i\partial p_j = \partial x_i/\partial p_j$, is negative
definite.  Actually it~is often possible to make sense of the above
formul\ae\ even when this is not the case, by analytic continuation.  It~is
also possible to avoid the problem of positive eigenvalues by taking the
Legendre transform with respect to a partial (incomplete) set of variables.
In~$n=2$ dimensions, this means with respect to a single variable~only.  For
~example, one could use the alternative integral representation
\begin{equation}
\label{eq:alternativex}
K(x,y) \exp [-W(x,y)/\epsilon] \sim \epsilon^{-1/2}
\int \wt K^{(x)}(p_x,y) \exp \left\{ \left[ -xp_x + \wt
W^{(x)}(p_x,y)\right]/\epsilon\right\}\, dp_x,
\end{equation}
where $\wt W^{(x)}= xp_x - W$ is regarded as a function of $p_x$ and~$y$,
and $K$~and~$\wt K^{(x)}$ are related by
\begin{equation}
K(x,y) \propto \wt K^{(x)}(p_x,y)\left/\sqrt{- \dfrac{\partial^2 \wt
W^{(x)}}{\partial p_x^2}(p_x,y)}\right.
= \wt K^{(x)}(p_x,y)\,\sqrt{- \dfrac{\partial^2
W}{\partial x^2\vphantom{p_x}}(x,y)}.
\label{eq:curioustransform}
\end{equation}
Here the correspondence between $(x,y)$ and~$(p_x,y)$ is given by
$p_x(x,y)=\partial W/\partial x$, or equivalently $x(p_x,y)=\partial \wt
W^{(x)}/\partial p_x$.

It is clear that the transformed prefactor~$\wt K$ (resp.\ $\wt
K^{(x)}$, etc.)\ in these integral representations, like $W$, $K$,
and~$\wt W$ (resp.\ $\wt W^{(x)}$, etc.), can be thought of as a
function on the Lagrangian manifold~$\cal M$.  Also, the momentum
integration can be viewed as an integration over~$\cal M$.  So~introducing
integral representations of this sort is really a way of replacing the
position-space WKB approximation $K(\ov x)\exp[-W(\ov x)]$
to~$\rho(\ov x)$ by~a smeared-out equivalent~one, or~ones, involving
integration over the manifold.  As~derived, these `momentum space'
approximations are accurate only to leading order as~$\epsilon\to0$, since
subdominant terms in~$\epsilon$ arising from the method of steepest descent
have been neglected.  But such terms could be incorporated, if~desired, by
adding $\epsilon$-dependent corrections to the transformed prefactor.

If the new formulations of the WKB approximation are equivalent to the old,
why have we introduced them?  The reason is that the equivalence holds only
at points~$\ov x$ at~which $K$~is finite.  At~singularities of~$K$, the
new formulations provide a means of computing the true $\epsilon\to0$
asymptotics of~$\rho$.  Moreover, they reveal how at~least some
singularities of~$K$ can be explained as {\em artifacts\/}, arising from
the way in~which $K$~is computed from~$\wt K$.  It~follows
from~(\ref{eq:alternativeall}) that if the determinant of the Hessian
matrix $\partial^2 W/\partial x_i\partial x_j = \partial p_i/\partial x_j$
diverges at some point~$\ov x$, then $\ov x$~will be a singularity
of~$K$ whenever $\wt K$~is nonzero at the corresponding momentum
$\ov p=\ov p(\ov x)$.  In~other words, singularities of~$K$ may
be more apparent than real: they can arise from points $(\ov x,\ov
p)$ on the manifold where $\wt K$~does not actually diverge.  A~similar
effect can arise from the representation~(\ref{eq:alternativex}), or from
any other diffraction integral representation.

The matrix $\partial p_i/\partial x_j$ is a matrix of partial slopes, which
specifies (to~first order) the shape of the manifold in the vicinity of the
point $(\ov x, \ov p) = \left(\ov x, \ov p(\ov x)\right)$.  Its~determinant
becomes infinite only when at~least one of its elements is infinite.  Such
a blowup occurs only at locations on the manifold where the
($n$-dimensional) tangent hyperplane to the manifold `turns vertical,' \ie,
points along a momentum direction in the $2n$-dimensional phase space.
This is precisely the behavior one sees at a {\em fold\/}, as~in
Figs.~\ref{fig:flow}(c), \ref{fig:flow}(d), and~\ref{fig:whorl}.  {\em The
folding over of the Lagrangian manifold can create singularities of~$K$.}
This is clearly the cause of the singular behavior of~$K$ at the on-axis
foci occurring in models with a bifurcated MPEP\hbox{}, though {\em not\/} of the
singular behavior at the nascent cusp occurring at criticality (which
cannot be transformed~away).

Whether or not a singularity of the prefactor~$K$ occurring at some point
$\ov x = {\ov x}^*$ is an artifact of this sort, by employing an
appropriate diffraction integral representation one may compute the true
weak-noise asymptotics of~$\rho({\ov x}^*)$.  One usually finds
leading-order behavior of the form $\const\times
\epsilon^{-\sigma}\exp[-W({\ov x}^*/\epsilon)]$, where $\sigma$~is
by~definition the singularity index of~${\ov x}^*$.  In~fact the
$\epsilon$-dependent prefactor~$\epsilon^{-\sigma}$ should appear at all
points~$\ov x$ within some $\epsilon$-dependent distance of~${\ov
x}^*$, which shrinks to zero as~$\epsilon\to0$.  Within this local region
an~asymptotically exact formula for~$\rho$, derived from the integral
representation, will be uniformly valid.  This asymptotic approximation
(non-WKB, at~least in the traditional sense) will match in the far~field to
the WKB approximation $K(\ov x)\exp[-W(\ov x)/\epsilon]$.

In Sections~\ref{sec:real} and~\ref{sec:nascent} we shall see how this
`glueing~in' procedure works, both in models with a bifurcated MPEP and in
models at criticality.  For the moment we note only that the construction
of a local approximation to~$\rho$, near the singular point~${\ov
x}^*$, depends crucially on the determination of the behavior of $\wt
W$ and~$\wt K$ near the corresponding point~${\ov p}^*$ in momentum
space.  The case when $\wt K$~is well-behaved (`slowly varying') in a
neighborhood of~${\ov p}^*$, and the singularity at~${\ov x} =
{\ov x}^*$ is an artifact, is the simplest.  Suppose that $\wt W =
\wt W({\ov p})$ can be expanded in a power series around ${\ov
p}={\ov p}^*$.  The matrix $\partial^2 \wt W/\partial p_i\partial
p_j$ must have a zero eigenvalue at~${\ov p}= {\ov p}^*$, since
otherwise the determinant of its inverse $\partial^2 W/\partial x_i\partial
x_j$ would not tend to infinity as~${\ov x}\to{\ov x}^*$, the
Lagrangian manifold would not turn vertical there, and the singularity
in~$K$ would not appear.  The term {\em catastrophe\/} is used to describe
what happens to the manifold at~$\ov x = {\ov x}^*$.  It~is a
standard result, due largely to Arnol'd~\cite{Arnold75}, that if the
manifold is smooth near~$({\ov x}^*,{\ov p}^*)$, the catastrophic
behavior at~${\ov x}={\ov x}^*$ can be captured by approximating
$\wt W=\wt W({\ov p})$ by one of a handful of polynomial
functions.  These are the `structurally stable' {\em elementary
catastrophes\/}.

A single example, illustrating the similarity to the Ginzburg-Landau theory
of phase transitions, will suffice.  In~$n$~dimensions, suppose that a
singularity at~${\ov x}={\ov x}^*$ arises as an artifact in the
above sense, and that ${\ov p}^*=\partial W/\partial{\ov
x}({\ov x}^*)$.  In~appropriate (linearly transformed) coordinates,
write
\begin{eqnarray}
{\ov z} &=& (z_1,\dots,z_n) = {\ov x}-{\ov x}^*\\
{\ov g} &=& (g_1,\dots,g_n) = {\ov p}-{\ov p}^*.
\end{eqnarray}
A particularly common sort of catastrophe (a~`cuspoid') would be described
locally by a single-variable Legendre transform of the form
\begin{equation}
\wt W^{(z_n)} (z_1,\dots,z_{n-1},g_n) = -\dfrac{a_{0} g_n^{n+2}}{n+2} 
-\dfrac{a_{1} z_1g_n^n}n - \dots -\dfrac{a_{n-1} z_{n-1}g_n^2}2
+R(z_1,\dots,z_{n-1})
\end{equation}
where $a_{0},\dots,a_{n-1}$ are constants, and $R(z_1,\dots,z_{n-1})$ is a
quadratic polynomial.  
Since $z_n = \partial \wt W^{(z_n)}/\partial g_n$, this 
expression implies
\begin{equation}
\label{eq:cuspoidequationofstate}
z_n = z_n(z_1,\dots,z_{n-1},g_n) = -a_{0} g_n^{n+1} - a_{1} z_1g_n^{n-1}
-\dots-a_{n-1} z_{n-1}g_n.
\end{equation}
The presence of a catastrophe at $(z_1,\dots,z_{n-1},g_n)=(0,\dots,0,0)$ is
signalled by the fact that $\partial^2\wt W/\partial g_n^2=\partial
z_n/\partial g_n$ equals zero there.

We~have already seen the $n=2$ version of
eq.~(\ref{eq:cuspoidequationofstate}) in Section~\ref{subsec:MAE3}, as a
phenomenological description of the shape of the manifold~$\cal M$ near an
on-axis focus.  Recall that we interpreted eq.~(\ref{eq:phenomenological}),
which is the $n=2$ version, in thermodynamic terms: as~the equation of
state of a substance undergoing a Ginzburg-Landau second-order phase
transition.  (\Eg,~$z_1$~is $T-T_c$, $z_2$~is a negative magnetic field,
and $g_2$~is magnetization.)  Equation~(\ref{eq:cuspoidequationofstate}) is
in~fact a {\em normal form\/} for the shape of a Lagrangian manifold near a
cuspoid singularity.  When~${n=2}$, the cuspoid is a {\em cusp\/}.
If~$n=1$, only the first term on the right-hand side
of~(\ref{eq:cuspoidequationofstate}) is present, and the cuspoid reduces
to a quadratic~{\em fold\/}.

In general, to~each possible polynomial expression (normal form) for the
Legendre-transformed action, there corresponds a non-WKB approximation
to~$\rho$ in a local region near~${\ov x}={\ov x}^*$, computed from
the appropriate diffraction integral.  These integrals serve to define the
{\em canonical diffraction functions\/} first explored by Maslov.  The
canonical diffraction functions include the classical Airy and Pearcey
functions, which arise from folds and cusps respectively~\cite{Berry76}.
We~shall study the cusp case further in the next section, as a warmup for
the study of the nascent cusp appearing at criticality.  The normal form
for the action near a nascent cusp will turn~out to be nonpolynomial, but
the Maslov-WKB technique will still apply.

We close this section by noting that diffraction integral representations
are also useful for incorporating symmetry constraints and boundary
conditions.  As~an example of this, consider behavior near the saddle point
of a double well model.  We~emphasized in Section~\ref{subsec:MAE1} that if
no bifurcation of the MPEP has occurred, the WKB tube of probability density
centered on the axis will be well~behaved as the saddle is approached.
In~particular, $w_2(x)=\partial^2 W/\partial y^2(x,0)$ will tend to
$2|u_1(0)|$ as~$x\to0^+$.  Since
\begin{equation}
\dfrac{\partial^2 W}{\partial x^2}(0,0) = \dfrac{\partial p_x}{\partial
x}(0,0) = -2v_0'(0)
\end{equation}
and $v_0(x)=u_x(x,0)$ is assumed to be smooth, in the absence of
bifurcations $W$~will to leading order be {\em locally quadratic\/} at the
saddle.  If~$\ov u(x,y)\approx(\lambda_x x,-|\lambda_y|y)$ is the
linearization of the drift at the saddle, we~have $u_1(0)=-|\lambda_y|$
and~${v_0'(0)=\lambda_x}$.  So,~near $(x,y)=(0,0)$,
\begin{equation}
W(x,y)\approx W(0,0)-\lambda_x x^2 + |\lambda_y| y^2.
\end{equation}
And
\begin{eqnarray}
\label{eq:doubletransform}
\wt W(p_x,p_y) &\approx&-W(0,0) -p_x^2/4\lambda_x + p_y^2/4|\lambda_y|\\
\wt  W^{(x)}(p_x,y) &\approx&-W(0,0) -p_x^2/4\lambda_x -|\lambda_y|y^2
\label{eq:partway}
\end{eqnarray}
will be the leading-order approximations to the Legendre-transformed actions.

Since the Hessian matrix $\partial^2\wt W/\partial p_i\partial p_j$ is not
negative definite, an integral representation of the
type~(\ref{eq:pointtointegraln}) is not appropriate.  But a representation of
the type~(\ref{eq:alternativex}) may be used.  In~the absence of
bifurcations $K$~and~$\wt K$ are well~behaved near the saddle, so
substituting (\ref{eq:partway}) into~(\ref{eq:alternativex}) yields
\begin{equation}
\label{eq:tired}
\rho(x,y) \sim
\const \times
\left\{\epsilon^{-1/2}\int\exp\left[-(xp_x +
p_x^2/4\lambda_x)/\epsilon\right]\,dp_x \right\}  e^{-|\lambda_y|y^2/\epsilon}.
\end{equation}
If $p_x$ here is integrated from $-\infty$ to~$\infty$, this approximation
will be even in~$x$.  It~will therefore serve as an approximation to the
{stationary} density~$\rho_0(x,y)$ near the saddle.  The integral may
be evaluated explicitly, and the approximation reduces to the standard
inverted Gaussian approximation
\begin{equation}
\rho_0(x,y) \sim \const\times e^{+\lambda_xx^2/\epsilon}
e^{-|\lambda_y|y^2/\epsilon}.
\end{equation}
But when approximating the {\em quasistationary\/} density~$\rho_1(x,y)$
near the saddle, one needs an approximate solution of the
Smoluchowski equation that is odd rather than even.  Such an approximate
solution is obtained by integrating $p_x$ from $0$~to~$\infty$ rather than
from $-\infty$~to~$\infty$.  If~this is in~fact done, eq.~(\ref{eq:tired})
reduces to~(\ref{eq:erfintegral}), the standard Kramers-type error function
approximation to the quasistationary density!

Although error function approximations originated (with~Kramers) in an
entirely different context, they fit naturally into the Maslov-WKB
framework.  We~conclude that diffraction integral representations can be
modified to incorporate the effects of symmetry constraints.
In~Section~\ref{subsec:nascent2} we shall use a similar half-range
integration in our integral representation for the quasistationary density
near a nascent cusp.

\section{{ Scaling Behavior Near a Cusp}}
\label{sec:real}
We can apply the Maslov-WKB method of the last section to symmetric double
well models in~which the MPEP has bifurcated, and the instanton
trajectories emerging from $S=(x_s,0)$ focus at a point~$(x_f,0)$, with
$0<x_f<x_s$.  As~we shall see, behavior near the focal point~$(x_f,0)$ is
best described in the language of critical phenomena.

The Maslov-WKB method was first applied to focusing (cusp) singularities in
two-dimensional models by Dykman {\em et~al.}~\cite{Dykman94}.  Their
analysis, which does not assume any sort of symmetry, specializes in the
case of symmetry about the $x$-axis to the following.  Assume that the
Legendre-transformed action $\wt W^{(y)}=yp_y-W$, regarded as a function of
$x$~and~$p_y$, may be asymptotically approximated near $(x,p_y)=(x_f,0)$ by
the cuspoid (codimension~$n=2$) normal form
\begin{equation}
\label{eq:newapprox}
\wt W^{(y)}(x,p_y) \sim
-\dfrac{a_0}4 p_y^4 -\dfrac{a_1}2 (x-x_f)p_y^2 - w_0(x).
\end{equation}
Here $a_0$ and~$a_1$ are positive constants, and $w_0(x)$ is simply
$W(x,0)$, \ie, $-\wt W^{(y)}(x,0)$.  Since $y(x,p_y)=\partial
\wt W^{(y)}/\partial p_y(x,p_y)$, this assumption is equivalent to
\begin{equation}
\label{eq:newapprox2}
y(x,p_y) \sim -a_0p_y^3 -a_1 (x-x_f) p_y
\end{equation}
which is the phenomenological (Ginzburg-Landau)
equation of state~(\ref{eq:phenomenological}), discussed at~length in
Section~\ref{subsec:MAE3}.  $\wt W^{(y)} = \wt W^{(y)}(x,p_y)$ can be
viewed as a Helmholtz free energy, just~as $W=W(x,y)$ can be viewed as a
Gibbs free energy.

The cuspoid form for~$\wt W^{(y)}$ is certainly
consistent with the folding of the Lagrangian manifold~$\cal M$, as~seen
(in~projection) in Figs.\ \ref{fig:flow}(c)~and~\ref{fig:flow}(d).  It~is
also consistent with the quantitative asymptotics of
Section~\ref{subsec:MAE3}.  Since $p_y=0$ corresponds to~$y=y(x,p_y)=0$,
(\ref{eq:newapprox2})~implies
\begin{eqnarray}
w_2(x)&=& \dfrac{\partial^2 W}{\partial y^2}(x,0) = \dfrac{\partial
p_y}{\partial y}(x,0) = \left[\dfrac{\partial y}{\partial
p_y}(x,0)\right]^{-1}\nonumber\\ &\sim& -a_1^{-1} (x-x_f)^{-1}, \qquad x\to x_f^+.
\label{eq:bardog}
\end{eqnarray}
This is precisely the near-focus blowup behavior of 
eqs.\ (\ref{eq:bazdogrealminus})--(\ref{eq:bazdogreal}), which we derived
analytically from the Riccati equation~(\ref{eq:w2eqn2}).  By comparing
(\ref{eq:bardog}) with~(\ref{eq:w2eqn2}), we~see that the constant $a_1$
must equal~$1/v_0(x=x_f)$, the {\em reciprocal speed\/} of the on-axis
instanton trajectory as it passes through the focus.  Since $x-x_f$ is
analogous to~$T-T_c$ and $w_2$~to a (negative) magnetic susceptibility, the
blowup of~(\ref{eq:bardog}) is analogous to the critical exponent~$\gamma$
of the focus, in thermodynamic language, equalling unity.

Dykman {\em et~al.}\ use a one-dimensional diffraction integral
representation, resembling~(\ref{eq:alternativex}) but with $x$~and~$y$
interchanged, to approximate the stationary probability density~$\rho_0$
near~$\ov x=(x_f,0)$.  A~crucial assumption is that the transformed
prefactor $\wt K^{(y)} = \wt K^{(y)}(x,p_y)$, which has no direct
thermodynamic interpretation, is well behaved (locally constant, or `slowly
varying') near $(x,p_y)=(x_f,0)$.  If~this is the case, and it may be
approximated by a constant, one can construct the Maslov-WKB approximation
\begin{eqnarray}
&&K(x,y)\exp[-W(x,y)/\eps]\\
&&\qquad\sim \eps^{-1/2} \int_{-\infty}^{\infty} \wt K^{(y)}(x,p_y)
\exp \left\{ \left[ -yp_y + \wt W^{(y)}(x,p_y)\right]/\eps 
\right\}\,dp_y\nonumber\\
&&\qquad\approx \eps^{-1/2} \wt K^{(y)}(x_f,0) e^{-W(x_f,0)/\eps}
\int_{-\infty}^\infty \exp \left\{ -\left[
\dfrac{a_0}4 p_y^4 + \dfrac{a_1}2 (x-x_f)p_y^2 + yp_y \right] 
/\eps \right\}\,dp_y.
\nonumber
\end{eqnarray}
In~terms of `stretched' variables $X\defeq(x-x_f)/\eps^{1/2}$ and
$Y\defeq y/\eps^{3/4}$ this becomes
\begin{equation}
\label{eq:DMSexpr}
\eps^{-1/4} \wt K^{(y)}(x_f,0)
e^{-W(x_f,0)/\eps}
e^{-[X W'(x_f,0)/\eps^{1/2} + (X^2/2)W''(x_f,0)]}
\,{\cal P}(a_1a_0^{-1/2} X, a_0^{-1/4} Y),
\end{equation}
where the primes denote derivatives with respect to~$x$.
Here the canonical diffraction function 
\begin{equation}
{\cal P}(u,v) \defeq \int_{-\infty}^{\infty} \exp
\left\{ - \left[ \dfrac{1}4t^4 + \dfrac{1}2 ut^2 + vt\right]\right\} \,dt
\end{equation}
is a modified (real) Pearcey function (cf.\ Paris~\cite{Paris91}).

The expression~(\ref{eq:DMSexpr}) is an asymptotically ($\eps\to0$) 
valid approximation to the stationary density~$\rho_0$ and
quasistationary density~$\rho_1$, on the $x-x_f=O(\eps^{1/2})$,
$y=O(\eps^{3/4})$ lengthscale near the cusp~$(x_f,0)$.  It~supplements the
WKB approximation, which is singular there.  One sees that on this
lengthscale, the pre-exponential factor in $\rho_0$ and~$\rho_1$ is
actually of magnitude~$O(\eps^{-1/4})$.  The {\em singularity index\/} of
the cusp equals~$1/4$, as in physical optics.

The absence of an `{\it i\/}' from the exponent gives rise to unusual
asymptotic behavior of the diffraction function.  The familiar Pearcey
fringes of physical optics are replaced by an exponential slope, which
becomes increasingly steep as~$\eps\to0$.  Beyond the cusp (\ie,
at~$x<x_f$, which is analogous to~$T<T_c$), the WKB approximation
$K(x,y)\exp[-W(x,y)/\eps]$ is again valid, but $W$~is no~longer
differentiable through the $x$-axis~\cite{MaierC}.  This is reflected in
the far-field asymptotics of the Pearcey function~${\cal P}$.  One can
show that in the far~field, \ie, as~$X=(x-x_f)/\eps^{1/2}\to-\infty$, the
expression~(\ref{eq:DMSexpr}) matches to a WKB approximation displaying
this nondifferentiability.  One can also show that the fold caustic
emanating from~$(x_f,0)$, as~in Fig.~\ref{fig:flow}(c), is {\em
nonphysical\/}.  It~arises from subdominant saddle points of the Pearcey
integral, and does not contribute to the leading weak-noise asymptotics for
$\rho_0$~and~$\rho_1$.  This is closely related to the fact that ``optimal
paths [\ie,~{\em physical\/} instanton trajectories] do~not encounter
caustics,'' as~Dykman {\em et~al.}~\cite{Dykman94} put~it.

\smallskip
Now the preceding Maslov-WKB treatment is satisfactory so~far as it goes.
But it leaves unresolved the issue of the validity of the Ginzburg-Landau
approximation.  The quartic normal form~(\ref{eq:newapprox}) for $\wt
W^{(y)}=\wt W^{(y)}(x,p_y)$, and the cubic equation of
state~(\ref{eq:newapprox2}) for its first derivative $y=y(x,p_y)$, model a
second-order phase transition with {\em mean~field\/} (\ie,~classical) {\em
critical exponents\/}.  Equivalently, they model the critical behavior of a
system which, though it~has a phase transition, has a {\em smooth
thermodynamic surface\/}.  In~the present context, assuming the local
validity of the Ginzburg-Landau approximation amounts to assuming that the
Lagrangian manifold~$\cal M$ is smooth through the point $(x,p_y)=(x_f,0)$.
Of~course the surface turns vertical there, causing $\partial p_y/\partial
y$ to diverge.  The assumption is that the singularity can be transformed
away by using $x$~and~$p_y$, rather than $x$~and~$y$, as independent
variables.

This assumption requires proof.  One could presumably justify~it by
analysing the smoothness (and~blowup) properties of solutions of the
Hamilton-Jacobi equation.  But we shall give a different, more physical
justification.  First, we shall model the local behavior of $W$
and~$\wt W^{(y)}$ by a scaling law, as in the modern theory of
critical phenomena.  Our treatment will serve as a warmup for
Section~\ref{sec:nascent}, where we shall analyse the much more complicated
(nonclassical) singularity appearing in models where the MPEP is beginning to
bifurcate.

To see that a scaling law is appropriate in models with a bifurcated
MPEP\hbox{}, consider the behavior of the on-axis transverse derivatives
$w_{2m} = \partial^{2m} W/\partial x^{2m}(x,0)$ as~$x\to x_f^+$.  We~know
by (\ref{eq:bazdogrealminus})--(\ref{eq:bazdogreal}) that $w_2$~diverges
as~$(x-x_f)^{-1}$.  The Riccati equation satisfied by~$w_2$ is only the
first of a hierarchy of ordinary differential equations, describing the
evolution of the functions~$w_{2m}$ as one moves along the on-axis
instanton trajectory from~$S$ (where $x=x_s$, and~$t=-\infty$) to the
saddle (where $x=0$, and~$t=+\infty$).  For~example, $w_4$~satisfies the
ODE~(\ref{eq:w4eqn}).  $w_2$~appears in each of the higher equations, and
its blowup will induce a blowup of $w_4$,~$w_6,\ldots$\ \ It~is not
difficult to show that
\begin{equation}
\label{eq:blowups}
w_{2m}(x) = \dfrac{\partial^{2m}W}{\partial x^{2m}}(x,0)
\sim \const \times (x-x_f)^{-(3m-2)}, \qquad x\to x_f^+,
\end{equation}
for $2m=2,4,6,\ldots$\ \ These blowup rates motivate the {\em scaling
Ansatz}
\begin{equation}
\label{eq:realscalingAnsatz}
W(x,y) \sim W(x,0) + (x-x_f)^2 h_{\pm} \left(\dfrac y {|x-x_f|^{3/2}}\right), 
\qquad
x\to x_f^\pm
\end{equation}
for the behavior of~$W$ near the cusp~$(x_f,0)$.  Here the exponents
$2$~and~$3/2$ are determined uniquely by the $m$-dependence of the blowup rates,
and the functions~$h_\pm(\cdot)$ of the scaling
variable $z\defeq y/|x-x_f|^{3/2}$ are not yet determined (though they must be
even).  This Ansatz is  assumed to be accurate to~$O((x-x_f)^2)$, when
$y=O(|x-x_f|^{3/2})$.    We~could equally well posit
\begin{equation}
\label{eq:scalingform}
W\left(x,z|x-x_f|^{3/2}\right) \sim W(x_f,0) + (x-x_f)W'(x_f,0) +
\dfrac{(x-x_f)^2}2 W''(x_f,0) + (x-x_f)^2 h_{\pm} (z)
\end{equation}
as~$x\to x_f^\pm$, since we are assuming the accuracy of the scaling Ansatz
only up to~$O((x-x_f)^{2})$.  The first three terms in this asymptotic
approximation are `regular'; the~scaling behavior appears only in the final,
singular term. 

The exponents $2$ and~$3/2$ are typical of a mean field theory.  One can
show that the scaling functions~$h_\pm$ are also those of a mean~field
theory.  They may be computed by substituting the scaling
Ansatz~(\ref{eq:scalingform}) into the Hamilton-Jacobi equation $H(\ov
x,\nabla W)=0$.  For this, one needs to rewrite the
Hamilton-Jacobi equation in~terms of the independent variables $x$~and~$z$.
Using the formula~(\ref{eq:Hamiltonian}) for~$H$, and the
expansions~(\ref{eq:velexp}), one finds
\begin{eqnarray}
H(\ov x,\ov p) &=& \dfrac{p_x^2}2 
+ \dfrac{p_y^2}2 + u_x(x,y)p_x + u_y(x,y)p_y\nonumber\\
&=&\dfrac{p_x^2}2 + \dfrac{p_y^2}2 + \left[v_0(x)+v_2(x)y^2+\cdots\right]p_x 
+\left[u_1(x)y + u_3y^3+\cdots\right]p_y \nonumber\\
&\approx& \dfrac{p_x^2}2 + \dfrac{p_y^2}2 + \Bigl[v_0(x_f) \pm
v_0'(x_f)\left|x-x_f\right|\Bigr]\,p_x 
\label{eq:astar}
\end{eqnarray}
up~to $O(|x-x_f|^{1})$ accuracy, since $y=z|x-x_f|^{3/2}$.  It~follows from
the scaling form~(\ref{eq:realscalingAnsatz}) that up~to $O(|x-x_f|^{1})$
accuracy,
\begin{eqnarray}
p_x(x,y) = \dfrac{\partial W}{\partial x}(x,y) &\sim&
-2v_0(x) \pm \left[2h_\pm(z) 
- (3/2)zh_\pm'(z)\right]\left|x-x_f\right| \nonumber\\
\label{eq:pxapprox}
&\approx& -2\Bigl[v_0(x_f) \pm v_0'(x_f)\left|x-x_f\right|\Bigr]
\pm \left[2h_\pm(z) - (3/2)zh_\pm'(z)\right]\left|x-x_f\right| \\
\label{eq:pyapprox}
p_y(x,y) = \dfrac{\partial W}{\partial y}(x,y) 
&\sim& \left|x-x_f\right|^{1/2} h'_\pm(z)
\end{eqnarray}
where we have used the fact (see Section~\ref{subsec:MAE2}) that
$W'(x,0)=w_0'(x)=-2v_0(x)$.  Substituting
(\ref{eq:pxapprox})--(\ref{eq:pyapprox}) into~(\ref{eq:astar}), and setting
the coefficient of~$|x-x_f|^1$ equal to zero, yields the~ODE
\begin{equation}
\label{eq:veryadhocminus}
(h_\pm')^2 = \pm v_0(x_f)\left[ 4h_\pm - 3z h_\pm'\right].
\end{equation}
It is easier to solve for $z$ as a function of~$h_\pm'$, than for $h_\pm$
as a function of~$z$.  One finds
\begin{equation}
\label{eq:veryadhoc}
z=z(h_\pm') = -C(h_\pm')^3 \mp v_0(x_f)^{-1}h_\pm'
\end{equation}
where $C$~is undetermined.  But $z=y/|x-x_f|^{3/2}$, and
by~(\ref{eq:pyapprox}), $h_\pm'\sim p_y/|x-x_f|^{1/2}$.  Rewriting~$z$ in~terms
of $y$ and~$|x-x_f|$, and $h'$ in~terms of $p_y$ and~$|x-x_f|$, yields
\begin{eqnarray}
\label{eq:doubleeqn}
y=y(x,p_y) &\sim& -Cp_y^3 \mp v_0(x_f)^{-1}\left|x-x_f\right|p_y \\
&=&-Cp_y^3 -v_0(x_f)^{-1} (x-x_f)p_y.\nonumber
\end{eqnarray}
If one identifies the model-dependent constant~$C$
with~$a_0$, this is precisely eq.~(\ref{eq:newapprox2}), the mean~field
(Ginzburg-Landau) equation of state!  It~is valid on both sides of the
on-axis focus, \ie, both when $x-x_f>0$ and when~$x-x_f<0$.

This derivation illustrates how one may go from the pattern of blowup rates
of the transverse derivatives $w_{2m}(x) = \partial^{2m} W/\partial y^{2m}(x,0)$
as $(x_f,0)$~is approached, to~a scaling form for~$W$, to~an equation of
state.  The singular behavior of the WKB prefactor~$K$ can be analysed
similarly (we~only summarize the analysis).  We~know by~(\ref{eq:bazdogreal})
that $K(x,0)$ diverges as~$(x-x_f)^{-1/2}$ when~$x\to x_f^+$.  
A~scaling form
\begin{equation}
\label{eq:KAnsatz}
K(x,y) \sim\const \times \left|x-x_f\right|^{-1/2} q_\pm
\left(\dfrac{y}{\left|x-x_f\right|^{3/2}}\right), \qquad x\to x_f^\pm,
\end{equation}
modelled after the scaling form~(\ref{eq:realscalingAnsatz}) for~$W$, may
be used to approximate~$K$ away from the $x$-axis.  This approximation
should be accurate to~$O(|x-x_f|^{-1/2})$ as~$x\to x_f^\pm$, when
$y=O(|x-x_f|^{3/2})$.  By~substituting the two scaling forms
(\ref{eq:realscalingAnsatz}) and~(\ref{eq:KAnsatz}) into the transport
equation~(\ref{eq:origKeqn}) for~$K$, and working to leading order
near~$(x_f,0)$, one can determine the scaling functions~$q_\pm=q_\pm(z)$.
It~is easily verified that collecting the $O(|x-x_f|^{-3/2})$ terms in the
transport equation yields the~ODE
\begin{equation}
\label{eq:stuffode}
\left[ 2h_\pm' \pm 3v_0(x_f) z\right] 
q_\pm^{\,\prime} + \left[ h_\pm'' \pm v_0(x_f)\right] q_\pm = 0,
\end{equation}
which $q_\pm=q_\pm(z)$ must satisfy.
Using elementary calculus, and the fact that $h_\pm=h_\pm(z)$ satisfies
the ODE~(\ref{eq:veryadhocminus}), one can show that
eq.~(\ref{eq:stuffode}) has solution
\begin{equation}
\label{eq:Kscalingfuncs}
q_\pm(z) = \const\times\sqrt{-h_\pm''(z)}.
\end{equation}
But since $z=y/|x-x_f|^{3/2}$, we know by~(\ref{eq:pyapprox}) that
\begin{equation}
\label{eq:pyapprox2}
h_\pm''(z) \sim |x-x_f| \dfrac{\partial p_y}{\partial y}
\left(x,y=z|x-x_f|^{3/2}\right). 
\end{equation}
Substituting (\ref{eq:Kscalingfuncs}) and~(\ref{eq:pyapprox2})
into the scaling form~(\ref{eq:KAnsatz}) for~$K$ reduces~it to
\begin{equation}
\label{eq:simplicity}
K(x,y) \sim \const\times 
\sqrt{-\dfrac{\vphantom{\partial^2}\partial p_y}{\partial y}(x,y)}.
\end{equation}
This asymptotic approximation is very simple, and has a profound consequence.
We~know that
the transformed prefactor $\wt K^{(y)}(x,p_y)$ can be obtained
from~$K(x,y)$ by dividing by a `Van~Vleck factor,' as~in~(\ref{eq:curioustransform}).
We~therefore have that
\begin{eqnarray}
\label{eq:seventwenty}
\wt K^{(y)}(x,p_y) &\propto& K(x,y)
\left/ \sqrt{-\dfrac{\partial^2 W}{\partial y^2}(x,y)}\right.\\
&\sim& \const,
\label{eq:seventwentyone}
\end{eqnarray}
since $\partial^2 W/\partial y^2 = \partial p_y / \partial y$.  This {\em
constant\/} asymptotic approximation is accurate to leading order as~$x\to
x_f$, when $y=O(|x-x_f|^{3/2})$.

We~have just deduced that on the appropriate lengthscale near the focus,
\ie, $x-x_f=o(1)$ and~$y=O\bigl(|x-x_f|^{3/2}\bigr)$, {\em the~transformed
WKB prefactor~$\wt K^{(y)}$ does not diverge\/}.  $\wt K^{(y)}$,~unlike the
prefactor $K$ itself, is asymptotically constant near the focus.  This was
the crucial assumption made by Dykman {\em et~al.}, and we see that like
the Ginzburg-Landau normal form for the Legendre-transformed action, it~is
justified by our scaling theory of local behavior.

We conclude that at~least in the case of a generic (cusp) singularity,
by~investigating the blowup rates of the transverse action derivatives as
the singularity is approached, one can derive scaling relations for
$W$~and~$K$, and ultimately construct a Maslov-WKB approximation to the
stationary probability density near the singularity.  This technique is not
restricted to singularities of the classical Ginzburg-Landau type.

\section{{Scaling Behavior Near a Nascent Cusp}}
\label{sec:nascent}
Finally, we can construct a scaling theory of weak-noise behavior near
the `nascent cusp' singularity appearing at the saddle point of any
symmetric double well model, at~the onset of bifurcation.  The construction
will closely parallel the construction of the last section. But several novel
features will appear.  We~shall find that  Legendre-transformed versions of
the action
are approximated, in the vicinity of a nascent cusp, by  {\em
nonpolynomial\/} normal forms.  Equivalently, the nascent cusp singularity,
unlike an on-axis focus, will prove to have {\em nonclassical\/} critical
exponents.  The exponents will depend continuously on the parameter
$\mu\defeq|\lambda_y|/\lambda_x$, which characterizes the linearized drift
field at the saddle.

The universal presence at~criticality of a nongeneric two-sided caustic
(which, as~shown in~Fig.~\ref{fig:nascent}, extends sideways from the
saddle point) will follow from the nonpolynomial normal forms for the
Legendre-transformed actions.  Indeed, one of the normal forms will supply
a {\em nonpolynomial unfolding\/} of the nongeneric caustic.  Moreover, the
fact that the critical exponents of the nascent cusp are model-dependent
and continuously varying will induce a continuously varying singularity
index, and a continuously varying prefactor exponent in the non-Arrhenius
weak-noise MFPT asymptotics.  To~see this, we shall have to go beyond the
WKB approximation, by applying the Maslov-WKB method.
In~Section~\ref{subsec:nascent1} we analyse the scaling properties of the
action and the WKB prefactor, and in Section~\ref{subsec:nascent15},
we~compare our scaling formul\ae\ with numerical data.
In~Section~\ref{subsec:nascent2} we~apply the Maslov-WKB method, and
construct weak-noise approximations to the stationary and quasistationary
probability densities near the saddle.

\subsection{Scaling in the WKB Approximation}
\label{subsec:nascent1}
Our scaling treatment of the nascent cusp begins with an investigation of
the blowup rates of the transverse action derivatives
$w_{2m}(x)=\partial^{2m}W/\partial y^{2m}(x,0)$, as~$x\to0^+$.  Up~to now
we have written down only the ODE's satisfied by~$w_2$ (\ie,~the Riccati
equation~(\ref{eq:w2eqn})) and~$w_4$ (\ie, eq.~(\ref{eq:w4eqn})).  The full
hierarchy of ODE's may be derived by substituting the Taylor series
$\sum_{m=0}^\infty w_{2m}(x)y^{2m}/(2m)!$ for $W(x,y)$ into the Hamilton-Jacobi
equation $H(\ov x, \nabla W)=0$, and separating~out the
coefficients of each power of~$y$.  One finds
\begin{eqnarray}
\label{eq:hierarchy}
\dot w_{2m} &=& -v_0 w_{2m}' \\
&= &
-\sum_{k=1}^m {{2m}\choose{2k-1}}[w_{2k}/2 + \tilde u_{2k-1}]w_{2m-2k+2}
-\sum_{j=1}^{m-1} {{2m}\choose{2j}}
[ w_{2j}'/2 + \tilde v_{2j}]w_{2m-2j}'
+2\tilde v_0\tilde v_{2m},
\nonumber
\end{eqnarray}
where $\tilde u_{2j+1}\defeq (2j+1)!\,u_{2j+1}$ and $\tilde v_{2j}\defeq
(2j)!\,v_{2j}$, and $u_{2j+1}$~and~$v_{2j}$ are the drift velocity
derivatives defined in~(\ref{eq:velexp}).  As~usual, the time derivative
here is with respect to transit time of the on-axis instanton trajectory,
which satisfies $\dot x=-v_0(x)$ as it moves from $S=(x_s,0)$ to the
saddle.  Since $v_0(x)\defeq u_x(x,0)$, this trajectory $t\mapsto x^*(t)$
moves anti-parallel to the drift.  And since $\ov u(x,y)\approx(\lambda_x
x,-|\lambda_y|y)$ near~$(0,0)$, $x^*(t)$~is approximated (as~$t\to+\infty$)
by~$\const\times e^{-\lambda_x t}$.

\begin{figure}
\begin{center}
\epsfxsize=4in		
\leavevmode\epsfbox{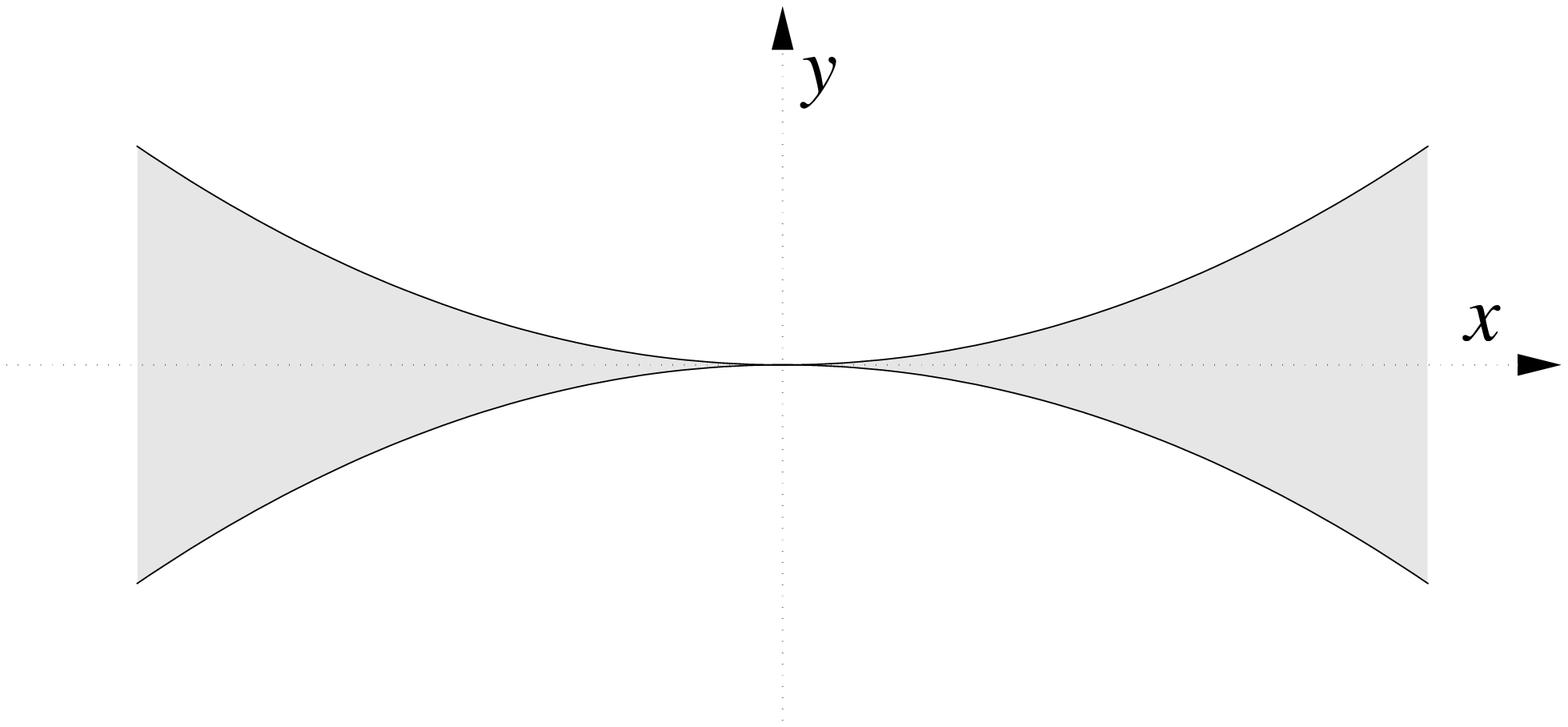}
\end{center}
\caption{A sketch of the near-axis Region~N\hbox{}, defined by
$|y|/|x|^{2\mu}\le\protect\const$.  The choice of constant is immaterial,
so~long as it~is positive.  Most of our expansions for the action~$W$ and
its~Legendre transforms, in double well models at~criticality, are valid
as~$(x,y)\to(0,0)$ from within Region~N\hbox{}.  The region could
equally well be defined by $|p_y|/|x|^{2\mu}\le\protect\const$,
$|y|/|p_x|^{2\mu}\le\protect\const$,
or~$|p_y|/|p_x|^{2\mu}\le\protect\const$.  To~leading order near~$(0,0)$,
the four definitions are equivalent.}
\label{fig:N}
\end{figure}

We showed in Section~\ref{subsec:MAE3}, by analysing the Jacobi equation
satisfied by the transverse soft mode, that 
\begin{equation}
\label{eq:atypical}
w_2(x)\sim \const\times x^2, \quad x\to0^+
\end{equation}
in any double well model at the onset of bifurcation (see 
eqs.\ (\ref{eq:z1})--(\ref{eq:z2}), and Fig.~\ref{fig:dls}).
If~the fact that $w_2\to0$ as~$x\to 0^+$
(\ie, as~$t\to+\infty$) is substituted into the general
ODE~(\ref{eq:hierarchy}), it~is easy to show, by integrating forward
in~time toward $t=+\infty$, that
\begin{equation}
\label{eq:nascentpattern}
w_{2m}(x) \sim \const\times x^{-(4m-4)\mu}, \quad x\to0^+,
\end{equation}
for $2m=4,6,8,\ldots$\ \ This pattern of blowup rates, as the saddle is
approached, motivates the scaling Ansatz (cf.~(\ref{eq:realscalingAnsatz}))
\begin{equation}
W(x,y)\sim W(x,0) + |x|^{4\mu}h(y/|x|^{2\mu}), \qquad x\to0.
\label{eq:firstscaling}
\end{equation}
Here $h(\cdot)$ is some (even) scaling function, as yet undetermined, and
the exponents $4\mu$ and~$2\mu$ are determined uniquely by the
$m$-dependence of the blowup rates of~(\ref{eq:nascentpattern}).
This Ansatz is assumed to be accurate to~$O(|x|^{4\mu})$ as~$x\to0$, when
$y=O(|x|^{2\mu})$. 
We~could equally well posit a finite-length asymptotic expansion for
$W(x,z|x|^{2\mu})$, namely
\begin{equation}
\label{eq:firstscalingextra}
W(x,z|x|^{2\mu}) \sim \left(
\sum_{k=0}^{\lfloor 4\mu\rfloor} 
\dfrac{\partial^{k}W}{\partial x^{k}}(0,0) \dfrac{x^{k}}{k!}
\right) + |x|^{4\mu}h(z), \qquad x\to0.
\end{equation}
Here $z\defeq y/|x|^{2\mu}$ is the scaling variable.  Only even powers
of~$x$ appear in the summation, and by~convention, here and below $\lfloor
4\mu\rfloor$~denotes the greatest {\em even\/} integer less than or equal
to~$4\mu$.  The expansion~(\ref{eq:firstscalingextra}) is assumed to be
accurate to~$O(|x|^{4\mu})$, at~any fixed value of~$z$.  It~can be
thought~of as an asymptotic development of~$W(\ov x)$, as~$\ov x\to\ov 0$
from within the near-axis region defined by the condition
$\left|z\right|\le\const$.  This condition defines a notch-shaped region,
which we call Region~N\hbox{}.  (See Fig.~\ref{fig:N}.)

It~follows by differentiating~(\ref{eq:firstscaling}) twice with respect
to~$y$ that $w_2(x)\sim h''(0)$ as~$x\to0$.  For consistency with the
`splayout' behavior $w_2(x)\sim \const\times x^2$ of~(\ref{eq:atypical}),
we~must have $h''(0)=0$.  Notice the slight discrepancy: the falloff rate
of~$w_2$ is not fully captured by the scaling Ansatz.  Actually, this is
unsurprising.  A~term proportional to~$x^2y^2$ in~$W$, such~as would arise
from the $O(x^2)$ falloff of~$w_2$ as~$x\to0$, would (in~terms of $x$ and
$z=y/|x|^{2\mu}$) be proportional to~$|x|^{4\mu+2}z^2$.  It~would therefore
be negligible in comparison to the scaling term~$|x|^{4\mu}h(z)$,
as~$x\to0$.  The scaling term captures the blowup as~$x\to0$ of
$w_4$,~$w_6$,~$w_8,\ldots$, but capturing the precise falloff rate of~$w_2$
would require a more refined analysis.  We~shall not attempt to include in
our Ansatz the `sub-scaling' terms that such an analysis would require.

The scaling function~$h(\cdot)$ may be computed by the technique used in
Section~\ref{sec:real}.  By~substituting the
expression~(\ref{eq:firstscalingextra}) into the Hamilton-Jacobi
equation~$H(\ov x,\nabla W)=0$, rewriting the Hamilton-Jacobi
equation in~terms of the independent variables $x$~and~$z$, and setting the
coefficient of~$|x|^{4\mu}$ equal to zero, one obtains an ODE for~$h=h(z)$.
This ODE turns~out to be (cf.~(\ref{eq:veryadhocminus}))
\begin{equation}
\label{eq:ODEforh}
(h')^2 = 2\left|\lambda_y\right|\left[ 4h - zh'\right].
\end{equation}
As~in Section~\ref{sec:real}, it~is easier to solve for $z$~as a function
of~$h'$, than for $h$ as a function of~$z$.  One finds
(cf.~(\ref{eq:veryadhoc}))
\begin{equation}
z=z(h') = h'/2|\lambda_y| + c(h')^{1/3},
\end{equation}
where $c$~is undetermined.  But $z=y/|x|^{2\mu}$ and~$p_y=\partial
W/\partial y\sim |x|^{2\mu}h'(z)$.  Rewriting $z$ in~terms of $x$~and~$y$,
and $h'$ in terms of $x$~and~$p_y$, yields an asymptotically accurate
equation of state (cf.~(\ref{eq:doubleeqn}))
\begin{equation}
\label{eq:eqstate1}
y=y(x,p_y) \sim p_y/2|\lambda_y| + c_{y;x,p_y} |x|^{4\mu/3} p_y^{1/3},
\end{equation}
where $c_{y;x,p_y}= c$.  In~practice the model-dependent constant~$c$
would be computed numerically, by fitting~(\ref{eq:eqstate1}) to the flow
field of instanton trajectories in the vicinity of the saddle.  We~must
have $c>0$, since the map $p_y\mapsto y$ is necessarily monotone
increasing near the saddle.  This is because `whorling,' as
in~Fig.~\ref{fig:whorl}, occurs only in models with a bifurcated
MPEP\hbox{}.  Whorling is absent at~criticality, \ie, at~the onset of
bifurcation.

The equation of state~(\ref{eq:eqstate1}) is certainly not of the classical
Ginzburg-Landau form.  By~anti-differentiating~it, we can obtain an equally
unusual approximation to the Legendre-transformed action $\wt
W^{(y)}=yp_y-W$, where $p_y=\partial W/\partial y$.  Since $y=\partial \wt
W^{(y)}/\partial p_y$, we~necessarily have
\begin{eqnarray}
\label{eq:newapprox3a}
\wt W^{(y)}(x,p_y) &\sim& -W(x,0) + p_y^2/4|\lambda_y| + C_{x,p_y}
|x|^{4\mu/3} p_y^{4/3}\\
&\sim& -\left(\sum_{k=0}^{\lfloor 4\mu\rfloor} 
\dfrac{\partial^{k}W}{\partial x^{k}}(0,0) 
\dfrac{x^{k}}{k!}\right)  + p_y^2/4|\lambda_y| + C_{x,p_y}
|x|^{4\mu/3} p_y^{4/3},
\label{eq:newapprox3b}
\end{eqnarray}
where $C_{x,p_y}= 3c/4$.  This asymptotic approximation should be
accurate to~$O(|x|^{4\mu})$, when $p_y=O(|x|^{2\mu})$ [\ie,~when
$y=O(|x|^{2\mu})$, or when $\ov x\to\ov 0$ from within Region~N].

The formula~(\ref{eq:newapprox3b}) can be called a {\em nonpolynomial
normal form\/} for the transformed action~$\wt W^{(y)}$ near the nascent
cusp.  Notice that as~$\ov x\to\ov 0$, the final, nonpolynomial term
$C_{x,p_y}|x|^{4\mu/3}p_y^{4/3}$ is significant in a relative sense only
within Region~N\hbox{}.  In~the far field of the $p_y=O(|x|^{2\mu})$
lengthscale, as~$x\to0$ it~is increasingly dominated by the $p_y^2$~term,
and the normal form reduces to a polynomial.  The
$C_{x,p_y}|x|^{4\mu/3}p_y^{4/3}$ term plays a much more important role in
the near~field.  One can think of~(\ref{eq:newapprox3b}) as providing an
{\em interpolation\/} between the non-polynomial asymptotic development
that is valid as~$\ov x\to\ov 0$ from within Region~N\hbox{}, and the
polynomial development that is valid as~$\ov x\to\ov 0$ from within its
far~field.  {\em The scaling behavior is visible only within Region~N.}

It~is worth noting that despite its asymptotic validity, the nonpolynomial
normal form~(\ref{eq:newapprox3b}) 
does not fully capture the $p_y\to0$ behavior of~$\wt W^{(y)}(x,p_y)$
at~{\em fixed\/}, nonzero~$x$.  If~the nonanalytic $p_y^{4/3}$~falloff were
exact, it~would follow by differentiating twice with respect to~$p_y$ that
$\partial^2 \wt W^{(y)}/\partial p_y^2$, \ie, $\partial y/\partial p_y$,
would diverge as~$p_y\to0$.  This would imply that $w_2=\partial
p_y/\partial y(y=0)$ would be {\em identically zero\/} at~any nonzero~$x$.
But we know that $w_2(x)\sim \const\times x^2$, $x\to0$.  The discrepancy
is due~to the fact that the nonzero~$w_2$ near~$x=0$ arises from
`sub-scaling' behavior that we are not attempting to model.  It~is not
difficult to see that at~fixed nonzero~$x$, the apparent nonanalyticity
at~$p_y=0$ must be `rounded' at a lengthscale $p_y=O(|x|^{2\mu+3})$, or
equivalently at~$y=O(|x|^{2\mu+1})$, to~yield consistency with the
$w_2(x)\sim \const\times x^2$ asymptotics.  However, on~the
$p_y=O(|x|^{2\mu})$ lengthscale the rounding becomes invisible as~$x\to0$.

With its continuously varying (in~general, irrational) exponent~$4\mu/3$,
the normal form for~$\wt W^{(y)}$ in Region~N looks quite different from
the normal forms of catastrophe theory~\cite{Arnold75,Berry76}.  Its~most
striking feature is the non-analyticity at~$(x,p_y)=(0,0)$, which can be
interpreted thermodynamically.  Recall that $\wt W^{(y)}$ (like~$\wt
W^{(x)}$, $\wt W$, and~$W$) can be viewed as a {\em thermodynamic
potential\/} on the thermodynamic surface (\ie,~Lagrangian manifold)~$\cal
M$.  In~fact, through its derivatives it~determines the shape of~$\cal M$.
So~at $(x,p_y)=(0,0)$, or equivalently at~$(x,y)=(0,0)$, the surface~$\cal
M$ will itself be non-analytic.  However, as~$\mu$~increases, $\wt
W^{(y)}$~becomes increasingly differentiable (with~respect to~$x$,
at~least) at~$x=0$.  The order of the `phase transition' appearing at the
saddle at~criticality is, therefore, an increasing function of~$\mu$.

We can Legendre-transform the normal form for~$\wt W^{(y)}$ to obtain a
normal form for the
double Legendre
transform $\wt W=\ourvec x\cdot\ourvec p\!-\!W= xp_x+\wt W^{(y)}$, as~a
function of $p_x$ and~$p_y$.
A~further Legendre transform will yield a normal form for the remaining
thermodynamic potential,~$\wt W^{(x)}$.  We~sketch only the first of these
two computations. 
Differentiating (\ref{eq:newapprox3a})--(\ref{eq:newapprox3b}) with respect
to~$x$, and using $p_x=\partial W/\partial x = - \partial \wt
W^{(y)}/\partial x$, yields
\begin{eqnarray}
p_x= p_x(x,p_y) &\sim& p_x(x,0) +  
c_{p_x;x,p_y} \left[\left|x\right|^{4\mu/3-1}\sgn x\right] p_y^{4/3}\\
&\sim& \left(-2\lambda_x x + \cdots 
+ \const\times x^{\lfloor 4\mu\rfloor-1}\right) +  
c_{p_x;x,p_y} \left[\left|x\right|^{4\mu/3-1}\sgn x\right] p_y^{4/3},
\label{eq:verylate}
\end{eqnarray}
where $c_{p_x;x,p_y}=-(4\mu/3)C_{x,p_y}= -\mu c$.  Here we have used the fact that
$W(x,0)=-2v_0(x)$, so that $W''(0,0)=-2\lambda_x$,~etc.  This approximation
is accurate to~$O(|x|^{4\mu-1})$ as~$x\to0$, when~$p_y=O(|x|^{2\mu})$
[\ie,~when $\ov x\to\ov 0$ from within Region~N].  If~$\mu>1/2$, it~is
easy to invert the series~(\ref{eq:verylate}) to approximate
$x=x(p_x,p_y)$.  The $-2\lambda_x x$ term is dominant, and inversion yields
\begin{eqnarray}
x = x(p_x,p_y) &\sim& x(p_x,0) + c_{x;p_x,p_y}
\left[\left|p_x\right|^{4\mu/3-1}\sgn p_x\right] p_y^{4/3}\\
&\sim& \left(  -p_x/2\lambda_x + \cdots + 
\const\times p_x^{\lfloor
 4\mu\rfloor-1} \right) + c_{x;p_x,p_y}\left
[\left|p_x\right|^{4\mu/3-1}\sgn p_x\right] p_y^{4/3},
\end{eqnarray}
where $c_{x;p_x,p_y}=-(2\lambda_x)^{-4\mu/3} c_{p_x;x,p_y} =
 (2\lambda_x)^{-4\mu/3}\mu c$.  Since $x(p_x,p_y) = \partial \wt W/\partial
 p_x(p_x,p_y)$, we must have
\begin{eqnarray}
\label{eq:newnewapproxa}
\wt W(p_x,p_y) &\sim& \wt W(p_x,0) + p_y^2/4|\lambda_y| +
C_{p_x,p_y}|p_x|^{4\mu/3}p_y^{4/3} \\
\label{eq:newnewapproxb}
&\sim& \left(-W(0,0) - p_x^2/4\lambda_x + \cdots + \const\times p_x^{\lfloor 
4\mu\rfloor}\right)+ p_y^2/4|\lambda_y| +
C_{p_x,p_y}|p_x|^{4\mu/3}p_y^{4/3} \\
&\approx& -W(0,0) - p_x^2/4\lambda_x + p_y^2/4|\lambda_y| +
C_{p_x,p_y}|p_x|^{4\mu/3}p_y^{4/3},
\label{eq:newnewapproxc}
\end{eqnarray}
where $C_{p_x,p_y}=(3/4\mu)c_{x;p_x,p_y}=(3c/4)(2\lambda_x)^{-4\mu/3}$.
The momentum-space normal forms (\ref{eq:newnewapproxa}) and~
(\ref{eq:newnewapproxb}) should be accurate to~$O(|p_x|^{4\mu})$
as~$p_x\to0$, when $p_y=O(|p_x|^{2\mu})$.  This is simply a momentum-space
version of the condition that $\ov x\to\ov 0$ from within Region~N.

It is useful to compare the truncated normal form~(\ref{eq:newnewapproxc})
with~(\ref{eq:doubletransform}), the quadratic approximation to~$\wt W$
that is valid near the saddle point in the absence of focusing.  We~see
that the fact that a double well model is `critical' modifies the double
Legendre transform~$\wt W$ near the saddle in a very simple way: it~adds
the final, nonpolynomial term.  In~a sense, the coefficient
$C_{p_x,p_y}$~measures the {\em strength\/} of the nascent cusp singularity
at the saddle.

\begin{figure}
\begin{eqnarray*}
\wt W(p_x,p_y) & \sim& \wt W(p_x,0) + p_y^2/4|\lambda_y| + C_{p_x,p_y}
|p_x|^{4\mu/3} p_y^{4/3}\\[-\jot]
&\sim& \left(-W(0,0) - p_x^2/4\lambda_x + \cdots + \const\times p_x^{\lfloor 
4\mu\rfloor}\right)+ p_y^2/4|\lambda_y| +
C_{p_x,p_y}|p_x|^{4\mu/3}p_y^{4/3}\\[\jot]
\wt W^{(x)}(p_x,y) & \sim& \wt W^{(x)}(p_x,0) -|\lambda_y|y^2 + C_{p_x,y}
|p_x|^{4\mu/3} y^{4/3}\\[-\jot]
&\sim& \left(-W(0,0) - p_x^2/4\lambda_x + \cdots + \const\times p_x^{\lfloor 
4\mu\rfloor}\right) -|\lambda_y|y^2 + C_{p_x,y}|p_x|^{4\mu/3} y^{4/3}\\[\jot]
\wt W^{(y)}(x,p_y) & \sim& \wt W^{(y)}(x,0) + p_y^2/4|\lambda_y| + C_{x,p_y}
|x|^{4\mu/3} p_y^{4/3}\\[-\jot]
&\sim& \left(-W(0,0) +\lambda_x x^2 + \cdots + \const\times x^{\lfloor 
4\mu\rfloor}\right)+ p_y^2/4|\lambda_y|+ C_{x,p_y}
|x|^{4\mu/3} p_y^{4/3}\\[\jot]
W(x,y) & \sim& W(x,0) + |x|^{4\mu} h(y/|x|^{2\mu})\\[-\jot]
&\sim&  \left(W(0,0) -\lambda_x x^2 + \cdots + \const\times x^{\lfloor 
4\mu\rfloor}\right)      +|x|^{4\mu} h(y/|x|^{2\mu})
\end{eqnarray*}
\caption[Normal forms for the thermodynamic potentials in the vicinity of a
nascent cusp.]{Normal forms for the thermodynamic potentials (the~Legendre
transforms of the action~$W$, and $W$ itself) in the vicinity of a nascent
cusp.  In~terms of the model-dependent constant~$c$,
$C_{p_x,p_y}=(3c/4)(2\lambda_x)^{-4\mu/3}$, $C_{p_x,y} =
(3c/4)(2\lambda_x)^{-4\mu/3}(2|\lambda_y|)^{4/3}$, and $C_{x,p_y}=3c/4$.
These asymptotic expansions are valid in Region~N\hbox{}, in critical double well
models with~$\mu>1/2$.  They are accurate to~$O(|x|^{4\mu})$,
or~equivalently to~$O(|p_x|^{4\mu})$.}
\label{fig:Wtable}
\end{figure}

The computation of the remaining thermodynamic potential,~$\wt W^{(x)}$, is
left to the reader.  In~Figure~\ref{fig:Wtable} we list the normal forms
for $\wt W$, $\wt W^{(x)}$, and~$\wt W^{(y)}$, as~well as the scaling form
for~$W$.  The expressions listed there are accurate to~$O(|x|^{2\mu})$,
\ie, to~$O(|p_x|^{2\mu})$, as~the nascent cusp is approached from within
Region~N\hbox{}.  In~Figure~\ref{fig:eqstatetable} we~list the four
possible equations of state for $x$~and~$y$.  They are accurate
to~$O(|x|^{2\mu-1})$, \ie, to~$O(|p_x|^{2\mu-1})$, in the same limit.

\begin{figure}
\begin{eqnarray*}
x(p_x,y) &\sim& x(p_x,0) 
+ c_{x;p_x,y} \left[\left|p_x\right|^{4\mu/3-1}\sgn p_x\right] y^{4/3}\\[-\jot]
&\sim& 
\left( -p_x/2\lambda_x + \dots + \const\times p_x^{\lfloor 4\mu\rfloor - 1}\right)
+ c_{x;p_x,y} \left[\left|p_x\right|^{4\mu/3-1}\sgn p_x\right] y^{4/3}\\[\jot] 
x(p_x,p_y) &\sim& x(p_x,0) 
+ c_{x;p_x,p_y}\left[\left|p_x\right|^{4\mu/3-1}\sgn p_x\right]p_y^{4/3}\\[-\jot]
&\sim&
\left( -p_x/2\lambda_x + \dots + \const\times p_x^{\lfloor 4\mu\rfloor - 1}\right)
+ c_{x;p_x,p_y}\left[\left|p_x\right|^{4\mu/3-1}\sgn p_x\right]p_y^{4/3}\\[\jot]
y(x,p_y) &\sim& p_y/2\left|\lambda_y\right| 
+ c_{y;x,p_y}  \left|x\right|^{4\mu/3} p_y^{1/3}\\[\jot]
y(p_x,p_y) &\sim& p_y/2\left|\lambda_y\right| 
+ c_{y;p_x,p_y}  \left|p_x\right|^{4\mu/3} p_y^{1/3}
\end{eqnarray*}
\caption[The four asymptotic equations of state in the vicinity of the
nascent cusp.]{The four asymptotic equations of state, which describe the
shape of the Lagrangian manifold~$\protect\cal M$ in the vicinity of a
nascent cusp.  It~follows by differentiating the normal forms listed in
Fig.~\ref{fig:Wtable} that $c_{x;I,J}=(4\mu/3)C_{I,J}$ and
$c_{y;I,J}=(4/3)C_{I,J}$.  So~$c_{x;p_x,p_y}=\mu c/(2\lambda_x)^{4\mu/3}$,
$c_{x;p_x,y}=\mu c(2|\lambda_y|)^{4/3} / (2\lambda_x)^{4\mu/3}$,
$c_{y;p_x,p_y}= c/(2\lambda_x)^{4\mu/3}$, and~$c_{y;x,p_y}=c$.  These
asymptotic expansions are valid in Region~N\hbox{}, in critical double well models
with~$\mu>1/2$.}
\label{fig:eqstatetable}
\end{figure}

We emphasize that the normal form~(\ref{eq:newnewapproxb})
for~$\wt W$, the normal form for~$\wt W^{(x)}$, and the equations of state
that follow from them, are valid only for critical models with~$\mu>1/2$.
The reason is that when~$\mu\le1/2$, the final term in~(\ref{eq:verylate}),
which when $p_y=O(|x|^{2\mu})$ is of magnitude $O(x^{4\mu-1})$, is at~least
as large as the $-2\lambda_xx$~term as~$x\to0$.  In~fact when $\mu<1/2$,
in~Region~N (except on the $x$-axis) the leading asymptotics of
$p_x=p_x(x,p_y)$ are not linear in~$x$.  This makes difficult the
computation of asymptotic approximations to $x=x(p_x,p_y)$ and~$\wt W= \wt
W(p_x,p_y)$.  For this reason we shall assume $\mu>1/2$ henceforth.

It is a reasonable conjecture that in critical models where the symmetrical
approximation~(\ref{eq:newnewapproxc}) to~$\wt W=\wt W(p_x,p_y)$ is valid,
it~is valid not merely near the $p_x$-axis (\ie,~in Region~N), but
uniformly as~$\ourvec p\to\ourvec0$.  One would like to substitute~it into
the Maslov-WKB diffraction integral~(\ref{eq:pointtointegraln}), so as to
obtain boundary layer approximations to the stationary and quasistationary
probability densities near the saddle point~$(0,0)$.  The approximation to
the quasistationary density would be a replacement for the usual
Kramers-type error function approximation,~(\ref{eq:erf}).  From~it, one could
derive an Eyring formula for the MFPT asymptotics, as in
Section~\ref{subsec:MAE2}. Unfortunately there is a problem.
If~$C_{p_x,p_y}=0$ and (\ref{eq:newnewapproxc})~becomes quadratic, the
Hessian matrix $\partial^2 \wt W/\partial p_i\partial p_j$ is clearly not
negative definite.  As~we noted in Section~\ref{sec:Maslov}, this precludes
the use of the two-dimensional diffraction
integral~(\ref{eq:pointtointegraln}).  The situation does not improve much
if $C_{p_x,p_y}$~is positive, so it~is preferable to use an alternative
integral representation.  The asymptotic approximation to the Legendre
transform $\wt W^{(x)} = xp_x - W = -yp_y + \wt W$, as~a function of $p_x$
and~$y$, is listed in the table in Figure~\ref{fig:Wtable}.  A~truncated
version of~it would be
\begin{equation}
\label{eq:newWapprox}
\wt W^{(x)}(p_x,y) \approx 
- W(0,0) - p_x^2/4\lambda_x - \left|\lambda_y\right|y^2
+ C_{y,p_x} |p_x|^{4\mu/3}y^{4/3} ,
\end{equation}
which is a nonpolynomial modification of the Gaussian
approximation~(\ref{eq:partway}).  This approximation is precisely what is
needed in the one-dimensional diffraction integral~(\ref{eq:alternativex}),
which is what we shall use instead of~(\ref{eq:pointtointegraln}).

The reader may wonder about the domain of validity of the
approximation~(\ref{eq:newWapprox}) to $\wt W^{(x)} = \wt W^{(x)}(p_x,y)$.
Is~it valid outside Region~N\spacefactor=1000?  In~Section~\ref{subsec:nascent15}
we present numerical evidence that it~is, in~fact, a~useful asymptotic
approximation near the $y$-axis, even at fixed, nonzero~$y$.  Indeed,
it~explains the mysterious `sideways' caustic of Fig.~\ref{fig:nascent}!
To~see this, differentiate~(\ref{eq:newWapprox}) with respect to~$p_x$ to
get
\begin{equation}
\label{eq:unfolding}
x= x(p_x,y) \approx -p_x/2\lambda_x + c_{x;p_x,y}  y^{4/3} \Bigl[|p_x|
^{(4\mu/3)-1}\sgn p_x\Bigr],
\end{equation}
which is a truncated version of the asymptotic expansion of $x=x(p_x,y)$
listed in Figure~\ref{fig:eqstatetable}.  If~$\mu<3/2$, the
formula~(\ref{eq:unfolding}) predicts that at any nonzero~$y$, the map
$p_x\mapsto x$ will {\em not be monotone\/}.  This is because the
coefficient $c_{x;p_x,y}$ is positive.  By~examination, if
\begin{equation}
|x| \simlt \const \times |y|^{ (3/2-\mu)^{-1}},\quad y\to0,
\end{equation}
then the inverse map $p_x=p_x(x,y)$ (and hence~$W=W(x,y)$) will be
multivalued.  This inequality defines a two-sided {\em nongeneric
caustic\/}, which emanates from~$(0,0)$ along the positive and negative
$y$-axes.

In the language of catastrophe theory, the formula~(\ref{eq:unfolding}) is
a (non-smooth) {\em unfolding\/} of the nongeneric caustic emanating from
the nascent cusp.  It~also resembles a thermodynamic equation of state in
the vicinity of a phase transition.  However, the thermodynamic
interpretation of the variables differs from the case of an on-axis focus,
as~analysed in the last section.  Here $|y|$~is analogous to~$T_c-T$,
for~example.  And by~examination, the thermodynamic critical
exponent~$\gamma$ is nonzero whenever $\mu<3/2$; it~equals
$(3/2-\mu)^{-1}$.  In~any event, the critical exponents of the nascent cusp
are clearly nonclassical: they depend continuously on the parameter~$\mu$.

We caution the reader that in {\em arbitrary\/} double well models
at~criticality, the nonpolynomial approximation~(\ref{eq:newWapprox}) and
the nonclassical equation of state~(\ref{eq:unfolding}) may not necessarily
describe the $p_x\to0$ behavior of $x=x(p_x,y)$ at fixed, nonzero~$y$.
If~$\mu<3/4$, the $y^{4/3}\left[|p_x|^{(4\mu/3)-1}\sgn p_x\right]$ term
in~(\ref{eq:unfolding}) would cause $x=x(p_x,y)$ to diverge as~$p_x\to0$,
at~any nonzero $y$~or~$p_y$.  Such a divergence would greatly distort the
shape of the Lagrangian manifold~$\cal M$.  So~we shall assume $\mu>3/4$
henceforth.  In~Section~\ref{subsec:nascent15} we present numerical evidence of
the need for the $\mu>3/4$ restriction, and also verify that the nongeneric
caustic is present if~and only if~$\mu<3/2$.  Incidentally, our numerical
results indicate that at~fixed ~nonzero~$y$, the apparent non-analyticity
at~$p_x=0$ is `rounded' at a sufficiently small ($\left|y\right|$-dependent)
lengthscale, as~$p_x\to0$.
This is analogous to the abovementioned rounding of~$\wt W^{(y)}(x,p_y)$,
and its derivative $y=y(x,p_y)$, as~$p_y\to0$ at~fixed nonzero~$x$.

There is also a problem with the nonpolynomial
approximation~(\ref{eq:newWapprox}) and the nonclassical equation of
state~(\ref{eq:unfolding}) when~$\mu\ge3$.  To~see this, note that a more
complete asymptotic expansion of~$x=x(p_x,y)$ near~$p_x=0$ would presumably
be of the form
\begin{equation}
\label{eq:zaptochar}
x=x(p_x,y) \approx \left( -p_x/2\lambda_x + \dots + \const\times p_x^{\lfloor
4\mu\rfloor - 1}\right) + c_{x;p_x,y} y^{4/3}
\left[\left|p_x\right|^{4\mu/3-1}\sgn p_x\right].
\end{equation}
Such an asymptotic expansion is listed in Fig.~\ref{fig:eqstatetable}, and
is certainly valid as~$\ov x\to\ov0$ from within Region~N\hbox{}.  If~a
similar expansion is valid near the $y$-axis, we~see that there will be a
crossover at~$\mu=3$ between two regimes.  When~$\mu=3$, the nonpolynomial
$c_{x;p_x,y} y^{4/3} \left[\left|p_x\right|^{4\mu/3-1}\sgn p_x\right]$ term
in~(\ref{eq:zaptochar}) becomes $c_{x,p_x,y} y^{4/3} p_x^3$.  This is
increasingly dominated by the $p_x^3$~term in~(\ref{eq:zaptochar})
as~$y\to0$.  In~fact when $\mu$~is raised above~$3$, at~small~$\left|y\right|$ the
leading corrections to the naive $x\sim -p_x/2\lambda_x$ behavior are
no~longer given by the nonpolynomial term, but rather by the $p_x^3$~term.
For this reason we shall assume for the remainder of our analysis that
$\mu<3$ as~well as~$\mu>3/4$.

To use the one-dimensional Maslov-WKB diffraction integral~(\ref{eq:alternativex})
as~promised, we need to approximate in the vicinity of the nascent cusp
at~$(p_x,y)=(0,0)$ not only the Legendre-transformed action $\wt
W^{(x)}(p_x,y)$, but also the transformed prefactor $\wt K^{(x)}(p_x,y)$.
It~may be approximated in a very similar way, which we only summarize.
By~(\ref{eq:z1})--(\ref{eq:z2}), $K(x,0)$~diverges in any critical model
as~$|x|^{-\mu}$, when~$x\to0$.  A~scaling form
\begin{equation}
\label{eq:Kscaling}
K(x,y)\sim |x|^{-\mu}q(y/|x|^{2\mu}),
\end{equation}
modelled after~(\ref{eq:KAnsatz}) and~(\ref{eq:firstscaling}), may be used
to approximate~$K$ away from the $x$-axis.  The
approximation~(\ref{eq:Kscaling}) should be accurate to~$O(|x|^{-\mu})$
as~$x\to0$, when~$y=O(|x|^{2\mu})$ [\ie,~as $\ov x\to\ov0$ from within
Region~N]\hbox{}.  By~substituting (\ref{eq:Kscaling}) and the scaling
form~(\ref{eq:firstscaling}) for~$W$ into the amplitude transport
equation~(\ref{eq:origKeqn}), and working to leading order near
$(x,y)=(0,0)$, one can determine the scaling function~$q=q(z)$.  The
procedure closely resembles the procedure used in Section~\ref{sec:real}.
It~is easily verified that collecting the $O(\left|x\right|^{-\mu})$ terms
in the transport equation yields the~ODE
\begin{equation}
\label{eq:stuffode2}
2\Bigl[ h' + \left| \lambda_y \right| z  \Bigr] q^{\,\prime} 
+ 
h''q=0,
\end{equation}
which $q=q(z)$ must satisfy.  (Cf.~(\ref{eq:stuffode}).)  Here $h=h(z)$ is
the scaling function for~$W$, which satisfies the ODE~(\ref{eq:ODEforh}).
Using elementary calculus, one can show that eq.~(\ref{eq:stuffode2}) has
solution
\begin{equation}
\label{eq:823}
q(z) = \const \times \left| h'(z) \right|^{-1/3} \sqrt{-h''(z)}.
\end{equation}
(Cf.~(\ref{eq:Kscalingfuncs}).)  But since $p_y=\partial W/\partial y\sim
\left|x\right|^{2\mu}h'$, one may write $|x|^{-2\mu}p_y$ for~$h'$, and
$\partial p_y/\partial y$ for~$h''$.  Substituting (\ref{eq:823}) into the
scaling form~(\ref{eq:Kscaling}), and performing the indicated rewriting,
yields
\begin{equation}
\label{eq:bizarre}
K(x,y)\sim \const \times |x|^{-\mu/3} |p_y|^{-1/3} \sqrt{-{{\partial
p_y}\over{\partial y}} (x,y)}.
\end{equation}
(Cf.~(\ref{eq:simplicity}).)  This asymptotic approximation is exact to
leading order as~$\ov x\to\ov 0$ from within Region~N.

The formula~(\ref{eq:bizarre}) facilitates the computation of the
transformed prefactor~$\wt K^{(y)}=\wt K^{(y)}(x,p_y)$.  It~may be computed
from~$K$ as in~(\ref{eq:seventwenty}), by dividing by the appropriate
`Van~Vleck factor.'  We~immediately find
\begin{eqnarray}
\label{eq:815a}
\wt K^{(y)}(x,p_y)&\propto& K(x,y)
\left/ \sqrt{- {{\partial^2 W}\over{\partial y^2}}(x,y)}\right.   \\
&\sim&\const\times |x|^{-\mu/3}|p_y|^{-1/3},
\label{eq:815b}
\end{eqnarray}
since $\partial^2 W/\partial y^2 = \partial p_y/\partial y$.
(Cf.~(\ref{eq:seventwenty})--(\ref{eq:seventwentyone}).)  The uncomplicated
asymptotic approximation~(\ref{eq:815b})  should be accurate
to~$O(|x|^{-\mu})$ as~$x\to0$, when~$p_y=O(|x|^{2\mu})$.  It~simply says
that $\wt K^{(y)}(x,a|x|^{2\mu})\sim\const \times |a|^{-1/3}
|x|^{-\mu}$ as~$x\to0$, for any nonzero~$a$.

The two remaining transformed prefactors, $\wt K$ and~$\wt K^{(x)}$, may be
computed from~$\wt K^{(y)}$ by dividing (or multiplying) by the appropriate
Van~Vleck factors.  (Cf.~(\ref{eq:alternativeall})
and~(\ref{eq:curioustransform}).)  For example,
\begin{equation}
\wt K(p_x,p_y) \propto \wt K^{(y)}(x,p_y) \sqrt
{- {{\partial^2 \wt W}\over{\partial p_x^2}}(p_x,p_y)}.
\end{equation}
The details are left to the reader.  One finds
\begin{eqnarray}
\wt K(p_x,p_y) &\sim&\const\times |p_x|^{-\mu/3}|p_y|^{-1/3},\\
\wt K^{(x)}(p_x,y) &\sim&\const\times |p_x|^{-\mu/3}|y|^{-1/3}.
\label{eq:newKapprox}
\end{eqnarray}
The transformed prefactor~$\wt K^{(x)}$ is the one we need for the
Maslov-WKB diffraction integral.  In~Section~\ref{subsec:nascent15},
we~examine the numerical evidence for the validity of this asymptotic
approximation to~$\wt K^{(x)}=\wt K^{(x)}(p_x,y)$, when $p_x\to0$ at~fixed
nonzero~$y$.

Remarkably, the formula~(\ref{eq:newKapprox}) predicts that $\wt
K^{(x)}(p_x,y)$ {\em diverges\/} at the location $(p_x,y)=(0,0)$ of the
nascent cusp.  This is different from the case of a generic (cusp)
singularity, treated in Section~\ref{sec:real}.  It~is also different from
the geometrical optics limit of physical optics, where the transformed
amplitude function near a singularity is normally a `slowly varying'
(\ie,~non-singular) function~\cite{Berry76}.  The fact that the transformed
prefactor $\wt K^{(x)}$~diverges at the nascent cusp is at~least as
important to the weak-noise behavior of critical double~well models as the
fact that the normal form for the transformed action~$\wt W^{(x)}$ is
nonpolynomial.

\subsection{Comparison with Numerics}
\label{subsec:nascent15}
We now summarize the numerical evidence for the validity, in double well
models at~criticality, of our nonpolynomial normal form for the
Legendre-transformed action $\wt W^{(x)}=\wt W^{(x)}(p_x,y)$, and our
approximation to the transformed WKB prefactor $\wt K^{(x)}=\wt
K^{(x)}(p_x,y)$.  We~shall see  that both are valid approximations near the
$y$-axis separatrix; in~particular, near the saddle point.  This justifies
their use in the Maslov-WKB method, which we shall employ in
Section~\ref{subsec:nascent2} to construct boundary layer approximations to
the stationary and quasistationary probability distributions of double well
models at criticality.

\begin{figure}
\vbox{
\hfil
\epsfxsize=3.2in		
\epsfbox{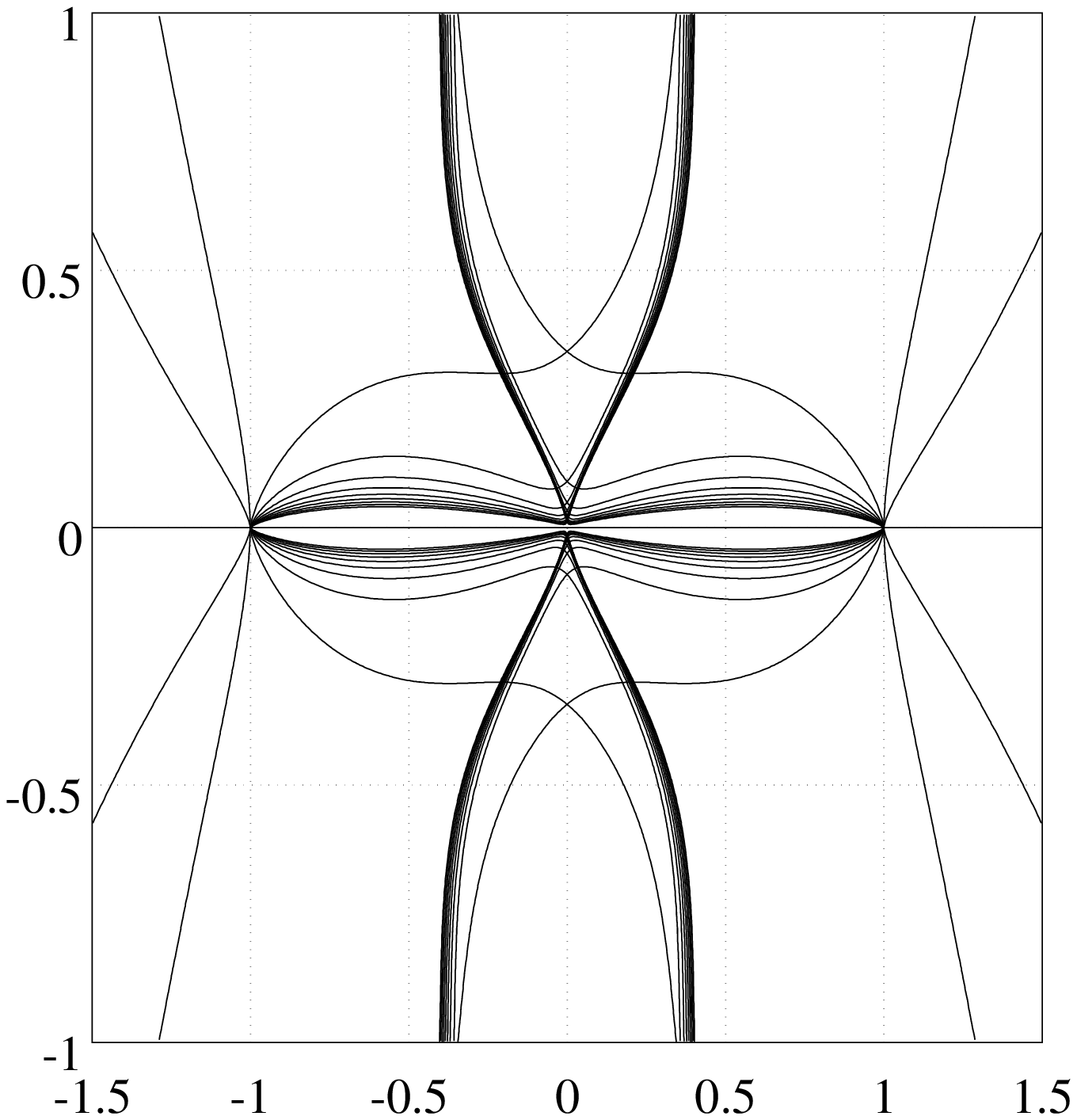}
\hfil
\epsfxsize=3.2in		
\epsfbox{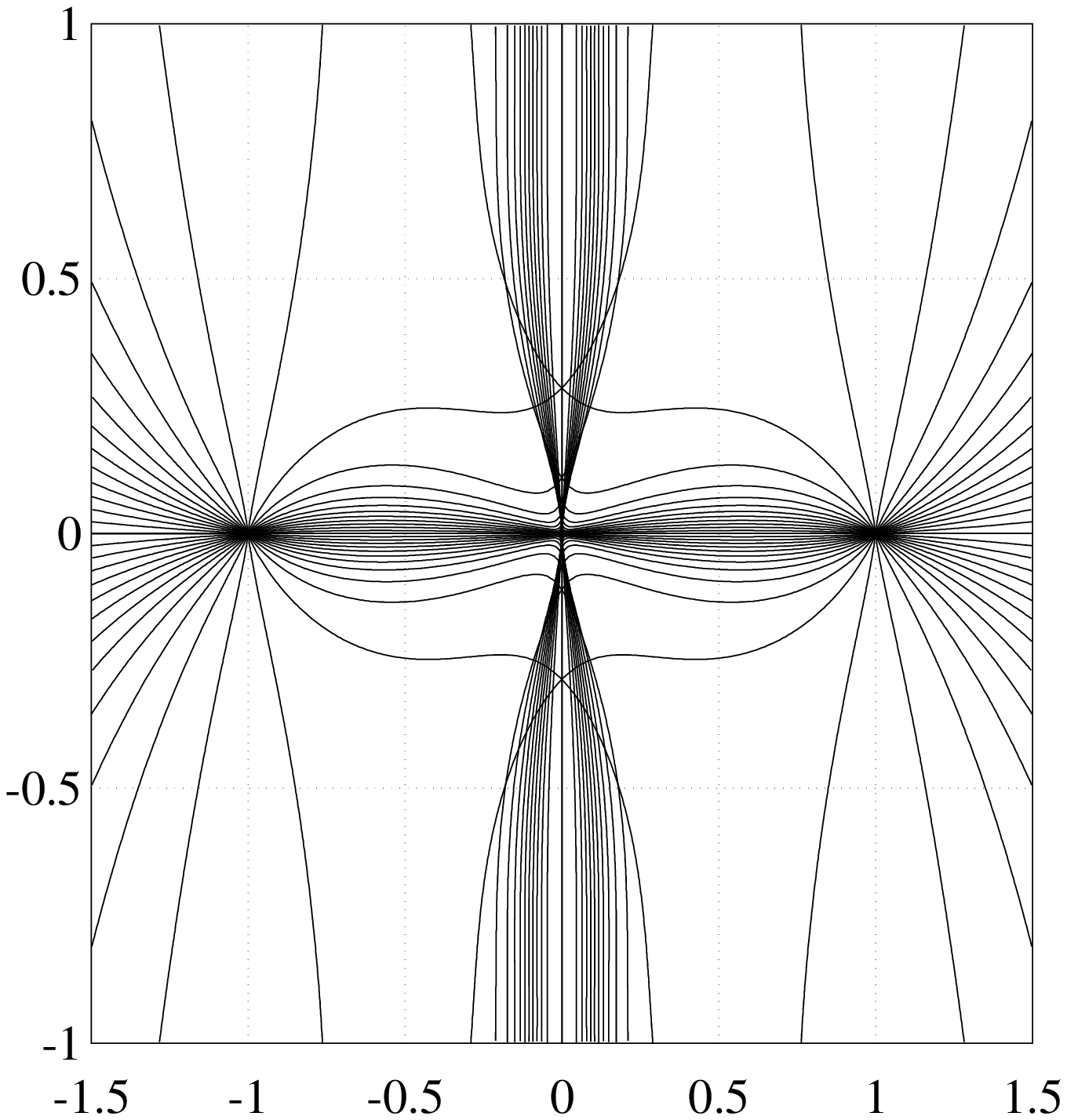}
}
\vskip0.5in
\vbox{
\hfil
\epsfxsize=3.2in		
\epsfbox{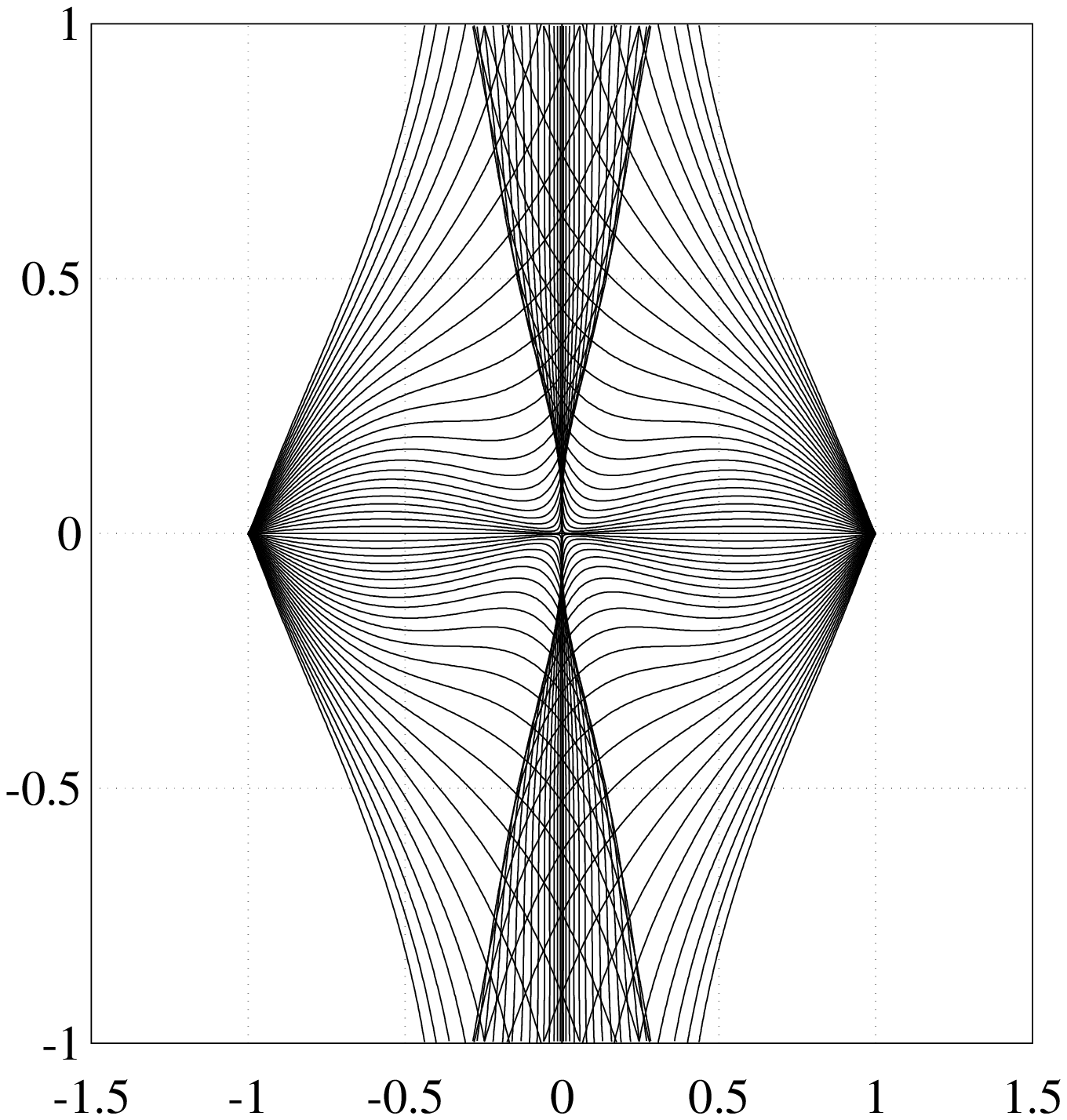}
\hfil
\epsfxsize=3.2in		
\epsfbox{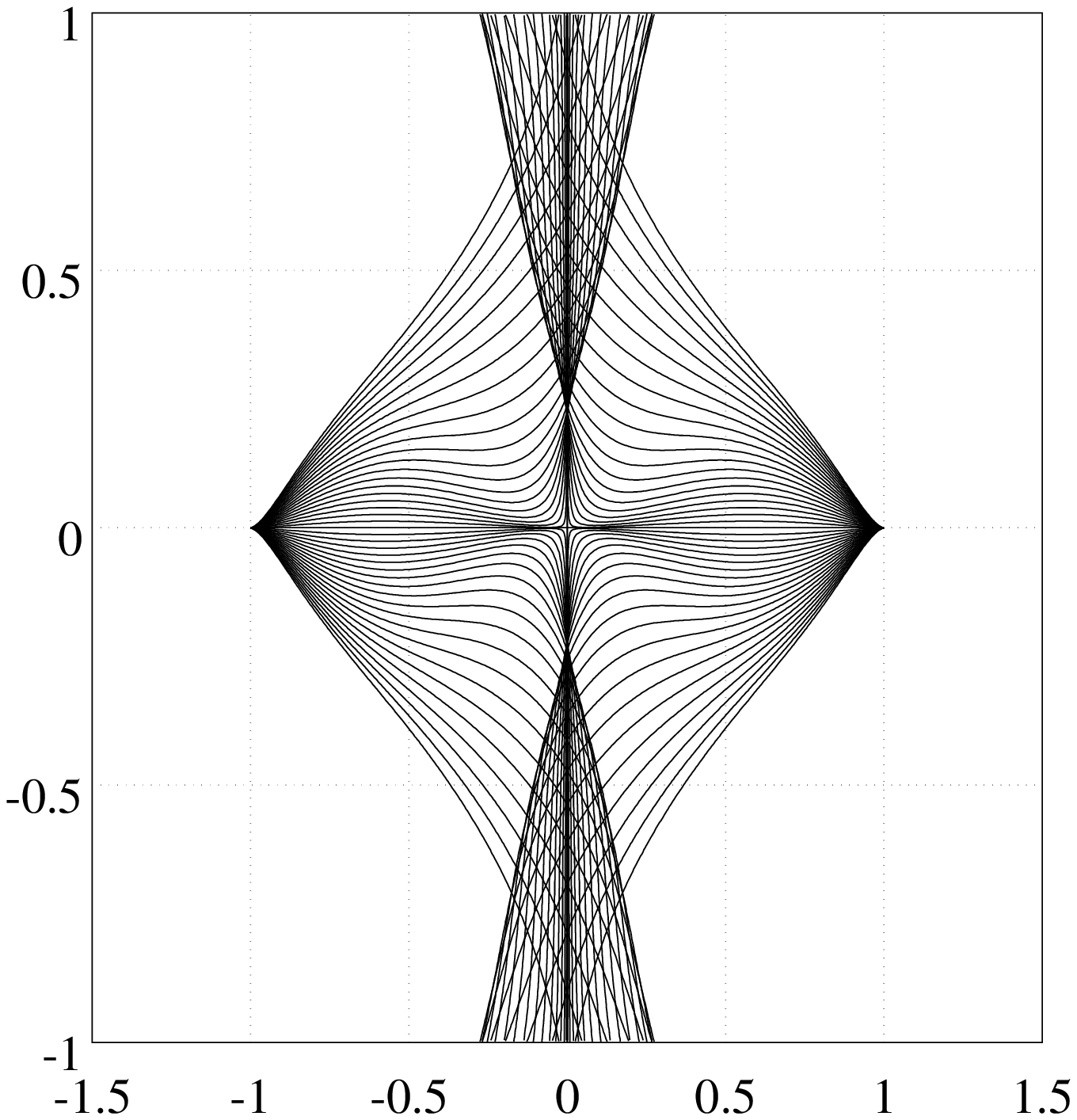}
}
\caption{The flow field of instanton trajectories emanating from both
stable fixed points $S$~and~$S'$, in critical versions of the standard
double well model~(\protect\ref{eq:standard}).  Parts (a), (b), (c),~(d) correspond
to models with $\mu=0.725$, $0.85$, $1.15$, and~$1.6$.  In~all
cases the parameter~$\alpha$ is set equal to the critical
value~$\alpha_c=2\mu(\mu+1)$, at~which the MPEP bifurcates.  The two-sided
nongeneric caustic of Fig.~\protect\ref{fig:nascent} is visible in parts
(b)~and~(c), but it~has separated into two generic caustics in part~(d).}
\label{fig:fourpart}
\end{figure}

We begin by examining the evidence for the nonpolynomial normal
form~(\ref{eq:newWapprox}) for $\wt W^{(x)}=\wt W^{(x)}(p_x,y)$.  Actually
we shall study the related nonpolynomial approximation~(\ref{eq:unfolding})
to its first derivative $x=x(p_x,y)$, \ie,
\begin{equation}
\label{eq:star}
x= x(p_x,y) \approx -p_x/2\lambda_x + c_{x;p_x,y}  y^{4/3} \Bigl[|p_x|
^{(4\mu/3)-1}\sgn p_x\Bigr].
\end{equation}
As explained above, we expect on theoretical grounds that this
approximation is generically valid near the saddle point, in critical
models in~which the quotient $\mu\defeq\left|\lambda_y\right|/\lambda_x$
satisfies $3/4<\mu<3$.  The formula~(\ref{eq:star}) predicts that at
nonzero~$y$, the correspondence $p_x\mapsto x$ is monotone if~$\mu\ge3/2$,
but non-monotone at nonzero~$y$ if~$\mu<3/2$.  When~$\mu<3/2$, the
correspondence $p_x\mapsto x$ is analogous to the correspondence
$m\mapsto -h$, in~a ferromagnet, between magnetization and
(negative) magnetic field.

It is easily checked that when
\begin{equation}
\label{eq:lastminuteineq}
\left|x\right| \simlt \const \times \left|y\right|^{(3/2-\mu)^{-1}}, \quad
y\to0, 
\end{equation}
the inverse map $x\mapsto p_x$ is three-valued rather than single-valued.
In~this region the three possible values for~$p_x$ are by~examination of
the same magnitude as~$x$, \ie,
\begin{equation}
p_x = O\left( \left|y\right|^{(3/2-\mu)^{-1}}\right), \quad y\to0.
\end{equation}
We~interpret the inequality~(\ref{eq:lastminuteineq}) as defining a
two-sided nongeneric caustic centered on the $y$-axis, in the interior of
which the action~$W$, and its gradient~$\ov p=\nabla W$, are
three-valued.  Recall that the Lagrangian manifold~$\cal M$, which is
traced~out by WKB bicharacteristics, comprises all points in phase space of
the form $(\ov x,\ov p(\ov x))$.  The three-valuedness of $W$ and~$\ov p$
within the caustic accordingly implies that there are three points on~$\cal
M$ `above' any point~$\ov x$ in the interior of the caustic.  We~have
already seen in Fig.~\ref{fig:nascent} that a nongeneric caustic
qualitatively agreeing with this prediction does indeed appear in the
$\mu=1$ standard double well model~(\ref{eq:standard}), at~criticality.

Figures \ref{fig:fourpart}(a)--\ref{fig:fourpart}(d) show the flow field of
instanton trajectories, \ie, projected bicharacteristics, in several more
critical variants of the standard double well model.  (At~criticality
$\alpha=\alpha_c=2\mu(\mu+1)$, by eq.~(\ref{eq:hardwon}).)  Figures
\ref{fig:fourpart}(b) and~\ref{fig:fourpart}(c), with $\mu=0.85$
and~$\mu=1.15$, illustrate the fact that the two-sided caustic of
Fig.~\ref{fig:nascent} appears at~criticality in any double well model
whose parameter~$\mu\defeq |\lambda_y|/\lambda_x$ satisfies $3/4<\mu<3/2$.
The caustic disappears, as~expected, in critical models with~$\mu\ge3/2$.
Figure~\ref{fig:fourpart}(d) shows what happens.  As~$\mu$~in the standard
model is raised above~$3/2$ (with $\alpha$~set equal
to~$\alpha_c=\alpha_c(\mu)$), the two-sided caustic separates into two
one-sided {\em generic\/} caustics, whose cusps move out along the positive
and negative $y$-axes, away~from the saddle.  In~any critical model
with~$\mu>3/2$, there is a portion of the separatrix near the saddle that
is not crossed by any instanton trajectory.

Figure~\ref{fig:fourpart}(a) illustrates the bizarre behavior that occurs
in critical models with~$\mu\le3/4$.  At~first glance it~seems that the
now~familiar two-sided caustic is present, but closer study reveals that
points in its interior are reached by only {\em two\/} instanton
trajectories, rather than three.  Apparently, in the $\mu\le3/4$ regime the
approximation~(\ref{eq:star}) breaks~down near the separatrix.
Empirically, when~$\mu\le3/4$ the $|p_x|^{(4\mu/3)-1}$ factor
in~(\ref{eq:star}) must be replaced by unity.  The $\mu\le3/4$ regime is
still under investigation, and we~shall not consider~it further in this
paper.

\begin{figure}
\vbox{
\hfil
\epsfxsize=3.0in		
\epsfbox{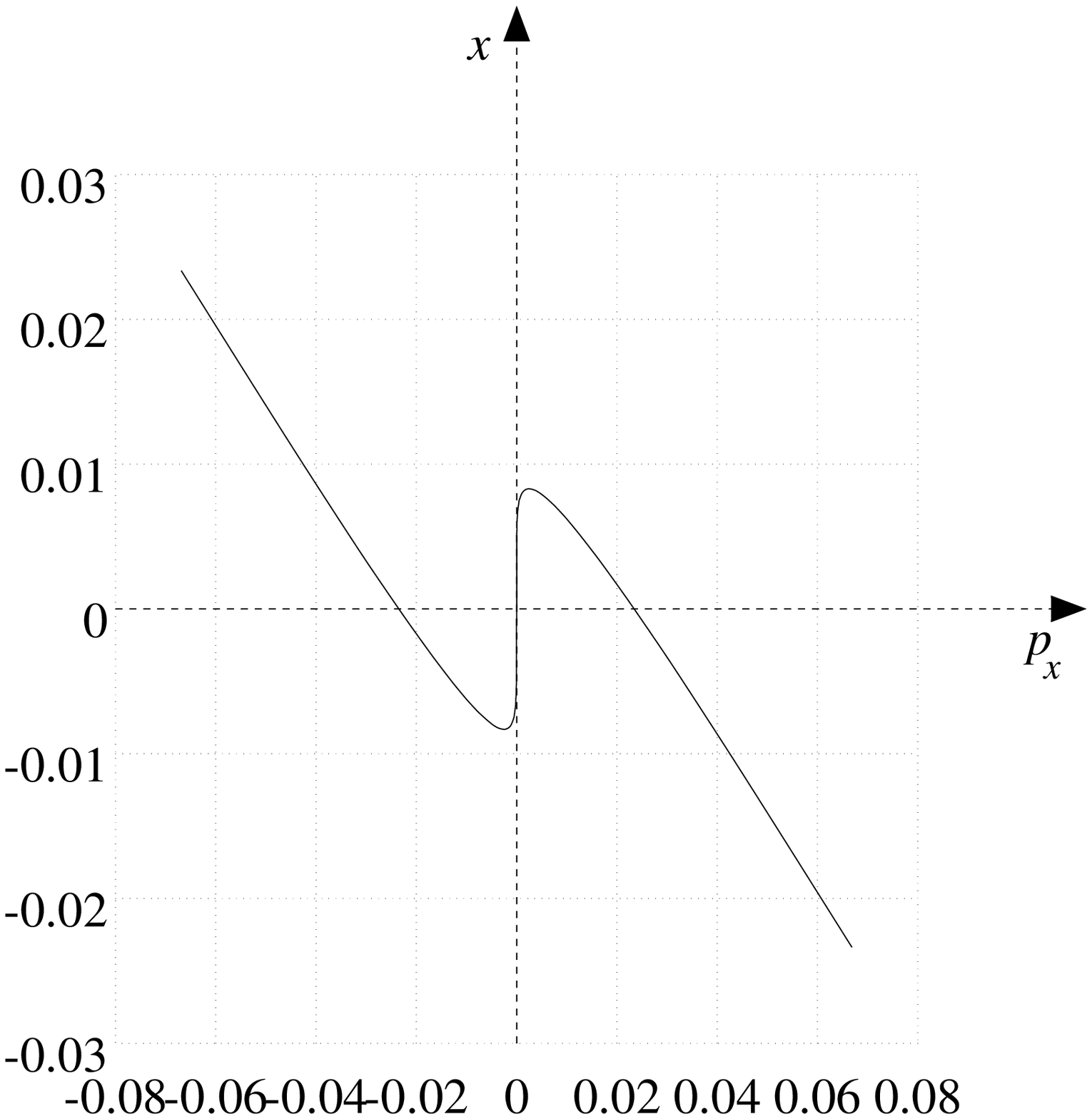}
\hfil
\epsfxsize=2.6in		
\epsfbox{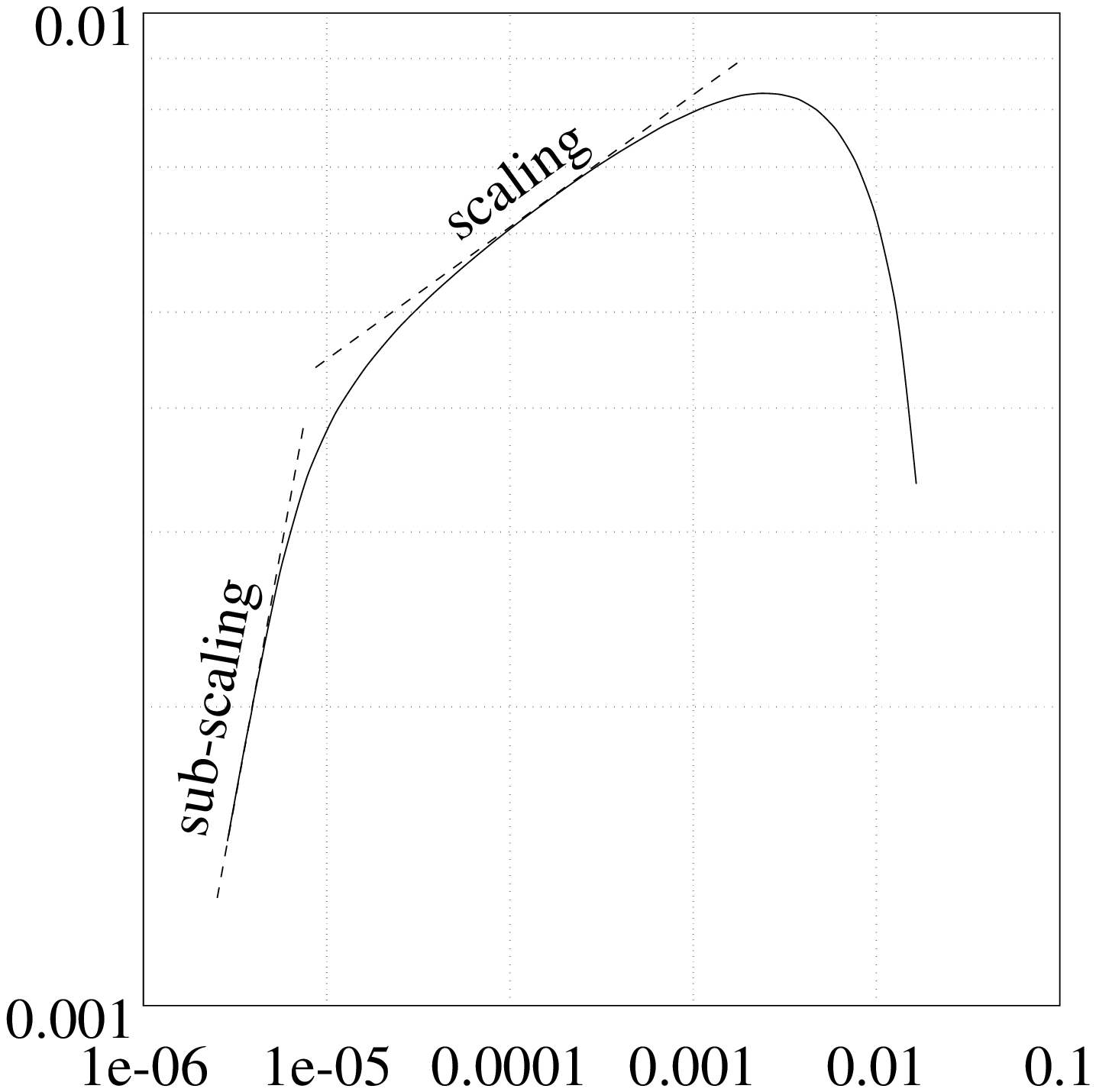}
}
\caption{A section through the nongeneric caustic appearing at~criticality
in the $\mu=0.85$ standard double well model, as shown in
Fig.~\protect\ref{fig:fourpart}(b).  This section is taken at~$y=0.05$.  Parts
(a)~and~(b) show a linear and a logarithmic plot of~$x=x(p_x,y=0.05)$.  The
scaling behavior ($x\propto |p_x|^{(4\mu/3)-1}$) and the sub-scaling
behavior ($x\propto p_x$) are both visible.  Crossover between the two
regimes occurs at~$p_x=O(|y|^{(5/2)(3/2-\mu)^{-1}})$, \ie, 
at~$p_x\doteq 10^{-5}$.}
\label{fig:hysteresis}
\end{figure}

We can now compare the predictions of our scaling theory with numerical
data.  Figure~\ref{fig:hysteresis}(a) is a section through the caustic of
Fig.~\ref{fig:fourpart}(b), \ie, a~cross-section through the corresponding
Lagrangian manifold.  It~shows the correspondence $p_x\mapsto x$,
at~$y=0.05$, in the $\mu=0.85$ standard model at~criticality.  The
qualitative shape of the curve certainly resembles the prediction of
formula~(\ref{eq:star}).  But before making a {\em quantitative\/}
comparison, we~need to discuss the interpretation of~(\ref{eq:star}).
It~was derived from an asymptotic development of the action about the
saddle point.  To~what extent does it describe the small-$|p_x|$
asymptotics of~$x=x(p_x,y)$ at~{\em fixed\/}, nonzero~$y$?  That is what is
plotted in Fig.~\ref{fig:hysteresis}.

 From a rigorous point of view, when $\mu<3/2$ the formula~(\ref{eq:star})
provides a two-term asymptotic expansion of~$x=x(p_x,y)$ as~$y\to0$, on the
$p_x = O( |y|^{(3/2-\mu)^{-1}})$ lengthscale on~which the nongeneric
caustic is visible.  This is strongly reminiscent of the `Region~N'
constraint of the last section.  There we began by approximating $W=W(x,y)$
in the notch-shaped region of Fig.~\ref{fig:N}.  Here we are approximating
$x=x(p_x,y)$, and by~extension its anti-derivative $\wt W^{(x)} = \wt
W^{(x)}(p_x,y)$, in~a region that is similarly notch-shaped, but is
centered on the $y$-axis rather than the $x$-axis.  We~shall not attempt to
expand $x$~and~$\wt W^{(x)}$ systematically, but the basic procedure is
plain.  If~we define a new scaling variable
\begin{equation}
\hat z \defeq p_x/\left|y\right|^{(3/2-\mu)^{-1}},
\end{equation}
then formula~(\ref{eq:star}) can be interpreted as comprising the first
two terms in an asymptotic development of $x(\hat z|y|^{(3/2-\mu)^{-1}}, y)$
as~$y\to0$.  The development should be valid at any fixed~$\hat z$.

Though this restatement is a bit pedantic, it suggests that on smaller
lengthscales than $p_x = O( |y|^{(3/2-\mu)^{-1}})$, the
formula~(\ref{eq:star}) might be invalid.  Actually there are strong
reasons for believing that the nonanalytic $|p_x|^{(4\mu/3)-1}\sgn p_x$
behavior as~$p_x\to0$ does not appear at~fixed, nonzero~$y$.  If~it~did,
$\partial x/\partial p_x(p_x=0)$ would diverge at~criticality, at~$y\neq0$,
in any model with~${3/4<\mu<3/2}$.  Equivalently, $\partial p_x/\partial
x(x=0)$, \ie, $\partial^2 W/\partial x^2(x=0)$, would be {\em identically
zero\/}, irrespective of the choice of nonzero~$y$.  But this prediction is
too simple: it~ignores the presence of `sub-scaling' terms.  We~noted in
the last section that at criticality, $W=W(x,y)$ should contain an $x^2y^2$
term, for consistency with the sub-scaling $w_2(x)\sim \const\times x^2$
behavior.  In~other words, $\partial^2 W/\partial x^2(x=0)$ near the saddle
should be nonzero and proportional to~$y^2$.  For consistency with this
prediction, the nonanalytic $|p_x|^{(4\mu/3)-1}\sgn p_x$ behavior
of~$x=x(p_x,y)$ must be `rounded' at sufficiently small~$|p_x|$.  It~is
easy to check that $p_x = O( |y|^{(5/2)(3/2-\mu)^{-1}})$ is the correct
lengthscale.  On~that lengthscale, one should find
$W(x,y)\approx\const\times x^2y^2$, \ie, $p_x\approx\const\times xy^2$,
or~$x\approx\const\times y^{-2} p_x$.

What we conclude from this discussion is that at~fixed, nonzero~$y$, the
nonpolynomial formula~(\ref{eq:star}) for $x=x(p_x,y)$ should be valid on
the `caustic lengthscale' $p_x = O( |y|^{(3/2-\mu)^{-1}})$, but that it
will break~down when $p_x$~is decreased to~$O( |y|^{(5/2)(3/2-\mu)^{-1}})$.
On~that smaller lengthscale, one expects a crossover to a linear regime,
where $x$~is proportional to~$p_x$.  We~can now proceed to our comparison
with numerics.  In~Fig.~\ref{fig:hysteresis}(b) we plot the correspondence
$p_x\mapsto x(p_x,y=0.05)$ of~Fig.~\ref{fig:hysteresis}(a) on a logarithmic
scale.  We~also fit two trendlines to~it: $x\propto |p_x|^{(4\mu/3)-1}$
and~$x\propto p_x$.  As~the two trendlines reveal, our theoretical analysis
is perfectly confirmed.  There is indeed a crossover to a linear,
sub-scaling regime when $p_x$~is decreased
to~$O(|y|^{(5/2)(3/2-\mu)^{-1}})$.  But at larger lengthscales, \eg,
$p_x=O( |y|^{(3/2-\mu)^{-1}})$, the fractional power
$|p_x|^{(4\mu/3)-1}\sgn p_x$ of the scaling formula~(\ref{eq:star}) is
clearly visible.  Similar crossover plots can be obtained for other
critical double~well models, whose parameter $\mu=|\lambda_y|/\lambda_x$
lies in the range $3/4<\mu<3/2$.

Our asymptotic approximation $\wt K^{(x)}(p_x,y) \sim \const\times
|p_x|^{-\mu/3}|y|^{-1/3}$ to the transformed prefactor~$\wt K^{(x)}$
at~criticality, derived in Section~\ref{subsec:nascent1}, can also be
numerically tested.  The approximation should be valid in the vicinity of
the $y$-axis separatrix, \ie, as~$p_x\to0$ at~fixed, nonzero~$y$.  There
are two separate cases:  $3/4<\mu<3/2$,~when a caustic is present,
and~$\mu\ge3/2$, when one is~not.  For simplicity we consider only the
latter.  When~$\mu\ge3/2$, it~follows from~(\ref{eq:star}) that
$x(p_x,y)\approx -p_x/2\lambda_x$ as~$p_x\to0$; the nonpolynomial term is
subdominant.  So~our asymptotic approximation to~$\wt K^{(x)}$ implies that
(cf.~(\ref{eq:curioustransform}))
\begin{eqnarray}
K(x,y) &\propto& \wt K^{(x)}(p_x,y) \sqrt{- \dfrac{\partial^2 W}{\partial
x^2} (p_x,y)} \\
&=&\wt K^{(x)}(p_x,y) \sqrt{- \dfrac{\partial p_x}{\partial x} (p_x,y)} \\
&\approx& \const\times \left|x\right|^{-\mu/3} \left|y\right|^{-1/3},
\end{eqnarray}
as~$x\to0$ at~fixed, nonzero~$y$.  This comparatively slow power-law
divergence as the separatrix is approached at (small) nonzero~$y$ is to be
contrasted with the $K(x,0)\sim\const\times |x|^{-\mu}$ divergence that
occurs when the saddle point is approached along the $x$-axis.
(See~(\ref{eq:z2}).)  It~is susceptible to numerical test.

\begin{figure}
\begin{center}
\epsfxsize=2.75in		
\leavevmode\epsfbox{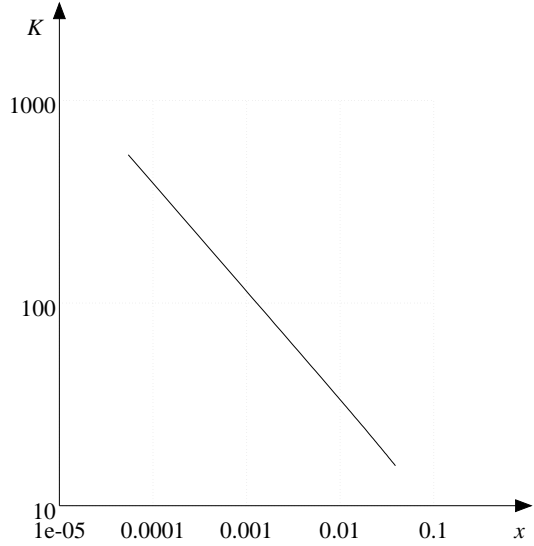}
\end{center}
\caption{Behavior of the WKB prefactor near the $y$-axis separatrix, in a
critical version of the $\mu=1.6$ standard double well model.
$K=K(x,y=0.05)$ is plotted on a logarithmic scale, revealing the scaling
behavior ($K\propto x^{-\mu/3}$ at nonzero~$y$).}
\label{fig:kdecay}
\end{figure}

In Fig.~\ref{fig:kdecay} we graph $K=K(x,y)$ as a function of~$x$,
at~$y=0.05$, for the critical version of the standard double well model
with~$\mu=1.6$.  (This is the same model whose instanton trajectories are
shown in Fig.~\ref{fig:fourpart}(d).)  The curve is fitted to high accuracy
by a power-law $\const\times x^{-0.533}$, \ie, $\const\times x^{-\mu/3}$.
This confirms the prediction of our scaling theory.  No~sub-scaling regime
is evident at small~$|x|$.  Similar plots can be obtained for the
near-separatrix behavior of~$K$ in critical models with other values
of~$\mu$.

We conclude that the asymptotic approximations to $\wt W^{(x)}=\wt
W^{(x)}(p_x,y)$ and $\wt K^{(x)}=\wt K^{(x)}(p_x,y)$ derived from our
scaling theory have a wide domain of validity, and may be employed in the
Maslov-WKB method.

\subsection{Scaling Beyond the WKB Approximation}
\label{subsec:nascent2}
We can now compute the Maslov-WKB boundary layer approximations to the
stationary density~$\rho_0$ and the quasistationary density~$\rho_1$ near
the nascent cusp at the saddle point, where the conventional WKB
approximation breaks down.  The boundary layer approximations are
determined by (\ref{eq:newWapprox}) and~(\ref{eq:newKapprox}), the
asymptotic approximations to the Legendre-transformed action~$\wt W^{(x)}$
and the transformed prefactor~$\wt K^{(x)}$ respectively.  As~discussed
above, these approximations should be valid when the eigenvalue ratio
$\mu\defeq |\lambda_y|/\lambda_x$ satisfies $3/4<\mu<3$.

Substituting (\ref{eq:newWapprox}) and~(\ref{eq:newKapprox})
into the one-dimensional diffraction integral~(\ref{eq:alternativex})
yields the rather complicated expression
\begin{eqnarray}
\label{eq:mess2}
&&K(x,y)\exp[-W(x,y)/\eps]\\
&&\qquad\sim \eps^{-1/2} \int \wt K^{(x)}(p_x,y)
\exp \left\{ \left[ -xp_x + \wt W^{(x)}(p_x,y)\right]/\eps \right\}\,dp_x\nonumber\\
&&\qquad\approx\const
\times \epsilon^{-1/2} e^{-W(0,0)/\epsilon}|y|^{-1/3}
e^{-|\lambda_y|y^2/\eps}\nonumber\\
&&\qquad\qquad\times  
\int |p_x|^{-\mu/3} \exp\left\{-\left.\left[ \dfrac{p_x^2}{4\lambda_x}
 - C_{p_x,y}y^{4/3}|p_x|^{4\mu/3} + xp_x \right]\right/\eps\right\}\,    dp_x
\nonumber
\end{eqnarray}
which requires a bit of explanation.  The first problem to be resolved is
the lengthscale near~$(x,y)=(0,0)$ on~which this diffraction integral
defines a valid Maslov-WKB approximation.  The $p_x^2$ and~$xp_x$ terms in
the argument of~$\exp(\cdot)$ are $O(1)$ when $p_x^2$ and~$xp_x$
are~$O(\eps)$; \ie, when $x$ and~$p_x$ are~$O(\eps^{1/2})$.  The term
$C_{p_x,y}y^{4/3}|p_x|^{4\mu/3}/\epsilon$ is $O(1)$ when, also,
$y=O(\eps^{3/4-\mu/2})$.  This is the case when~$\mu<3/2$, at~least.
If~$\mu\ge3/2$ then the $C_{p_x,y}y^{4/3}|p_x|^{4\mu/3}/\epsilon$ term is
negligible whenever $y=o(1)$.  We~conclude that in the weak-noise
($\eps\to0$) limit of models with~$\mu<3/2$, the diffraction
integral~(\ref{eq:mess2}) defines a valid Maslov-WKB approximation on the
$x=O(\eps^{1/2})$, $y=O(\eps^{3/4-\mu/2})$ lengthscale near the saddle
point.  This is precisely the caustic lengthscale
$x=O(|y|^{(3/2-\mu)^{-1}})$, \ie, $p_x=O(|y|^{(3/2-\mu)^{-1}})$, of~the
last section.  If~$\mu\ge 3/2$, so~that no~caustic is present, then
$y=O(\eps^{3/4-\mu/2})$ must be replaced by~$y=o(1)$.  On~the appropriate
lengthscale, the diffraction integral defines a {\em noncanonical\/}
diffraction function.

The diffraction integral, being one-dimensional, cannot resolve the
singularity at~$y=0$.  This is because it~does not include an integration
over~$p_y$.  So~one cannot expect the Maslov-WKB approximation to be valid
at arbitrarily small~$y$.  The stationary and quasistationary densities in
critical models are expected to be tightly concentrated on the
$y=O(\eps^{1/2})$ transverse lengthscale near the saddle point, as we saw
in (\ref{eq:nearH})--(\ref{eq:erf}) (which apply in the absence of
focusing).  At~most, the Maslov-WKB approximation will be valid in the {\em
far~field\/} of the $y=O(\epsilon^{1/2})$ lengthscale, where the factor
$e^{-\left|\lambda_y\right|y^2/\eps}$ is exponentially small.  So~it will not be
directly comparable to (\ref{eq:nearH})--(\ref{eq:erf}).  But it proves to
be very useful nonetheless.  Define `stretched' variables $X\defeq x/\eps^{1/2}$
and~$Y\defeq y/\eps^{1/2}$; also, change the integration variable
to~$P_x=p_x/\eps^{1/2}$.  The approximation becomes
\begin{equation}
\const \times \epsilon^{-(\mu+1)/6} e^{-W(0,0)/\epsilon}\left|Y\right|^{-1/3}
e^{-\left|\lambda_y\right|Y^2} 
\int \left|P_x\right|^{-\mu/3} e^{-XP_x} e^{-P_x^2/4\lambda_x}
\,dP_x,
\label{eq:mess3}
\end{equation}
since when $p_x,y=O(\epsilon^{1/2})$, the term
$C_{p_x,y}y^{4/3}\left|p_x\right|^{4\mu/3}/\eps$ is negligible.

At~this point we must explain how to interpret the integration over~$P_x$,
or~$p_x$.  The stationary density
$\rho_0(x,y)=\rho_0(X\eps^{1/2},Y\eps^{1/2})$ must be even in~$X$, and the
quasistationary density $\rho_1(x,y)=\rho_1(X\eps^{1/2},Y\eps^{1/2})$ must
be odd.  To~get approximations with these symmetry properties, we may
integrate~$P_x$ from $-\infty$~to~$\infty$ and $0$~to~$\infty$
respectively.  We~remarked at the end of Section~\ref{sec:Maslov} that
performing a half-range integration is one way of incorporating an
antisymmetry constraint, and that is the technique we shall use.

As~summarized in Abramowitz and Stegun (Ref.~\cite{Abramowitz65},~\S19.5),
definite integrals resembling~(\ref{eq:mess3}) define {\em parabolic
cylinder functions\/}.  Evaluating the integral~(\ref{eq:mess3}), with the
two possible choices of the range of integration (full~range and
half~range), yields the  Maslov-WKB approximations
\begin{eqnarray}
\label{eq:tired1}
\rho_0(x,y) & \sim&  \const   \times 
\eps^{-(\mu+1)/6}F_0(\lambda_x^{1/2} x/\epsilon^{1/2})
e^{+\lambda_xx^2/\epsilon}\left|y/\eps^{1/2}
\right|^{-1/3}e^{-\left|\lambda_y\right|y^2/\epsilon},\\
\rho_1(x,y) & \sim&  \const   \times 
\eps^{-(\mu+1)/6}F_1(\lambda_x^{1/2} x/\epsilon^{1/2})
e^{+\lambda_xx^2/\epsilon}\left|y/\eps^{1/2}
\right|^{-1/3}e^{-\left|\lambda_y\right|y^2/\epsilon}
\label{eq:tired2}
\end{eqnarray}
to the stationary and quasistationary probability densities, on the
$O(\eps^{1/2})$ lengthscale near the saddle point.  Here the so-called
boundary layer functions~$F_i=F_i(Z)$, where $Z\defeq\lambda_x^{1/2}X =
\lambda_x^{1/2}x/\eps^{1/2}$, are defined by
\begin{equation}
F_i(Z) \defeq y_{i+1}(1/2-\mu/3,2^{1/2}Z)e^{-Z^2/2}, 
\end{equation}
in the notation of Abramowitz and Stegun.
$y_1(1/2-\mu/3,\bullet)$~and~$y_2(1/2-\mu/3,\bullet)$ are even and odd
parabolic cylinder functions, respectively.  We~could equally well define
the boundary layer
functions~$F_i$ in~terms of an {\em Hermite function of non-integer
index\/}, by
\begin{equation}
\label{eq:arep}
F_0(Z)\defeq \left[H_{(\mu/3)-1}(Z) e^{-Z^2}\right]_{\mtext{\scriptsize 
even}},\qquad
F_1(Z)\defeq \left[H_{(\mu/3)-1}(Z) e^{-Z^2}\right]_{\mtext{\scriptsize odd}}.
\end{equation}
Here $[\cdot]_{\mtext{\scriptsize even}}$ and~$[\cdot]_{\mtext{\scriptsize 
odd}}$ signify even and odd
parts, under the reflection~$Z\mapsto -Z$.  The definitions~(\ref{eq:arep})
are meaningful whenever the index $n\defeq (\mu/3)-1$ is not an integer,
in~which case the Hermite function $H_n(Z)$ is not a conventional Hermite
polynomial, and is neither even nor~odd.  But since we are assuming
$3/4<\mu<3$, this is always the case. 
Irrespective of the choice of definitions,
\begin{eqnarray}
F_0(Z) &\sim& \const\times |Z|^{-\mu/3},\nonumber\\
F_1(Z) &\sim& \const\times |Z|^{-\mu/3}\sgn Z,
\label{eq:bdyas}
\end{eqnarray}
as~$Z\to \pm\infty$.  (Cf.~Ref.~\cite{Abramowitz65},~\S19.8.)

We~stress that the Maslov-WKB approximations to $\rho_0$ and~$\rho_1$ are
strictly valid only in the {\em transverse far~field\/}, \ie, as
$Y=y/\epsilon^{1/2}\to\pm\infty$.  But they make very clear how critical
double well models differ from non-critical double~well models.
By~comparing (\ref{eq:tired1})--(\ref{eq:tired2}) with
(\ref{eq:nearH})--(\ref{eq:erf}), we see that at~criticality, the boundary
layer functions $F_0(\cdot)$~and~$F_1(\cdot)$ replace the boundary layer
functions $1$~and $\erf(\cdot)$ respectively.  The approximations
(\ref{eq:tired1})--(\ref{eq:tired2}) are guaranteed to match to the
standard WKB approximation $K(\ov x)\exp[-W(\ov x)]$, as~one moves in a
transverse direction away from the saddle point.  For~example, the
$|Z|^{-\mu/3}$ falloff of eqs.~(\ref{eq:bdyas}) will match to the
$|x|^{-\mu/3}$ prefactor falloff of eqs.\ (\ref{eq:bizarre})
and~(\ref{eq:815b}), which is seen, \eg, in Figure~\ref{fig:kdecay}.

The Maslov-WKB approximations to $\rho_0$ and~$\rho_1$, and the
nonpolynomial normal form for the Legendre-transformed action that
engendered them, have several striking consequences for double well models
at criticality.
\begin{itemize}
\item A nongeneric caustic, emerging sideways from the nascent cusp at the
saddle.  In~Section~\ref{subsec:nascent1} we~predicted from the
nonpolynomial normal form for~$\wt W$ that when $3/4<\mu<3/2$, a~caustic is
located at~$|x|\simlt\const\times |y|^{(3/2-\mu)^{-1}}$.  Our prediction
was confirmed by Figure~\ref{fig:fourpart}.  This caustic has an
unusual (continuously varying) exponent.  It~is {\em nongeneric\/}, in the
sense of singularity theory.
\item An unusual (continuously varying) singularity index.  As~$\eps\to0$,
the falloff of the stationary density~$\rho_0$ at the saddle point~$(0,0)$
is not pure exponential, on~account of the $\eps^{-(\mu+1)/6}$ prefactor in
the Maslov-WKB approximation~(\ref{eq:tired1}).  This is interpreted as a
statement that the nascent cusp has singularity index $s=s(\mu)=(\mu+1)/6$,
as~mentioned in Section~\ref{sec:bifurcation}.  It~too is nongeneric, in
the sense of singularity theory.
\item Non-Arrhenius MFPT asymptotics.  If~one computes the rate at~which
the quasistationary density~$\rho_1$ is absorbed on the separatrix near the
saddle, the $\eps^{-(\mu+1)/6}$ prefactor in the Maslov-WKB
approximation~(\ref{eq:tired2}) will appear in the $\eps\to0$ asymptotics.
Equivalently, the exponentially decaying eigenvalue
$\lambda_1=\lambda_1(\eps)$ of the Smoluchowski operator will have an
asymptotic $\eps^{-(\mu+1)/6}$ prefactor, as~well as
the usual  Arrhenius factor [\ie,~$\exp(-\Delta W/\eps)$].
And the MFPT will be asymptotic
to $\const\times \eps^{+(\mu+1)/6}\exp(+\Delta W/\eps)$, as~$\eps\to0$.
At~criticality, the weak-noise growth of the MFPT is {\em slower than pure
exponential\/}.
\item A non-Gaussian limiting exit location distribution.  In~the absence
of MPEP bifurcation, for a symmetric double well model the location of the
point of exit from either of the two wells would have an asymptotic
Gaussian distribution, on the transverse $O(\eps^{1/2})$ lengthscale near
the saddle.  In~fact, its density would fall~off
as~$e^{-\left|\lambda_y\right| y^2/\eps}$.  We~see from the Maslov-WKB
approximation to~$\rho_1$ that at~criticality, the exit location density on
the separatrix
includes {\em scaling corrections\/}.  In~the transverse far~field it~falls
off as $|y|^{-1/3}e^{-\left|\lambda_y\right| y^2/\eps}$, rather
than~$e^{-\left|\lambda_y\right| y^2/\eps}$.  It~has a {\em non-Gaussian
tail\/}.
\end{itemize}
These phenomena look natural from the point of view of the theory of
critical phenomena, though the stochastic escape problem has not previously
been considered from that point of view.

\section{{Discussion}}
\label{sec:discussion}
We can now step back and review our results.  We~began with a WKB treatment
of the weak-noise asymptotics of stationary (and quasistationary) solutions
of the Smoluchowski equation.  The WKB analysis led to instanton
trajectories, which have a physical interpretation as most probable
weak-noise fluctuational paths.  The instanton trajectories turned~out to
be zero-energy trajectories of an associated Hamiltonian dynamical system.
This is because the phase~space versions of the instanton trajectories
(\ie,~WKB bicharacteristics) trace~out a Lagrangian manifold in phase
space.  In~double well models the onset of bifurcation is associated with
the fleeting appearance of an unusual singularity (a~nascent cusp) in the
shape of this manifold, as the parameters of the model are varied.

There is a formal analogy between the Lagrangian manifold of a dynamical
system perturbed by weak noise, and the thermodynamic surface of a
condensed matter system.  This analogy led~us to construct a scaling theory
of the shape of the Lagrangian manifold near the nascent cusp.  To~date,
most work on Lagrangian manifolds has assumed that they are {\em smooth\/},
and that any apparent singularities in their shape can be transformed away
by a change of coordinates.  This is analogous to assuming that
thermodynamic surfaces are real analytic, and that non-analyticities in
thermodynamic behavior (\ie,~phase transitions) can be transformed away by
working in~terms of the appropriate thermodynamic potential.  Equivalently,
it~is analogous to assuming that all phase transitions have classical
critical exponents.  Our scaling theory makes it~clear that the nascent
cusp singularity is a genuine {\em point of non-smoothness\/} of the
Lagrangian manifold.  In~thermodynamic terms, it~has nonclassical, indeed
continuously varying, critical
exponents.

\begin{figure}
\begin{center}
\epsfxsize=3.75in		
\leavevmode\epsfbox{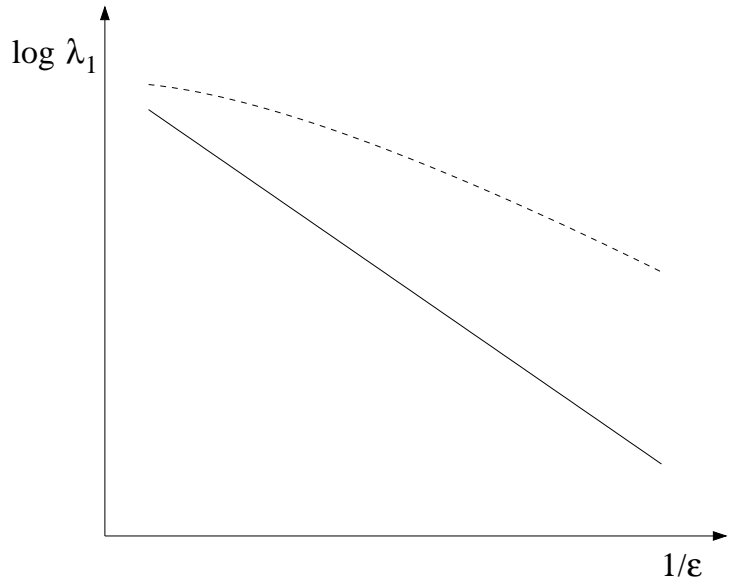}
\end{center}
\caption[A~sketch, on a logarithmic scale, of the rate of noise-activated
inter-well hopping~$\lambda_1$ as a function of the reciprocal noise
strength~$1/\eps$.]{A~sketch, on a logarithmic scale, of the rate of
noise-activated inter-well hopping~$\lambda_1$ as a function of the
reciprocal noise strength~$1/\eps$.  Off~criticality (solid curve),
$\lambda_1$~displays a pure exponential falloff as~$\eps\to0$, \ie,
$\lambda_1\sim\const\times\exp(-\Delta W/\eps)$.  At~the onset of
bifurcation (dashed curve), $\lambda_1\sim\const\times\eps^{-s}\exp(-\Delta
W/\eps)$ instead, where $s$~is the singularity index of the nascent cusp
appearing at the saddle point.}
\label{fig:arrhenius}
\end{figure}

Applying the Maslov-WKB method to the nascent cusp yielded several
interesting predictions, which we summed~up in the four bulleted items at
the end of the last section.  One normally expects that in a double well
system perturbed by weak noise of strength~$\eps$, the rate of inter-well
hopping $\lambda_1=\lambda_1(\eps)$ will be asymptotic to a {\em
constant\/} multiple of the Arrhenius factor $\exp(-\Delta W/\eps)$, where
$\Delta W$~is an effective barrier height.  Also, one expects that the
distribution of exit locations (from either well) will asymptotically become
a Gaussian of $O(\eps^{1/2})$ standard deviation, centered on the saddle
point between the two wells.  The Maslov-WKB method predicts that 
at~criticality, both these phenomena are strongly altered.
In~particular, the factor $\exp(-\Delta W/\eps)$ must be replaced by
$\eps^{-s}\exp(-\Delta W/\eps)$, where $s=(\mu+1)/6$ is the singularity index
of the nascent cusp.  (As~we noted in Section~\ref{subsec:MAE25}, the
singularity index is a sort of critical exponent.)
  In~Fig.~\ref{fig:arrhenius} we sketch an Arrhenius
plot, showing this anomalous (non-Arrhenius) behavior.

In mathematical terms, the nascent cusp appearing at the onset of
bifurcation is a {\em nongeneric\/} singularity, \ie, a~singularity
different from any of the now classical singularities of catastrophe
theory.  As~shown in Fig.~\ref{fig:nascent}, in~many double well models
it~induces an unusual caustic in the flow field of instanton trajectories.
This caustic is itself nongeneric, in~that its exponent is not equal
to~$3/2$.  As~we have seen (see,~\eg, Fig.~\ref{fig:hysteresis}),
its~presence quantitatively confirms the validity of our scaling theory.
It~is remarkable that such nongeneric phenomena are a {\em generic\/}
feature of singly parametrized symmetric double~well models.

At~least as developed in this paper, our scaling theory is a scaling theory
of weak-noise behavior near the nascent cusp, precisely at~criticality.
It~would be useful to treat as~well models that are nearly critical, but
not exactly~so.  Such models should display a crossover from non-Arrhenius
behavior to Arrhenius behavior at sufficiently weak noise strength.
By~developing a {\em joint scaling theory\/}, one~of the variables in~which
measures the distance from criticality, it~should be possible to analyse
this phenomenon.  We~expect that it is possible to derive a `Ginzburg
criterion'~\cite{Ma76}, expressing how~close to criticality any given
double well model should~be, for the non-Arrhenius behavior of
Fig.~\ref{fig:arrhenius} to be visible.  Work on~this is under way.

We briefly mention two geometric features of models `off criticality' that
cry~out for a theoretical explanation.  A~nongeneric caustic appears near
the separatrix not~only at~criticality, but in many non-critical models
as~well~\cite{Dykman94}.  Also, as~criticality is approached (\eg,~as
$\alpha\to\alpha_c^-$ in the standard model of~(\ref{eq:standard})),
it~frequently happens that the nascent cusp is formed by a collision of two
generic cusps, which move along the separatrix toward the saddle point.
These phenomena can presumably be explained by an appropriate {\em joint
unfolding\/}, but that is for the future.

We close by mentioning a possible extension of a more theoretical~sort.
In~this paper we have focused exclusively on the asymptotic solutions of
the time-independent weak-noise Smoluchowski equation. There is reason to
believe that nongeneric singularities resembling the nascent cusp can
occur, and are perhaps even widespread, in~the asymptotic solutions of
other singularly perturbed elliptic partial differential equations.  Most
WKB treatments of singularly perturbed elliptic PDE's (see,~\eg,
Duistermaat~\cite{Duistermaat74}) assume that each WKB characteristic
(\ie,~instanton trajectory) eventually leaves any bounded region of space.
This assumption is violated in the Smoluchowski equation for any
double~well model, since the MPEP(s) terminate on the saddle point, rather
than extending to infinity.  We~expect that when it is violated in other
PDE's, analogous nongeneric singularities in formal asymptotic solutions
can occur.  The nongeneric singular phenomena that we have seen in this
paper may simply be representatives of a larger class.


\end{document}